\documentclass[12pt]{article}
\usepackage{amsmath,amssymb,amsfonts,color,graphicx,cite,color,feynarts}
\usepackage{epsfig,psfrag,rotating,soul}
\usepackage{rotfloat}
\input paperdef

\graphicspath{{figs/}}

\evensidemargin \oddsidemargin
\allowdisplaybreaks

\begin{document}
\thispagestyle{empty}

\def\thefootnote{\fnsymbol{footnote}}

\begin{flushright}
IFT-UAM/CSIC-14-045 \\
FTUAM-14-18

\end{flushright}

\vspace{0.5cm}

\begin{center}

{\large\sc {\bf 
Updated Constraints on General Squark Flavor Mixing}}

\vspace{1cm}

{\sc
M.~Arana-Catania$^{1}$%
\footnote{email: miguel.arana@uam.es}%
, S.~Heinemeyer$^{2}$%
\footnote{email: Sven.Heinemeyer@cern.ch}%
~and  M.J.~Herrero$^{1}$%
\footnote{email: maria.herrero@uam.es}%
}

\vspace*{.7cm}

{\sl
$^1$Departamento de F\'isica Te\'orica and Instituto de F\'isica Te\'orica,
IFT-UAM/CSIC\\
Universidad Aut\'onoma de Madrid, Cantoblanco, Madrid, Spain

\vspace*{0.1cm}

$^2$Instituto de F\'isica de Cantabria (CSIC-UC), Santander, Spain

}

\end{center}

\vspace*{0.1cm}

\begin{abstract}
\noindent
We explore the phenomenological implications on non-minimal flavor
violating (NMFV) processes from squark flavor mixing within the Minimal
Supersymmetric Standard Model. We work under the model-independent
hypothesis of general flavor mixing in the squark sector, being
parametrized by a complete set of dimensionless $\deABij$ ($A,B=L,R$;
$i,j=u,c,t$ or $d,s,b$; $i\neq j$) parameters. The present upper
bounds on the most relevant 
NMFV processes, together with the requirement of compatibility in the
choice of the MSSM parameters with the recent LHC and $(g-2)_\mu$ data,
lead to updated constraints on all squark flavor mixing
parameters. 

\end{abstract}

\def\thefootnote{\arabic{footnote}}
\setcounter{page}{0}
\setcounter{footnote}{0}

\newpage


\section{Introduction}

Non-Minimal Flavor Violating (NMFV) processes in the scalar quark sector
of the Minimal Supersymmetric Standard Model
(MSSM)~\cite{mssm,Haber:1989xc,Gunion:1984yn,Gunion:1986nh}, provide 
important probes  to new physics involving non-vanishing flavor
mixing between the three generations. Within the Standard Model (SM),
the only source of flavor violation comes from the CKM matrix, 
$V_{\rm CKM}$, and thus
in general leads to small contributions. 
Within the MSSM there are clear candidates to produce
flavor mixings with important phenomenological implications.
The possible presence of soft Supersymmetry
(SUSY)-breaking parameters in the squark sector, which are
off-diagonal in flavor space (mass parameters as well as trilinear
couplings) are
the most general way to introduce squark flavor mixing within the
MSSM. The off-diagonality in the squark mass matrix reflects the   
misalignment (in flavor space) between quark and squark mass
matrices, that cannot be diagonalized 
simultaneously. This misalignment can be produced from various
origins, but we will not rely on any particular one in this work. 
For instance, these off-diagonal squark mass matrix entries can be
generated by renormalization effects from the CKM 
matrix, which can be obtained by means of the Renormalization
Group Equations 
(RGE) running from a high energy scale, where gauge coupling unification
is achieved, down to the low energies where the NMFV effects
are explored.

In this work we will not investigate the possible dynamical origin
of this squark-quark misalignment, nor the particular predictions for
the off-diagonal squark soft SUSY-breaking mass terms in specific
SUSY models, but instead we parametrize the general
non-diagonal entries in the squark mass matrices
in terms of generic soft SUSY-breaking terms, and we explore
here their phenomenological implications on various precision
observables. In 
particular, we explore the consequences of these general squark mass matrices
on the light MSSM Higgs boson mass, $\Mh$, as well as on the three most
prominent $B$-physics observables, \bsg, \bmm\ and \dmbs. Specifically, we 
parametrize the non-diagonal squark mass matrix entries in terms of a
complete set of generic dimensionless parameters, $\deABij$ ($A,B=L,R$;
$i,j=u,c,t$ or $d,s,b$) where $L,R$ refer to the 
``left-'' and ``right-handed'' SUSY partners of the corresponding
quark degrees of freedom
and $i,j$ ($i \neq j$) are the involved generation indexes. For the
presentation of our theoretical framework and notation we follow
closely our previous work~\cite{mhNMFV} on this same subject, which
was done previous to the Higgs discovery.

The main aspect of this work is setting updated bounds on the
allowed values of the $\deABij$'s in this
model-independent parametrization of general squark
flavor mixing.
In particular, this is done in view of the collected data at
LHC\cite{LHCHiggslast,LHCSusy}, which has 
provided very important information and constraints for the MSSM,
including the absence of SUSY particle experimental signals and
the discovery of a Higgs boson with a mass close to $125 - 126 \gev$.
We work consistently in MSSM scenarios that are compatible with
LHC data. It should be noted that the analyzed scenarios have relatively
heavy SUSY spectra, which are naturally in
agreement with the present MSSM particle mass bounds (although
substantially lower masses, especially in the electroweak sector, are
allowed by LHC data). Furthermore the analyzed scenarios are chosen such
that the light $\cp$-even MSSM Higgs mass is around $125 - 126 \gev$ and
thus in agreement with the Higgs boson discovery~\cite{LHCHiggs}.
In addition we
require that our selected MSSM scenarios give a
prediction for the muon anomalous magnetic moment, $(g-2)_\mu$, in
agreement with current data~\cite{Bennett:2006fi}.

The paper is organized as follows: first we review the main features of
the MSSM with general squark flavor mixing and set the relevant
notation for the $\deABij$'s in ~\refse{sec:nmfv}. The description
of the numerical scenarios that we choose here is also done in this
section. The selection of 
relevant precision observables and flavor observables 
we are working with are
presented in ~\refse{sec:obs}. A summary on the present
experimental bounds on NMFV, that will be used in our analysis are also
included in this section.  \refse{sec:results} contains the main results
of our numerical analysis and present the updated constraints found on the
$\deABij$'s. Our conclusions are summarized in
\refse{sec:conclusions}.


\section{Calculational basis for Non-Minimal Flavor Violation}
\label{sec:nmfv}

\subsection{Theoretical set-up}

We work in SUSY scenarios with the same particle content as the
MSSM, but with 
general flavor mixing hypothesis in the squark sector. Within these
SUSY-NMFV scenarios, besides the usual flavor violation originated by
the CKM matrix of the quark sector, the  general flavor mixing in the
squark mass matrices additionally generates flavor violation from the
squark sector. 
These squark flavor mixings are usually described in terms of a set of
dimensionless parameters $\deXYij$ ($X,Y=L,R$; $i,j=u,c,t$ or $d,s,b$).
In this section we summarize the main features of the squark flavor
mixing within the SUSY-NMFV scenarios and set the notation. 
The more theoretical background, including the derivation from the super
potential can be found in \citere{mhNMFV}.

The usual procedure to introduce general flavor mixing in the squark
sector is to include the non-diagonality in flavor space once the quarks
have been rotated to the physical basis, 
namely, in the so-called Super-CKM basis. Thus, one usually writes
the $6\times 6$ non-diagonal mass matrices, $\cM_{\tilde u}^2$
and $\cM_{\tilde d}^2$, referred to the Super-CKM basis, being ordered
respectively as $(\SupL, \SchaL, \StopL, \SupR, \SchaR, \StopR)$ and
$(\SdownL, \SstrL, \SbotL, \SdownR, \SstrR, \SbotR)$, and write them in
terms of left- and right-handed blocks $M^2_{\tilde q \, AB}$ 
($\tilde q= \tilde u, \tilde d;$ $A,B=L,R$), which are
non-diagonal $3\times 3$ matrices,  
\begin{equation}
\cM_{\tilde q}^2 =\left( \begin{array}{cc}
M^2_{\tilde q \, LL} & M^2_{\tilde q \, LR} \\[.3em] 
M_{\tilde q \, LR}^{2 \, \dagger} & M^2_{\tilde q \,RR}
\end{array} \right), \qquad \tilde q= \tilde u, \tilde d~,
\label{eq:blocks-matrix}
\end{equation} 
 where:
 \begin{alignat}{5}
 M_{\tilde u \, LL \, ij}^2 
  = &  m_{\tilde U_L \, ij}^2 + \left( m_{u_i}^2
     + (T_3^u-Q_u\sin^2 \theta_W ) M_Z^2 \cos 2\beta \right) \delta_{ij},  \notag\\
 M^2_{\tilde u \, RR \, ij}
  = &  m_{\tilde U_R \, ij}^2 + \left( m_{u_i}^2
     + Q_u\sin^2 \theta_W M_Z^2 \cos 2\beta \right) \delta_{ij} \notag, \\
  M^2_{\tilde u \, LR \, ij}
  = &  \left< \cH_2^0 \right> {\cal A}_{ij}^u- m_{u_{i}} \mu \cot \beta \, \delta_{ij},
 \notag, \\
 M_{\tilde d \, LL \, ij}^2 
  = &  m_{\tilde D_L \, ij}^2 + \left( m_{d_i}^2
     + (T_3^d-Q_d \sin^2 \theta_W ) M_Z^2 \cos 2\beta \right) \delta_{ij},  \notag\\
 M^2_{\tilde d \, RR \, ij}
  = &  m_{\tilde D_R \, ij}^2 + \left( m_{d_i}^2
     + Q_d\sin^2 \theta_W M_Z^2 \cos 2\beta \right) \delta_{ij} \notag, \\
  M^2_{\tilde d \, LR \, ij}
  = &  \left< \cH_1^0 \right> {\cal A}_{ij}^d- m_{d_{i}} \mu \tb \, \delta_{ij}~,
\label{eq:SCKM-entries}
\end{alignat}
with, $i,j=1,2,3$, $Q_u=2/3$, $Q_d=-1/3$, $T_3^u=1/2$ and
$T_3^d=-1/2$. 
$\sin^2\theta_W = 1 - \MW^2/\MZ^2$
with $M_{W,Z}$ denoting the masses of the $W$~and $Z$~boson mass,
respectively, 
and $(m_{u_1},m_{u_2}, m_{u_3})=(m_u,m_c,m_t)$, $(m_{d_1},m_{d_2},
m_{d_3})=(m_d,m_s,m_b)$. $\mu$ is the usual Higgsino mass term and
  $\tb=v_2/v_1$ 
with  $v_1=\left< \cH_1^0 \right>$ and $v_2=\left< \cH_2^0
\right>$ being the two vacuum expectation values of the corresponding
neutral Higgs boson in the Higgs $SU(2)_L$ doublets, 
$\cH_1= (\cH^0_1\,\,\, \cH^-_1)$ and $\cH_2= (\cH^+_2 \,\,\,\cH^0_2)$.

It should be noted that the non-diagonality in flavor comes from the
values of $m_{\tilde U_L \, ij}^2$, $m_{\tilde U_R \, ij}^2$, 
$m_{\tilde D_L \, ij}^2$, $m_{\tilde D_R \, ij}^2$, $\cA_{ij}^u$
and $\cA_{ij}^d$ for $i \neq j$.

The general squark flavor mixing is introduced via the
non-diagonal terms in the soft breaking squark mass matrices and
trilinear coupling matrices, which are defined here as:

\begin{equation}  
m^2_{\tilde U_L}= \left(\begin{array}{ccc}
 m^2_{\tilde Q_{1}} & \de_{12}^{LL} m_{\tilde Q_{1}}m_{\tilde Q_{2}} & 
 \de_{13}^{LL} m_{\tilde Q_{1}}m_{\tilde Q_{3}} \\
 \de_{21}^{LL} m_{\tilde Q_{2}}m_{\tilde Q_{1}} & m^2_{\tilde Q_{2}}  & 
 \de_{23}^{LL} m_{\tilde Q_{2}}m_{\tilde Q_{3}}\\
 \de_{31}^{LL} m_{\tilde Q_{3}}m_{\tilde Q_{1}} & 
 \de_{32}^{LL} m_{\tilde Q_{3}}m_{\tilde Q_{2}}& m^2_{\tilde Q_{3}} 
\end{array}\right)~,
\label{mUL}
\end{equation}
 
\noindent
\begin{equation}
m^2_{\tilde D_L}= V_{\rm CKM}^\dagger \, m^2_{\tilde U_L} \, V_{\rm CKM}~,
\label{mDL}
\end{equation}
 
\noindent 
\begin{equation}  
m^2_{\tilde U_R}= \left(\begin{array}{ccc}
 m^2_{\tilde U_{1}} & \de_{uc}^{RR} m_{\tilde U_{1}}m_{\tilde U_{2}} & 
 \de_{ut}^{RR} m_{\tilde U_{1}}m_{\tilde U_{3}}\\
 \de_{{cu}}^{RR} m_{\tilde U_{2}}m_{\tilde U_{1}} & m^2_{\tilde U_{2}}  & 
 \de_{ct}^{RR} m_{\tilde U_{2}}m_{\tilde U_{3}}\\
 \de_{{tu}}^{RR}  m_{\tilde U_{3}} m_{\tilde U_{1}}& 
 \de_{{tc}}^{RR} m_{\tilde U_{3}}m_{\tilde U_{2}}& m^2_{\tilde U_{3}} 
\end{array}\right)~,
\end{equation}

\noindent 
\begin{equation}  
m^2_{\tilde D_R}= \left(\begin{array}{ccc}
 m^2_{\tilde D_{1}} & \de_{ds}^{RR} m_{\tilde D_{1}}m_{\tilde D_{2}} & 
 \de_{db}^{RR} m_{\tilde D_{1}}m_{\tilde D_{3}}\\
 \de_{{sd}}^{RR} m_{\tilde D_{2}}m_{\tilde D_{1}} & m^2_{\tilde D_{2}}  & 
 \de_{sb}^{RR} m_{\tilde D_{2}}m_{\tilde D_{3}}\\
 \de_{{bd}}^{RR}  m_{\tilde D_{3}} m_{\tilde D_{1}}& 
 \de_{{bs}}^{RR} m_{\tilde D_{3}}m_{\tilde D_{2}}& m^2_{\tilde D_{3}} 
\end{array}\right)~,
\end{equation}

\noindent 
\begin{equation}
v_2 {\cal A}^u  =\left(\begin{array}{ccc}
 m_u A_u & \de_{uc}^{LR} m_{\tilde Q_{1}}m_{\tilde U_{2}} & 
 \de_{ut}^{LR} m_{\tilde Q_{1}}m_{\tilde U_{3}}\\
 \de_{{cu}}^{LR}  m_{\tilde Q_{2}}m_{\tilde U_{1}} & 
 m_c A_c & \de_{ct}^{LR} m_{\tilde Q_{2}}m_{\tilde U_{3}}\\
 \de_{{tu}}^{LR}  m_{\tilde Q_{3}}m_{\tilde U_{1}} & 
 \de_{{tc}}^{LR} m_{\tilde Q_{3}} m_{\tilde U_{2}}& m_t A_t 
\end{array}\right)~,
\label{v2Au}
\end{equation}

\noindent 
\begin{equation}
v_1 {\cal A}^d  =\left(\begin{array}{ccc}
 m_d A_d & \de_{ds}^{LR} m_{\tilde Q_{1}}m_{\tilde D_{2}} & 
 \de_{db}^{LR} m_{\tilde Q_{1}}m_{\tilde D_{3}}\\
 \de_{{sd}}^{LR}  m_{\tilde Q_{2}}m_{\tilde D_{1}} & m_s A_s & 
 \de_{sb}^{LR} m_{\tilde Q_{2}}m_{\tilde D_{3}}\\
 \de_{{bd}}^{LR}  m_{\tilde Q_{3}}m_{\tilde D_{1}} & 
 \de_{{bs}}^{LR} m_{\tilde Q_{3}} m_{\tilde D_{2}}& m_b A_b 
\end{array}\right)~.
\label{v1Ad}
\end{equation}

\noindent
In all this work, for simplicity, we are assuming that all $\deABij$
parameters are real, therefore, hermiticity of 
$\cM_{\tilde q}^2$ implies $\de_{ij}^{AB}= \de_{ji}^{BA}$.
It should be noted that  
we have used a common notation for the
$\del{LL}{ij}$'s with $i,j=1,2,3$ in the $\tilde U_L$ and $\tilde D_L$
sectors, due to the $SU(2)_L$ gauge invariance that relates 
$m^2_{\tilde U_L}$ and $m^2_{\tilde D_L}$ via $V_{\rm CKM}$, 
as given in \refeqs{mUL} and (\ref{mDL}).

\medskip
The next step is to rotate the squark states from the Super-CKM basis, 
${\tilde q}_{L,R}$, to the physical basis. 
If we set the order in the Super-CKM basis as above, 
$(\SupL, \SchaL, \StopL, \SupR, \SchaR, \StopR)$ and  
$(\SdownL, \SstrL, \SbotL, \SdownR, \SstrR, \SbotR)$, 
and in the physical basis as
${\tilde u}_{1,..6}$ and ${\tilde d}_{1,..6}$, respectively, these last
rotations are given by two $6 \times 6$ matrices, $R^{\tilde u}$ and
$R^{\tilde d}$,  
\BE
\VL  \tiu_{1} \\ \tiu_{2}  \\ \tiu_{3} \\
                                    \tiu_{4}   \\ \tiu_{5}  \\\tiu_{6}   \VR
  \; = \; R^{\tiu}  \VL \SupL \\ \SchaL \\\StopL \\ 
  \SupR \\ \SchaR \\ \StopR \VR ~,~~~~
\VL  \tid_{1} \\ \tid_{2}  \\  \tid_{3} \\
                                   \tid_{4}     \\ \tid_{5} \\ \tid_{6}  \VR             \; = \; R^{\tid}  \VL \SdownL \\ \SstrL \\ \SbotL \\
                                      \SdownR \\ \SstrR \\ \SbotR \VR ~,
\label{newsquarks}
\end{equation} 
yielding the diagonal mass-squared matrices as follows,
\BEA
{\rm diag}\{m_{\tiu_1}^2, m_{\tiu_2}^2, 
          m_{\tiu_3}^2, m_{\tiu_4}^2, m_{\tiu_5}^2, m_{\tiu_6}^2 
           \}  & = &
R^{\tiu}  \;  {\cal M}_{\tiu}^2   \; 
 R^{\tiu \dagger}    ~,\\
{\rm diag}\{m_{\tid_1}^2, m_{\tid_2}^2, 
          m_{\tid_3}^2, m_{\tid_4}^2, m_{\tid_5}^2, m_{\tid_6}^2 
          \}  & = &
R^{\tid}  \;   {\cal M}_{\tid}^2   \; 
 R^{\tid \dagger}    ~.
\EEA 

The corresponding Feynman rules in the physical basis for the vertices
including NMFV squarks had been 
implemented into the program packages
{\tt FeynArts}/{\tt FormCalc}~\cite{feynarts,formcalc} 
extending the previous MSSM model
file~\cite{famssm}. The Feynman rules of the NMFV MSSM that are
relevant for the present work can be found in \cite{mhNMFV}.


\subsection{Numerical scenarios}
\label{sec:scenarios}
Regarding our choice of MSSM parameters for our forthcoming numerical
analysis of the NMFV constraints, we have proceeded within two frameworks,
both compatible with present data, that we briefly describe in the following.  

\subsubsection{Framework 1}

In the first framework, we have selected six specific points in the MSSM
parameter space, S1, \ldots, S6, as examples of points that are allowed by
present data, including recent LHC searches and the measurements of the
muon anomalous magnetic moment. In \refta{tab:spectra} the
values of the various MSSM parameters as well as the values of the
predicted MSSM mass spectra are summarized, with all $\deABij = 0$.
They were evaluated with
the program \fh~\cite{feynhiggs,mhiggsAEC}. For simplicity, and to
reduce the number of 
independent MSSM input parameters we have assumed equal soft masses for
the squarks of the first and second generations (similarly for the
sleptons),  equal soft masses for the left and right squark sectors
(similarly for the sleptons, where $\tilde L$ denotes the
``left-handed'' slepton sector, whereas $\tilde E$ denotes
the ``right-handed'' charged slepton sector)
and also equal trilinear couplings for
the stop, $A_t$,  and sbottom squarks, $A_b$. In the slepton sector we
just consider the stau trilinear coupling, $A_\tau$. The other trilinear
sfermion couplings are set to zero value. Regarding the soft
SUSY-breaking parameters for the gaugino
masses, $M_i$ ($i=1,2,3$),  we assume an approximate GUT relation. The
pseudoscalar Higgs mass $\MA$, and the $\mu$ parameter are also taken as
independent input parameters. In summary, the six points 
S1, \ldots, S6  are
defined in terms of the following subset of ten input MSSM parameters
(plus the $\deABij$, which will be analyzed below):

\pagebreak

\BEA
m_{\tilde L_1} &=& m_{\tilde L_2} \; ; \; m_{\tilde L_3} \;  
(\mbox{with~} m_{\tilde L_{i}} = m_{\tilde E_{i}}\,\,,\,\,i=1,2,3)~, \non \\
m_{\tilde Q_1} &=& m_{\tilde Q_2} \; ; \; m_{\tilde Q_3} \; 
(\mbox{with~} m_{\tilde Q_i} = m_{\tilde U_i} = m_{\tilde D_i}\,\,,\,\,i=1,2,3)~,
                                                               \non \\
A_t&=&A_b\,\,;\,\,A_\tau~, \nonumber \\
M_2&=&2 M_1\, =\,M_3/4 \,\,;\,\,\mu \nonumber~, \\
\MA&\,\,;\,\, &\tb~.
\EEA

\begin{table}[h!]
\begin{tabular}{|c|c|c|c|c|c|c|}
\hline
 & S1 & S2 & S3 & S4 & S5 & S6 \\\hline
$m_{\tilde L_{1,2}}$& 500 & 750 & 1000 & 800 & 500 &  1500 \\
$m_{\tilde L_{3}}$ & 500 & 750 & 1000 & 500 & 500 &  1500 \\
$M_2$ & 500 & 500 & 500 & 500 & 750 &  300 \\
$A_\tau$ & 500 & 750 & 1000 & 500 & 0 & 1500  \\
$\mu$ & 400 & 400 & 400 & 400 & 800 &  300 \\
$\tb$ & 20 & 30 & 50 & 40 & 10 & 40  \\
$\MA$ & 500 & 1000 & 1000 & 1000 & 1000 & 1500  \\
$m_{\tilde Q_{1,2}}$ & 2000 & 2000 & 2000 & 2000 & 2500 & 1500  \\
$m_{\tilde Q_{3}}$  & 2000 & 2000 & 2000 & 500 & 2500 & 1500  \\
$A_t$ & 2300 & 2300 & 2300 & 1000 & 2500 &  1500 \\\hline
$m_{\tilde l_{1}}-m_{\tilde l_{6}}$ & 489-515 & 738-765 & 984-1018 & 474-802  & 488-516 & 1494-1507  \\
$m_{\tilde \nu_{1}}-m_{\tilde \nu_{3}}$& 496 & 747 & 998 & 496-797 & 496 &  1499 \\
$m_{{\tilde \chi}_1^\pm}-m_{{\tilde \chi}_2^\pm}$  & 375-531 & 376-530 & 377-530 & 377-530  & 710-844 & 247-363  \\
$m_{{\tilde \chi}_1^0}-m_{{\tilde \chi}_4^0}$& 244-531 & 245-531 & 245-530 & 245-530  & 373-844 & 145-363  \\
$M_{h}$ & 126.6 & 127.0 & 127.3 & 123.1 & 123.8 & 125.1  \\
$M_{H}$  & 500 & 1000 & 999 & 1001 & 1000 & 1499  \\
$M_{A}$ & 500 & 1000 & 1000 & 1000 & 1000 & 1500  \\
$M_{H^\pm}$ & 507 & 1003 & 1003 & 1005 & 1003 & 1502  \\
 $m_{\tilde u_{1}}-m_{\tilde u_{6}}$& 1909-2100 & 1909-2100 & 1908-2100 & 336-2000 & 2423-2585 & 1423-1589  \\
$m_{\tilde d_{1}}-m_{\tilde d_{6}}$ & 1997-2004 & 1994-2007 & 1990-2011 & 474-2001 & 2498-2503 &  1492-1509 \\
$m_{\tilde g}$ &  2000 & 2000 & 2000 & 2000 & 3000 &  1200 \\
\hline
\end{tabular}
\caption{
Selected points in the MSSM parameter space (upper part)
and their corresponding spectra (lower part), with all $\deABij = 0$. 
All mass parameters and trilinear couplings are given in GeV.} 
\label{tab:spectra}
\end{table}

The specific values of these ten MSSM parameters in \refta{tab:spectra},
to be used in the forthcoming NMFV analysis, are chosen to provide
different  
patterns in the various sparticle masses, but all leading to rather
heavy spectra, thus they are naturally in agreement with the
absence of SUSY signals at LHC. In particular  
all points lead to rather heavy squarks and gluinos above $1200\gev$ and
heavy sleptons above $500\gev$ (where the LHC limits would also
  permit substantially lighter scalar leptons). 
The values of $\MA$ within the interval
$(500,1500)\gev$, $\tb$ within the interval $(10,50)$ and a large
$A_t$ within $(1000,2500)\gev$ are fixed such that a light Higgs boson~$h$
within the LHC-favored range $(123, 128)\gev$ is obtained%
\footnote{
This range takes into account experimental uncertainties as well as
theoretical uncertainties, where the latter would permit an even larger
interval~\cite{mhiggsAEC,ehowp}. However, for the phenomenological
analyses later we will use a 
correspondingly wider range.}%
~in the Minimal Flavor Violation (MFV) limit.%
\footnote{
Here, by MFV limit we mean setting all $\deABij$'s to zero.}%
~It should also be noted that the large chosen values of $\MA \ge 500 \gev$
place the Higgs sector of our scenarios in the so called decoupling
regime~\cite{Haber:1989xc},  
where the couplings of~$h$ to gauge bosons and fermions are close to
the SM Higgs couplings, and the heavy~$H$ couples like the
pseudoscalar~$A$, and all heavy Higgs bosons are close in mass.
Increasing $\MA$  the heavy
Higgs bosons tend to decouple from low energy physics and the light~$h$
behaves like the SM Higgs boson. 
This type of MSSM Higgs sector seems
to be in good agreement with recent LHC
data~\cite{LHCHiggslast}. 
We have checked with the code {\tt HiggsBounds}~\cite{higgsbounds}
that the Higgs sector is in agreement with the LHC searches.
Particularly, the so far absence of gluinos at
LHC, forbids too low $M_3$ and, therefore, given the  assumed GUT
relation, forbids also a too low $M_2$. Consequently, the
values of $M_2$ and $\mu$ are fixed as to get gaugino masses compatible
with present LHC bounds. 
Finally, we have also required that all our points lead to a prediction
of the anomalous magnetic moment of the muon in the MSSM that can fill
the present discrepancy between the Standard Model prediction and the
experimental value. Specifically, we use \citeres{Bennett:2006fi} and 
\cite{Davier:2010nc} to extract the size of this discrepancy,
see also \citere{gm2-Jegerlehner}:
\begin{equation}
(g-2)_\mu^{\rm exp}-(g-2)_\mu^{\rm SM}= (30.2 \pm 9.0) \times 10^{-10}.
\label{gminus2}  
\end{equation}
We then require that the SUSY contributions from charginos and
neutralinos in the MSSM to one-loop level,  
$(g-2)_\mu^{\rm SUSY}$,  be within the interval defined by $3 \sigma$
around the central value in \refeq{gminus2}, namely:     
\begin{align}
 (g-2)_\mu^{\rm SUSY} &\in  (3.2 \times 10^{-10},57.2 \times 10^{-10}) ~.
\label{gminus2interval}  
\end{align} 


\subsubsection{Framework 2}
\label{sec:f2}

In the second framework, several possibilities for the
MSSM parameters have been considered, leading to simple patterns
of SUSY masses with specific 
relations among them and where the number of input parameters is strongly
reduced.  As in framework 1, the scenarios selected in
framework~2 lead to predictions of $(g-2)_\mu$ and
$\Mh$ (for all deltas equal to zero) that are compatible with present data
over a large part of the parameter space. To simplify the analysis of
the limits of the deltas, we will focus in scenarios where the
mass scales of the SUSY QCD sector that are relevant for the NMFV
processes  are all set  
relative to one mass scale, generically called here $\mQCD$.
These include the squark soft masses, the trilinear soft
squark couplings and the gluino soft mass, $M_3$. 
Similarly, also the mass scales in the SUSY electroweak sector are set
in reference to one common value, $\mEW$. 
These include the slepton soft masses, the gaugino soft
masses, $M_2$ and $M_1$, and the $\mu$ parameter.
It should also be noted that these latter mass
parameters are the relevant ones for $(g-2)_\mu$. 
To further simplify the scenarios, we will relate $\mQCD$ and $\mEW$.
The remaining
relevant parameter in both NMFV and for the $\Mh$ prediction 
is $\tb$. Since we wish to explore a wide range in $\tb$,
from 5 to 40, $\MA$ is fixed to $1000 \gev$ to ensure the agreement
with the present bounds in the $(\tb, \MA)$ plane from LHC
searches~\cite{CMSpashig12050,Carena:2013qia}.  

Finally, to reduce even further the number of
input parameters we will assume again an approximate GUT relation among
the gaugino soft masses, $M_2=2 M_1\, =\,M_3/4$ and the $\mu$ parameter
will be set equal to $M_2$. Regarding the (diagonal)
trilinear couplings, they will
all be set to zero except those of the stop and sbottom sectors, being
relevant for $\Mh$, and that will be simplified to $A_t=A_b$. 
All parameters are thus either fixed or set relative to $\mQCD$, where
the different relative settings exhibit certain mass patterns of the
MSSM. These kind of scenarios have
the advantage of reducing considerably the number of input parameters
respect to the MSSM and, consequently, making easier the analysis of
their phenomenological implications.  
Similar scenarios have been analyzed in the context of Lepton Flavor
Violation observables in \citere{Arana-Catania:2013nha}.
 
For the forthcoming numerical analysis we consider the following
specific scenarios:
\begin{itemize}
\item[{\bf (a)}]
\begin{align}
 m_{\tilde L}&= m_{\tilde E}=\mEW,\nonumber \\ 
 M_2&= \mEW \nonumber := 1/2 \, \mQCD,\\ 
 m_{\tilde Q} &= m_{\tilde U}=m_{\tilde D}=\mQCD,\nonumber \\ 
 A_t &= 1.3 \, \mQCD, \nonumber \\
M_3 &= 2 \mQCD,
\label{Sa}
\end{align}
\item[{\bf (b)}] 
\begin{align}
 m_{\tilde L} &= m_{\tilde E}=\mEW,\nonumber \\ 
 M_2 &=  1/5 \, \mEW := 1/10 \, \mQCD, \nonumber \\ 
 m_{\tilde Q} &= m_{\tilde U}=m_{\tilde D}=\mQCD,\nonumber \\ 
 A_t &= \mQCD, \nonumber \\
M_3 &= 2/5 \, \mQCD,
\label{Sb}
\end{align}
\item[{\bf (c)}] 
\begin{align}
 m_{\tilde L} &= m_{\tilde E}=\mEW,\nonumber \\ 
 M_2 &=  \mEW \nonumber := 1/4 \, \mQCD, \non \\ 
 m_{\tilde Q} &= m_{\tilde U}=m_{\tilde D}=\mQCD,\nonumber \\ 
 A_t &= \mQCD, \nonumber \\
M_3 &= \mQCD,
\label{Sc}
\end{align}
\item[{\bf (d)}] 
\begin{align}
  m_{\tilde L} &= m_{\tilde E}=\mEW,\nonumber \\ 
 M_2 &= 1/3 \, \mEW \nonumber := 1/3 \, \mQCD, \non \\ 
 m_{\tilde Q} &= m_{\tilde U}=m_{\tilde D}=\mQCD,\nonumber \\ 
 A_t &= \mQCD, \nonumber \\
M_3 &= 4/3 \, \mQCD. 
\label{Sd}
\end{align}
\end{itemize}
Here we have simplified the notation for the soft sfermion masses, by
using $m_{\tilde L}$ for 
$m_{\tilde L}=m_{\tilde L_{1}}=m_{\tilde L_{2}}=m_{\tilde L_{3}}$, etc.

In the forthcoming numerical analysis of the limits of 
the deltas within these scenarios, the most relevant parameters 
$\mQCD \equiv \msusy$ and $\tb$ will be varied within the
intervals: 
\begin{align}
1000 \gev \leq \msusy \leq 3000 \gev, \quad 
5 \leq \tb \leq 40~.
\end{align}
The main results in this framework~2 will be presented in the 
($\msusy$, $\tb$) plane. In the final analysis we will show the
compatibility with $(g-2)_\mu$, but 
focus on the consequences of the changes in $\Mh$ induced by non-zero
values for the deltas.


\subsubsection{Selected \boldmath{$\deABij$} mixings}

Finally, for our purpose in this paper, we need to select the squark
mixings and to set the range of values for the explored
$\deABij$'s. In principle, we work in a complete basis, that is we
take into account the full set of 21 $\deABij$'s.  
However, since the mixing between the first and second/third generation is
already very restricted, we focus here on the deltas that mix only
second and third generation (although our numerical code can handle any
kind of deltas).
For simplicity, we will assume real values for these flavor squark
mixing parameters. Concretely, the scanned interval
in our estimates of NMFV rates will be:  
\begin{align}
-1 \le \delta^{AB}_{ij} \le +1 
\end{align}
The above scan interval is simply meant to cover all possible ranges. Here
we do not take into account, for instance, constraints 
on $\del{LR,RL}{ij}$'s from the requirement of vacuum 
stability~\cite{Casas:1996de} or vacuum meta-stability~\cite{Park:2010wf},
which could invalidate large
values for these deltas, corresponding to large $\cA_{ij}$-terms.


\section{The precision observables}
\label{sec:obs}

In this section we briefly review the current status of the precision
observables that we consider in our NMFV analysis.
Since we are mainly interested
in the phenomenological consequences of the flavor mixing between the
third and second generations we will focus%
\footnote{
We have checked that electroweak precision observables, where NMFV
  effects enter, for instance, via $\De\rho$~\cite{mhNMFVearly}, do not
  lead to relevant additional constraints on the allowed parameter space. Our results on this constraint are in agreement with \citere{Cao1}.}%
~on the lightest Higgs boson mass in the (NMFV) MSSM and the 
following three B
meson observables: 1) Branching ratio of the $B$ radiative decay \bsg, 
2) Branching ratio of the $B_s$ muonic decay \bmm, 
and 3) $B_s-{\bar B_s}$ mass difference \dmbs.  Another
$B$ observable of interest in the present context is \bsll.
However, we have not included this in our study, because the
predicted rates in NMFV-SUSY scenarios for this observable are closely
correlated with those from \bsg\ due to the dipole
operators dominance in the photon-penguin diagrams mediating \bsll\ 
decays.  It implies that the restrictions on the flavor
mixing $\deXYij$ parameters from \bsll\ are also expected to be correlated with those
from the radiative decays.    

The summary of the relevant features for our analysis of these four
observables is given in the following.


\subsection{The lightest Higgs boson mass \boldmath{$\Mh$}}

In the 
Feynman diagrammatic approach that we are following here, the
higher-order corrected  
$\cp$-even Higgs boson masses are derived by finding the
poles of the $(h,H)$-propagator 
matrix. The inverse of this matrix is given by
\BE
\left(\Delta_{\rm Higgs}\right)^{-1}
= - i \ML p^2 -  \mHtree^2 + \hSi_{HH}(p^2) &  \hSi_{hH}(p^2) \\
     \hSi_{hH}(p^2) & p^2 -  \mhtree^2 + \hSi_{hh}(p^2) \MR~.
\label{higgsmassmatrixnondiag}
\end{equation}
Determining the poles of the matrix $\Delta_{\rm Higgs}$ in
\refeq{higgsmassmatrixnondiag} is equivalent to solving
the equation
\begin{equation}
\left[p^2 - \mhtree^2 + \hSi_{hh}(p^2) \right]
\left[p^2 - \mHtree^2 + \hSi_{HH}(p^2) \right] -
\left[\hSi_{hH}(p^2)\right]^2 = 0\,.
\label{eq:proppole}
\end{equation}

The NMFV parameters enter into the one-loop prediction of the various
(renormalized) Higgs-boson self-energies, where details can be found in
\citere{mhNMFV}. Numerically the results have been obtained using the
code \fh~\cite{feynhiggs,mhiggsAEC}, which contains the complete set of one-loop
NMFV corrections.%
\footnote{
Not yet taken into account are the logarithmically resummed
corrections~\cite{Mh-logresum}, which could be relevant for the largest values
of $\msusy$ as analyzed below.
}

The current experimental average for the (SM) Higgs boson mass
is~\cite{pdg-www}, 
\begin{align}
\MH^{\rm exp} &= 125.6 \pm 0.3 \gev~.
\label{MHexp}
\end{align}
The intrinsic theoretical uncertainty is taken to be~\cite{mhiggsAEC,ehowp}
\begin{align}
\de\Mh^{\rm th} = \pm 3 \gev~,
\label{Mhintr}
\end{align}
and both uncertainties combined give an estimate of the total 
uncertainty of $\Mh$ in the MSSM.


\subsection{\boldmath{\bsg}}
\label{sec:bsg}

For a more detailed description of the inclusion of NMFV effects into the
prediction of $B$-physics observables in general, and for \bsg\ in
particular, we refer the reader to \citere{mhNMFV} and references therein.

The relevant effective
Hamiltonian for this decay is given in terms of the Wilson coefficients $C_i$ and operators $O_i$ by:

\noindent \begin{equation}
\mathcal{H}_{\rm eff}=-\frac{4G_{F}}{\sqrt{2}}\VCKM^{ts*}\VCKM^{tb}
           \sum_{i=1}^{8}(C_{i}O_{i}+C'_{i}O'_{i}). 
\end{equation}
Where the primed operators can be obtained from the unprimed ones by
replacing $L \leftrightarrow R$.
The complete list of operators can be found, for instance, in
\citere{Gambino:2001ew}.   
In the context of SUSY scenarios with the MSSM particle content and
assuming NMFV, only four of these operators are relevant
(we have omitted the color indices here for brevity):

\noindent 
\begin{align}
O_{7} &= \frac{e}{16\pi^{2}}m_{b}
         \left(\bar{s}_{L}\sigma^{\mu\nu}b_{R}\right)F_{\mu\nu}~, \\
O_{8} & = \frac{g_{3}}{16\pi^{2}}m_{b}
         \left(\bar{s}_{L}\sigma^{\mu\nu}T^{a}b_{R}\right)G_{\mu\nu}^{a}~, \\
O'_{7} &= \frac{e}{16\pi^{2}}m_{b}
          \left(\bar{s}_{R}\sigma^{\mu\nu}b_{L}\right)F_{\mu\nu}~, \\
O'_{8} &= \frac{g_{3}}{16\pi^{2}}m_{b}
          \left(\bar{s}_{R}\sigma^{\mu\nu}T^{a}b_{L}\right)G_{\mu\nu}^{a}.
\end{align}

We have included in our analysis the most relevant loop contributions to
the Wilson coefficients%
\footnote{The RGE-running of the Wilson coefficients
is done in two steps: The first one is
from the SUSY scale down to the electroweak scale, and the second
one is from this electroweak scale down to the $B$-physics scale. 
For the first step, we use the LO-RGEs for the
relevant Wilson coefficients as in \cite{Degrassi:2000qf} and fix six
active quark flavors in this running. 
For the second running we use the NLO-RGEs
as in \cite{Hurth:2003dk} and fix, correspondingly, five active quark
flavors. For the charged Higgs sector, as in \citere{mhNMFV}, we use the
NLO formulas for the Wilson coefficients of
\citere{Borzumati:1998tg}.}%
, concretely: 
1)~loops with Higgs bosons (including the resummation of large $\tb$
effects~\cite{Isidori:2002qe}),  
2)~loops with charginos and 
3)~loops with gluinos. 
It should be noted that, at one loop order, the gluino 
loops do not contribute in MFV scenarios, but they are very relevant
(dominant in many cases) in the present NMFV scenarios.

The total branching ratio for this decay is finally estimated by adding
the new contributions from the SUSY and Higgs sectors to the  
SM rate. More specifically, we use eq.(42) of \cite{Hurth:2003dk} for
the estimate of  \bsg\ in terms of the ratios of 
the Wilson coefficients $C_{7,8}$ and  $C'_{7,8}$ (including all the
mentioned new contributions) 
divided by the corresponding $C_{7,8}^{\rm SM}$ in the SM. 

For the numerical estimates of \bsg\ (and the other $B$-physics
observables) we use the {\tt FORTRAN}
subroutine {\tt BPHYSICS} (modified as to include the contributions from
$C'_{7,8}$ which were not included in its original version) included in
the {\tt SuFla} code, that incorporates all the above mentioned
ingredients~\cite{sufla}.   

In order to obtain the updated limits on the NMFV parameters, the
following experimental measurement of
\bsg~\cite{hfag:rad}%
\footnote{We have added the various contributions to the experimental
  error in quadrature.}%
, and its prediction within the SM~\cite{Misiak:2009nr} have been used:

\begin{align}
\label{bsgamma-exp}
\bsg_{\rm exp} &= (3.43 \pm 0.22)\times10^{-4}~, \\
\bsg_{\rm SM} &= (3.15 \pm 0.23)\times10^{-4}~.
\label{bsgamma-SM}
\end{align}


\subsection{\boldmath{\bmm}}
\label{sec:bmm}

The relevant effective Hamiltonian for this process is
\cite{Chankowski:2000ng,Bobeth:2002ch}:  

\noindent 
\begin{equation}
\mathcal{H}_{\rm eff}=-\frac{G_{F}\alpha}{\sqrt{2} \pi}
                    \VCKM^{ts*}\VCKM^{tb}\sum_{i} (C_{i}O_{i}+C'_{i} O'_{i}), 
\end{equation}
where the operators $O_i$ are given by:
\begin{align}
{O}_{10} &= \left(\bar{s}\ga^{\nu}P_Lb\right)
           \left(\bar{\mu}\ga_{\nu}\ga_5\mu\right)~,
& {O}_{10}^{\prime} &= \left(\bar{s}\ga^{\nu}P_Rb\right)
                     \left(\bar{\mu}\ga_{\nu}\ga_5\mu\right)~, \non \\
{O}_{S} &= m_b\left(\bar{s}P_Rb\right)
              \left(\bar{\mu}\mu\right)~, 
& {O}_{S}^{\prime} &= m_s\left(\bar{s} P_Lb \right)
                   \left(\bar{\mu}\mu\right)~, \non \\
{O}_{P} &= m_b\left(\bar{s} P_Rb \right)
              \left(\bar{\mu}\ga_5\mu\right)~, 
& {O}_{P}^{\prime} &= m_s\left(\bar{s} P_Lb \right)
                     \left(\bar{\mu}\ga_5\mu\right)~.
\label{bsm:Ops}
\end{align}
We have again omitted the color indices here for brevity.

The prediction for the decay rate is expressed by: 
\begin{align}
\bmm &= \frac{G_F^2\alpha^2 m_{B_s}^2 f_{B_s}^2\tau_{B_s}}{64 \pi^3}
       \lvert \VCKM^{ts*}\VCKM^{tb}\rvert^2\sqrt{1-4\hat{m}_{\mu}^2}
       \non \\
&\times\left[\left(1-4\hat{m}_{\mu}^2\right)\lvert F_S\rvert^2
            +\lvert F_P+2\hat{m}_{\mu}^2 F_{10}\rvert^2\right],
\label{bsm:br}
\end{align}
where $\hat{m}_{\mu}=m_{\mu}/m_{B_s}$ and the 
$F_i$ are given by
\begin{align}
F_{S,P}&=m_{B_s}\left[\frac{C_{S,P}m_b-C_{S,P}^{\prime}m_s}{m_b+m_s}\right],
&F_{10}=C_{10}-C_{10}^{\prime}.
\nonumber
\end{align}

In the context of NMFV MSSM, with no preference for large $\tb$ 
values, there are in general three types of one-loop diagrams that contribute 
to the previous $C_i$ Wilson coefficients for this $B_s \to \mu^+ \mu^-$
decay: 
1)~Box diagrams, 
2)~$Z$-penguin diagrams and 
3)~neutral Higgs boson $\phi$-penguin diagrams, where $\phi$ denotes the
three neutral MSSM Higgs bosons, $\phi = h, H, A$ (again large resummed
$\tb$ effects have been taken into account).
In our numerical estimates 
we have included what are known to be the dominant contributions to
these three types of diagrams \cite{Chankowski:2000ng}: chargino
contributions to box and $Z$-penguin diagrams and chargino and gluino
contributions to $\phi$-penguin diagrams.   

The present experimental value
for this observable \cite{Chatrchyan:2013bka,Aaij:2013aka}, 
and the prediction within the SM \cite{Buras:2012ru} are given by 
\begin{align}
\label{bsmumu-exp}
\bmm_{\rm exp} &= (3.0^{+1.0}_{-0.9})\times 10^{-9}~, \\
\bmm_{\rm SM} &= (3.23\pm 0.27)\times 10^{-9}~.
\label{bsmumu-SM}
\end{align}


\subsection{\boldmath{\dmbs}} 
\label{sec:dmbs}

The relevant effective Hamiltonian for $B_s-{\bar B_s}$ mixing and, hence, for 
the $B_s/{\bar B_s}$ mass difference \dmbs\ is: 

\begin{equation}
\mathcal{H}_{\rm eff}=
\frac{G_F^2}{16\pi^2}M_W^2 
\left(\VCKM^{tb*}{}\VCKM^{ts}\right)^2
\sum_{i}C_i O_i.
\label{Ham}
\end{equation}
Within the NMFV MSSM the following operators are relevant (now including
the color indices explicitly):
\begin{align}
\label{SMOps}
O^{VLL} &= (\bar{b}^{\alpha}\ga_{\mu}P_L s^{\alpha})
          (\bar{b}^{\beta}\ga^{\mu}P_L s^{\beta})~, \\
\label{Ops1}
O^{LR}_{1} &= (\bar{b}^{\alpha}\ga_{\mu}P_L s^{\alpha})
             (\bar{b}^{\beta}\ga^{\mu}P_R s^{\beta})~, 
& O^{LR}_{2} &= (\bar{b}^{\alpha}P_L s^{\alpha})
               (\bar{b}^{\beta}P_R s^{\beta})~, \\
O^{SLL}_{1} &= (\bar{b}^{\alpha}P_L s^{\alpha})
              (\bar{b}^{\beta}P_L s^{\beta}),
& O^{SLL}_{2} &= (\bar{b}^{\alpha}\sigma_{\mu\nu}P_L s^{\alpha})
               (\bar{b}^{\beta}\sigma^{\mu\nu}P_L s^{\beta})~,
\label{Ops2}
\end{align}
and the corresponding operators $O^{VRR}$ and
$O^{SRR}_{i}$ that can be obtained by replacing $P_L \leftrightarrow P_R$
in~\refeq{SMOps} and~\refeq{Ops2}.
The mass difference $\Delta M_{B_s}$ is then evaluated by taking the matrix
element
\begin{align}
\dmbs &= 2\lvert\langle\bar{B}_s\lvert\mathcal{H}_{\rm eff}
          \rvert B_s\rangle\rvert,
\label{delmb}
\end{align}
where $\langle\bar{B}_s\lvert\mathcal{H}_{\rm eff}\rvert B_s\rangle$ is given
by
\begin{align}
\langle\bar{B}_s\lvert\mathcal{H}_{\rm eff}\rvert B_s\rangle=&
\frac{G_F^2}{48\pi^2}M_W^2 m_{B_s} f^2_{B_s}
\left(\VCKM^{tb*} \VCKM^{ts}\right)^2
\sum_{i}P_i C_i\left(\mu_W\right).
\label{matel}
\end{align}
Here $m_{B_s}$ is the $B_s$ meson mass, and
$f_{B_s}$ is the $B_s$ decay constant.
The coefficients $P_i$ contain the effects due to RGE running between
the electroweak scale $\mu_W$ and $m_b$ as well as the relevant hadronic
matrix element. We use the coefficients $P_i$ from the lattice
calculation \cite{Becirevic:2001xt}:  
\begin{align}
P^{VLL}_1=&0.73,
&P^{LR}_1=&-1.97,
&P^{LR}_2=&2.50,
&P^{SLL}_1=&-1.02,
&P^{SLL}_2=&-1.97.
\label{pcoef}
\end{align}

In the context of the NMFV MSSM, besides the SM contributions, 
there are in general three types of one-loop diagrams that contribute:
1)~Box diagrams, 
2)~$Z$-penguin diagrams and 
3)~double Higgs-penguin diagrams (again including the resummation of
large $\tb$ enhanced effects).
In our numerical estimates we have included what are known to be the
dominant contributions to these three types of diagrams in scenarios
with non-minimal flavor violation (for a review see, for instance,
\cite{Foster:2005wb}): gluino contributions to box 
diagrams, chargino contributions to box and $Z$-penguin diagrams, and 
chargino and gluino contributions to double $\phi$-penguin diagrams. 

For the numerical estimates we have modified the {\tt BPHYSICS} subroutine
included in the {\tt SuFla} code~\cite{sufla} which incorporates all the
ingredients that we have pointed out above, except the contributions
from gluino boxes which we have added, see \citere{mhNMFV} for a
detailed discussion on these contributions. 

The experimental result~\cite{hfag:pdg} and the SM
prediction (using the NLO expression of \cite{Buras:1990fn} and the
error estimate of \cite{Golowich:2011cx}) used to obtain our updated
bounds on the NMFV parameters are given by:
\begin{align}
\label{deltams-exp}
{\dmbs}_{\rm exp} &= (116.4 \pm 0.5) \times 10^{-10} \mev~, \\
{\dmbs}_{\rm SM} &= (117.1^{+17.2}_{-16.4}) \times 10^{-10} \mev~.
\label{deltams-SM}
\end{align}


\section{Numerical results}
\label{sec:results}

In this section we present our numerical results. First we analyze
the six scenarios of framework~1, exploring $\deABij \neq 0$,
with respect to the flavor observables
and derive the corresponding bounds on the deltas. 
In a second step we
will show which corrections to the Higgs boson masses can be found in
these scenarios, but bounds on the deltas are only derived from ``too
large'' corrections to the lightest Higgs boson mass, as will be defined
and discussed below. These ``too large'' corrections to $\Mh$ indicate
that the light Higgs boson mass itself can serve as an additional
observable constraining further the deltas, which can therefore
complement the previous constraints from $B$-physics observables.
The heavy Higgs boson masses, on the other hand, 
depend (to a good approximation) linearly on $\MA$ and can
thus easily avoid bounds by an appropriate choice of $\MA$.
Finally, having the new restrictions from $\Mh$ in mind, 
we then focus next on the simple scenarios of framework 2, where we have
performed a systematic study in the ($\msusy$, $\tb$) plane to conclude
on the maximum allowed deltas that are compatible with both the
$B$-physics data and the  
present Higgs mass value. In this analysis we will consider also the
compatibility with the $(g-2)_\mu$ data.


\subsection{Framework 1: flavor observables}

\begin{figure}[htb!] 
\centering
\includegraphics[width=6.75cm,height=8cm,angle=270]{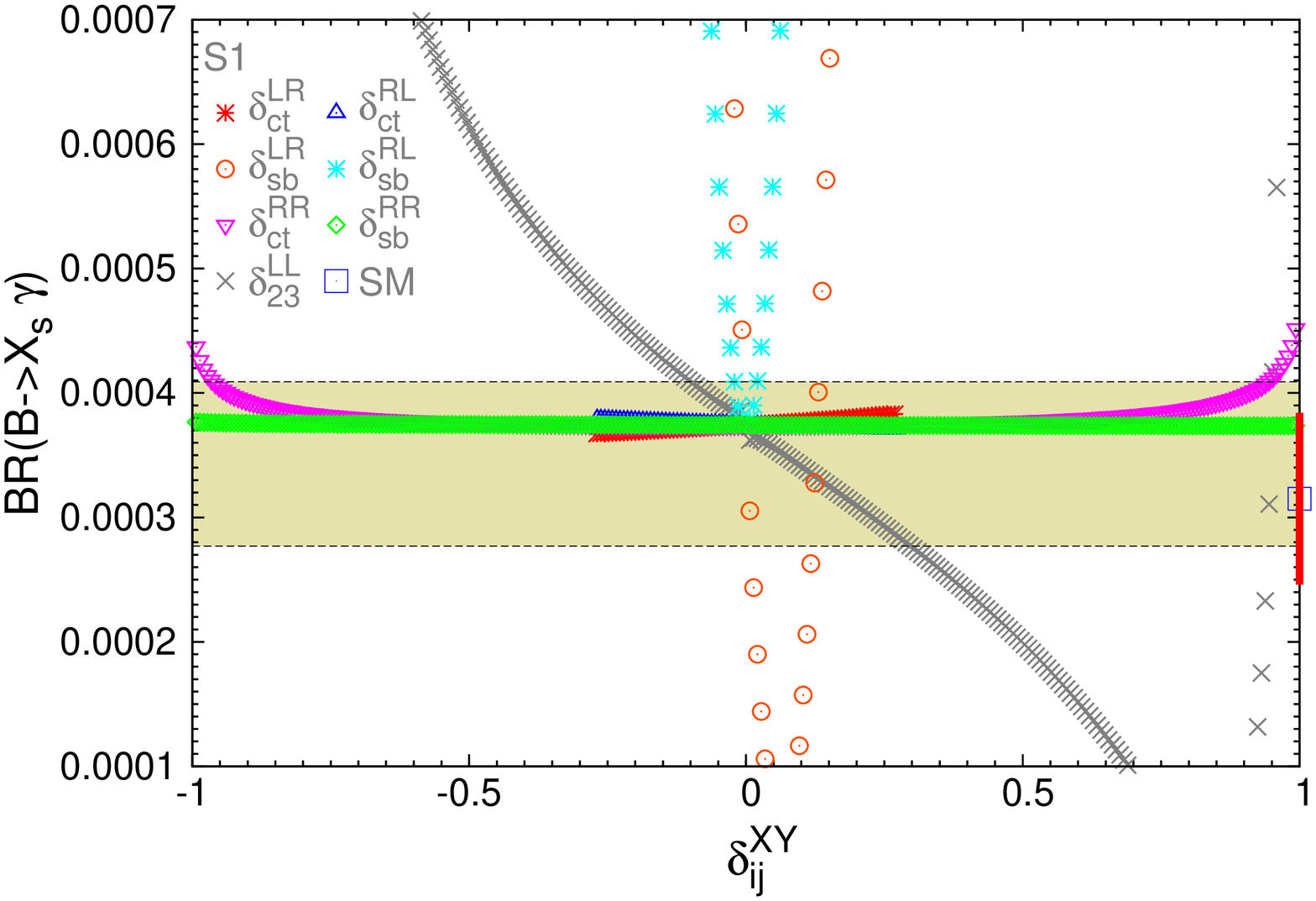}
\includegraphics[width=6.75cm,height=8cm,angle=270]{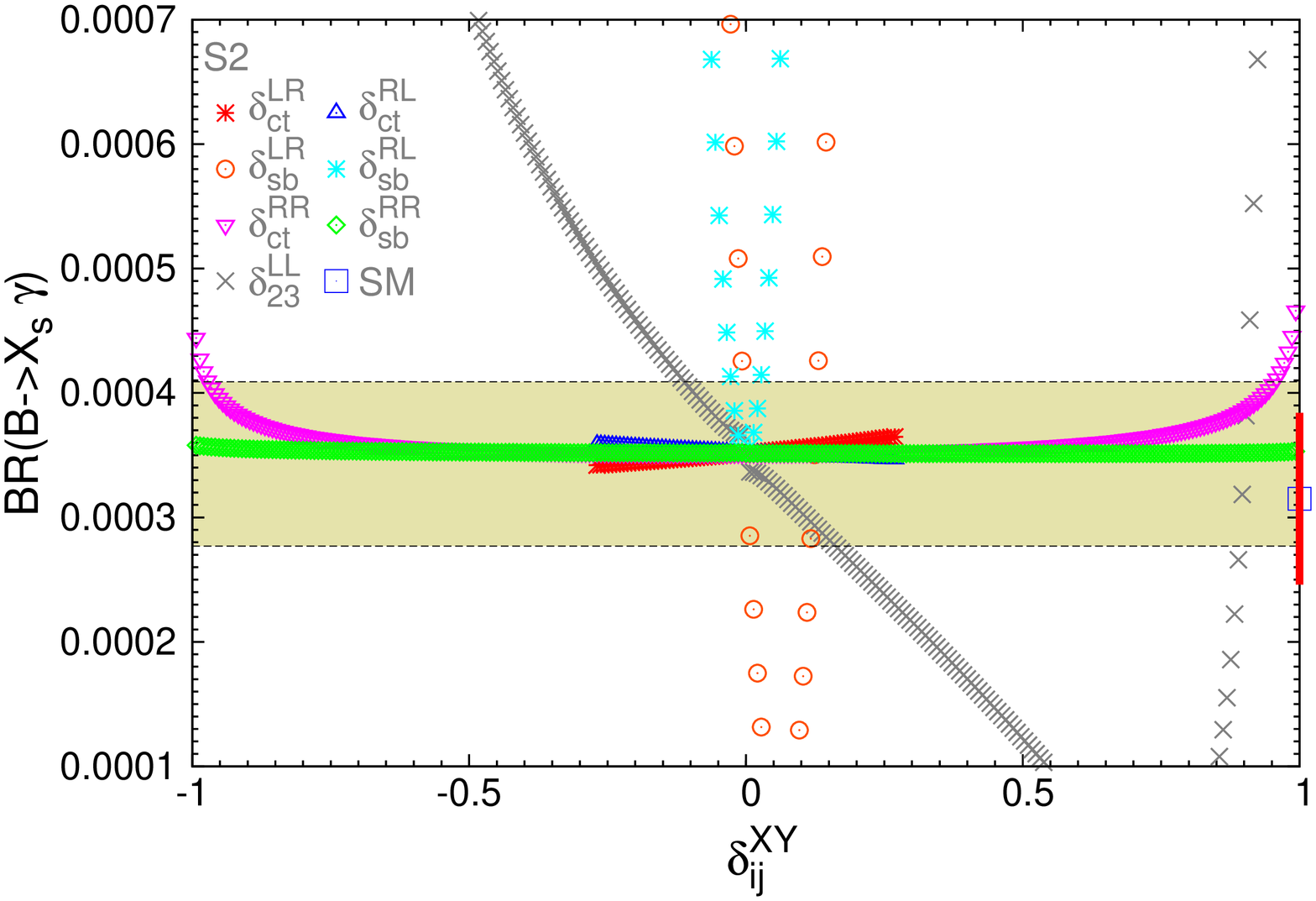}\\
\includegraphics[width=6.75cm,height=8cm,angle=270]{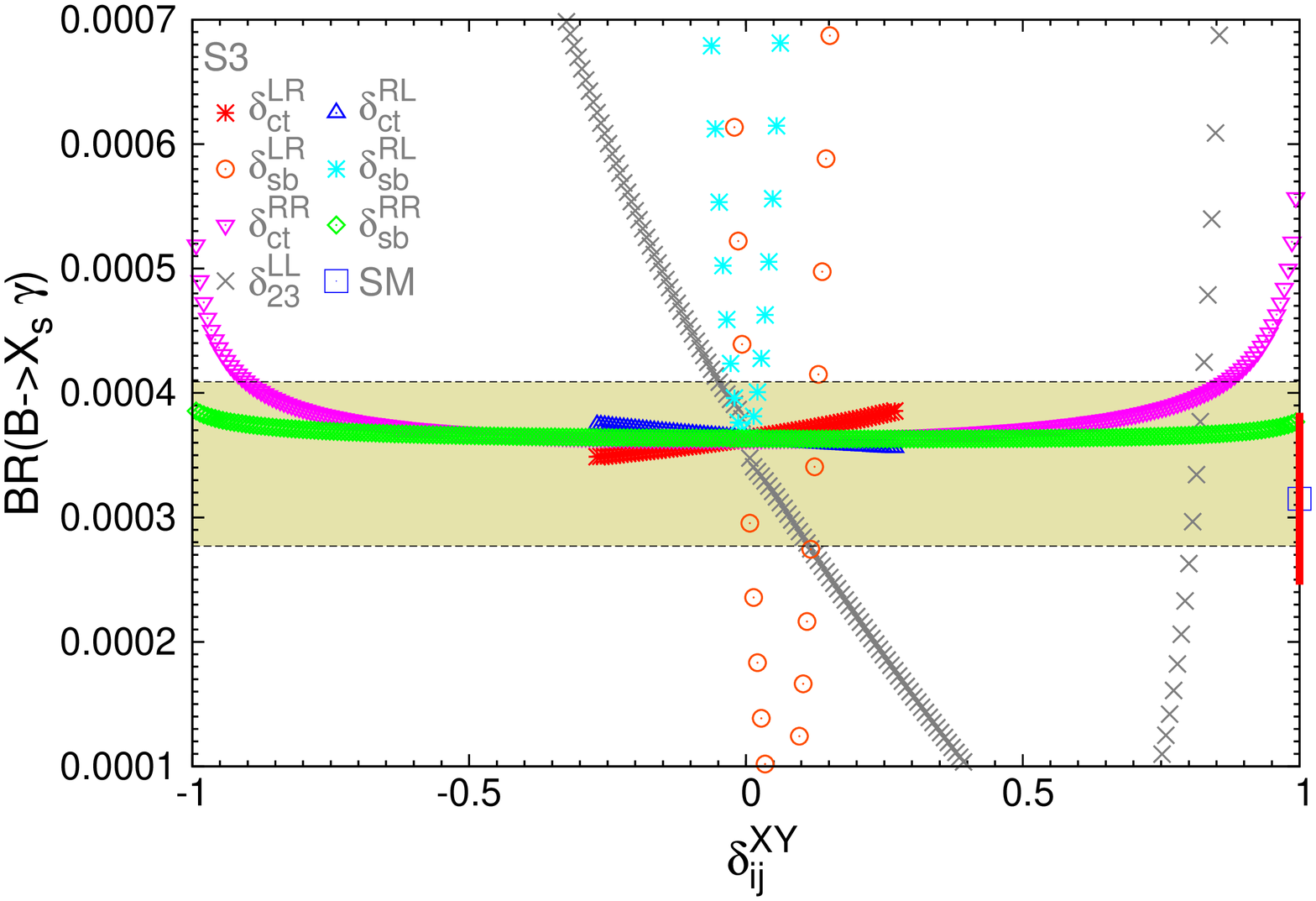}
\includegraphics[width=6.75cm,height=8cm,angle=270]{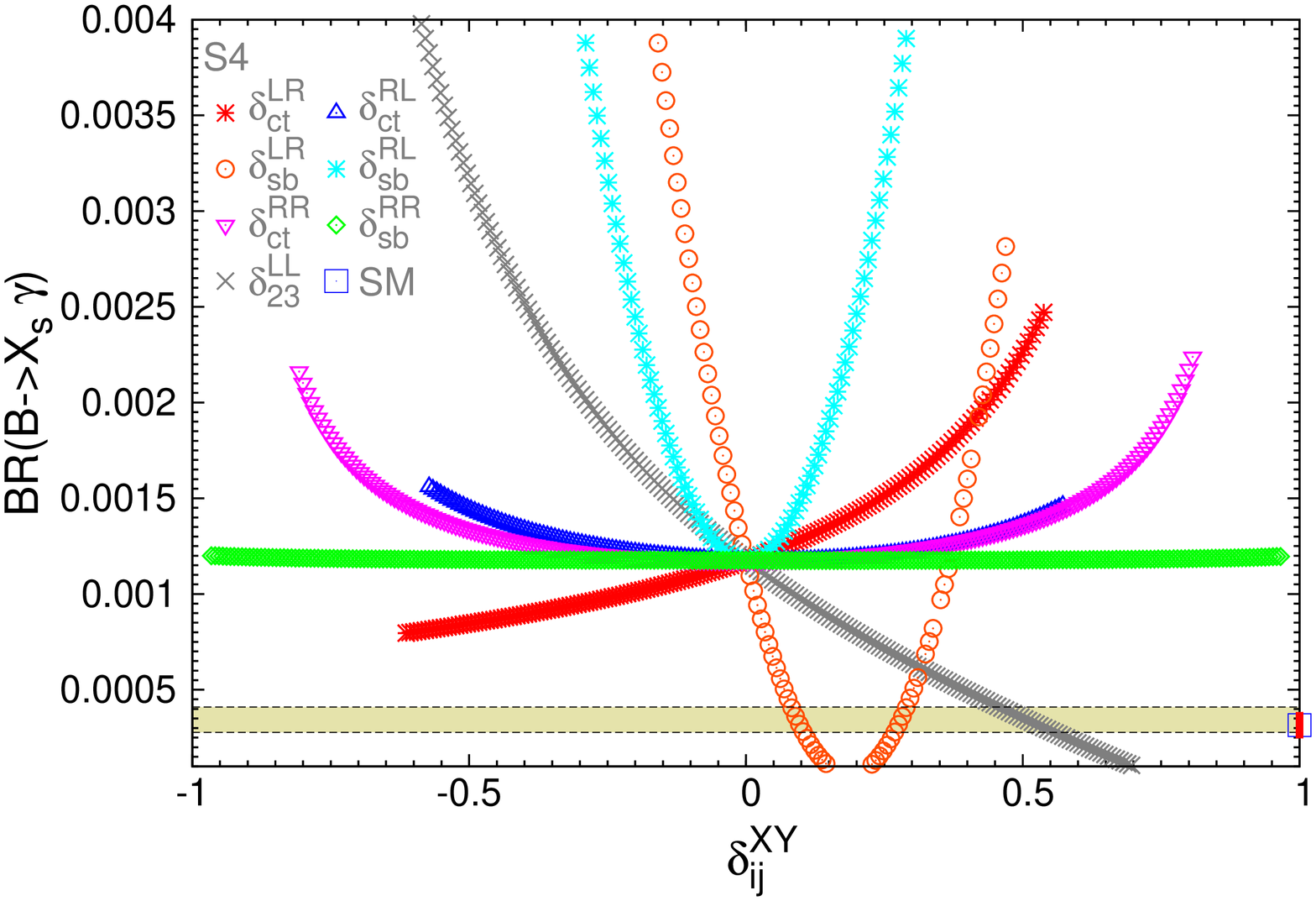}\\
\includegraphics[width=6.75cm,height=8cm,angle=270]{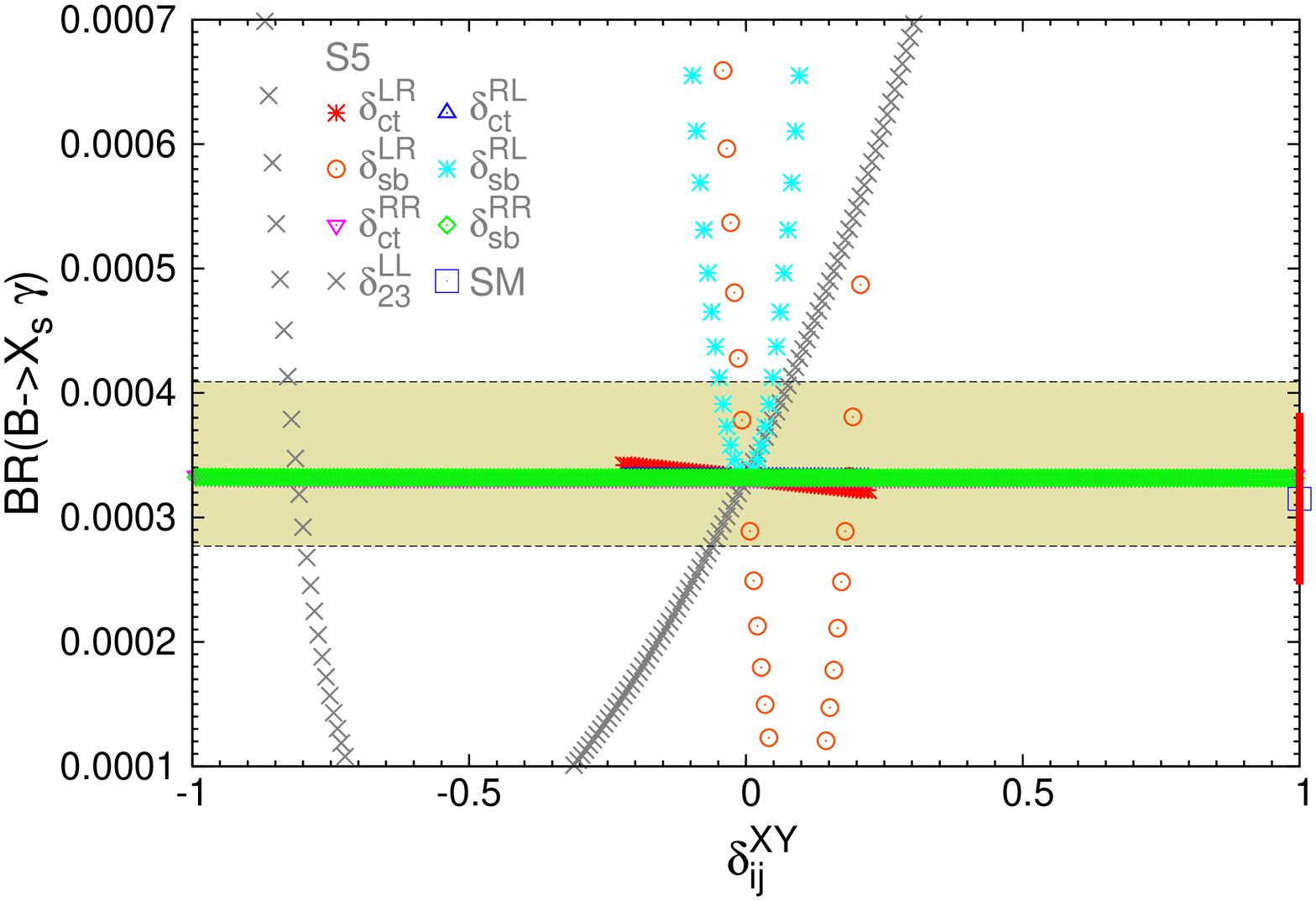}
\includegraphics[width=6.75cm,height=8cm,angle=270]{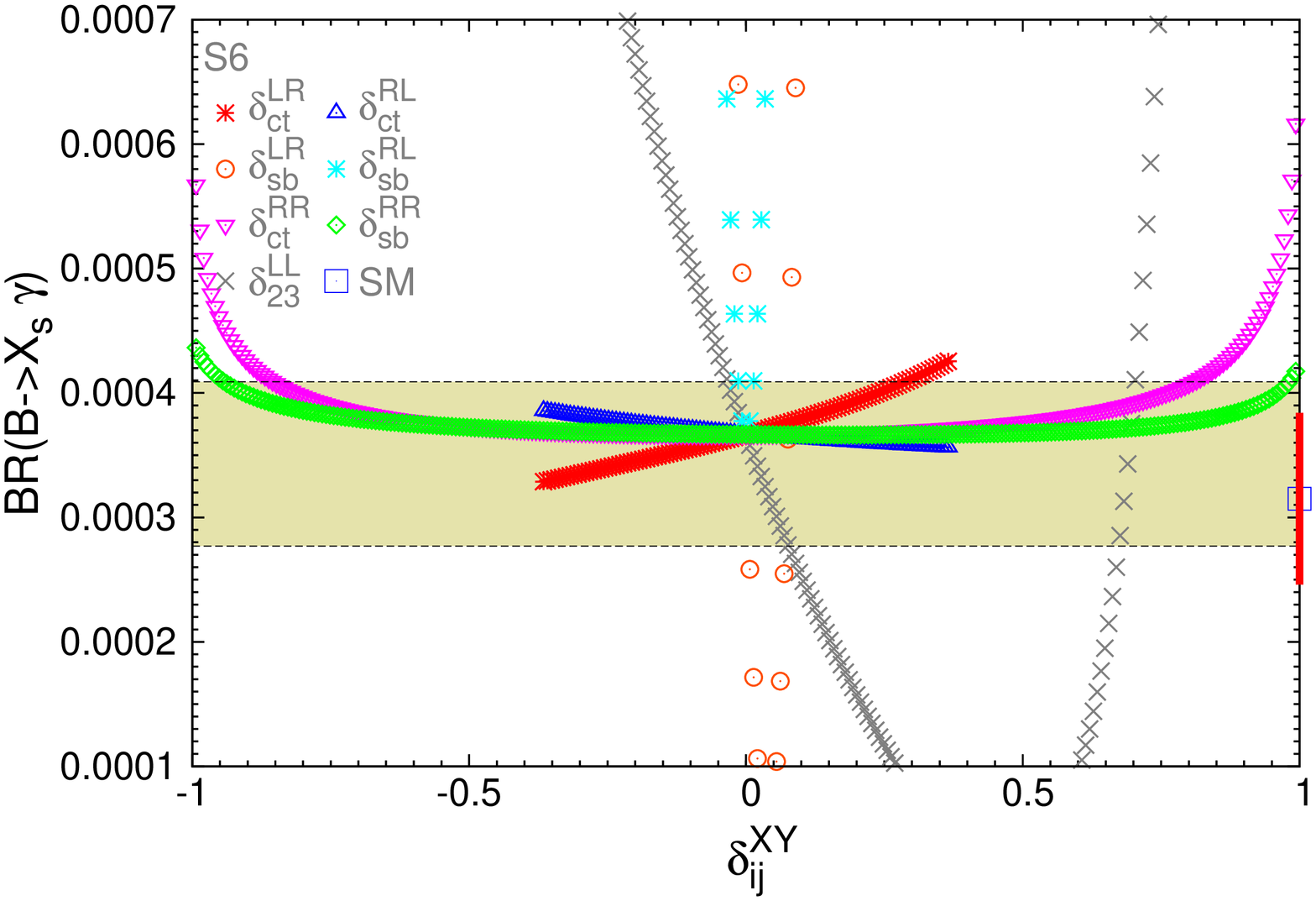}
\caption{Sensitivity to the NMFV deltas in \bsg\ for the points S1\ldots S6.
  The experimental allowed $3\sigma$ area is the horizontal
  colored band. The SM prediction and the theory uncertainty $\Dtheo(\bsg)$
  (red bar) is displayed on the right axis.}  
\label{fig:bsg}
\end{figure}

\begin{figure}[htb!] 
\centering
\includegraphics[width=6.75cm,height=8cm,angle=270]{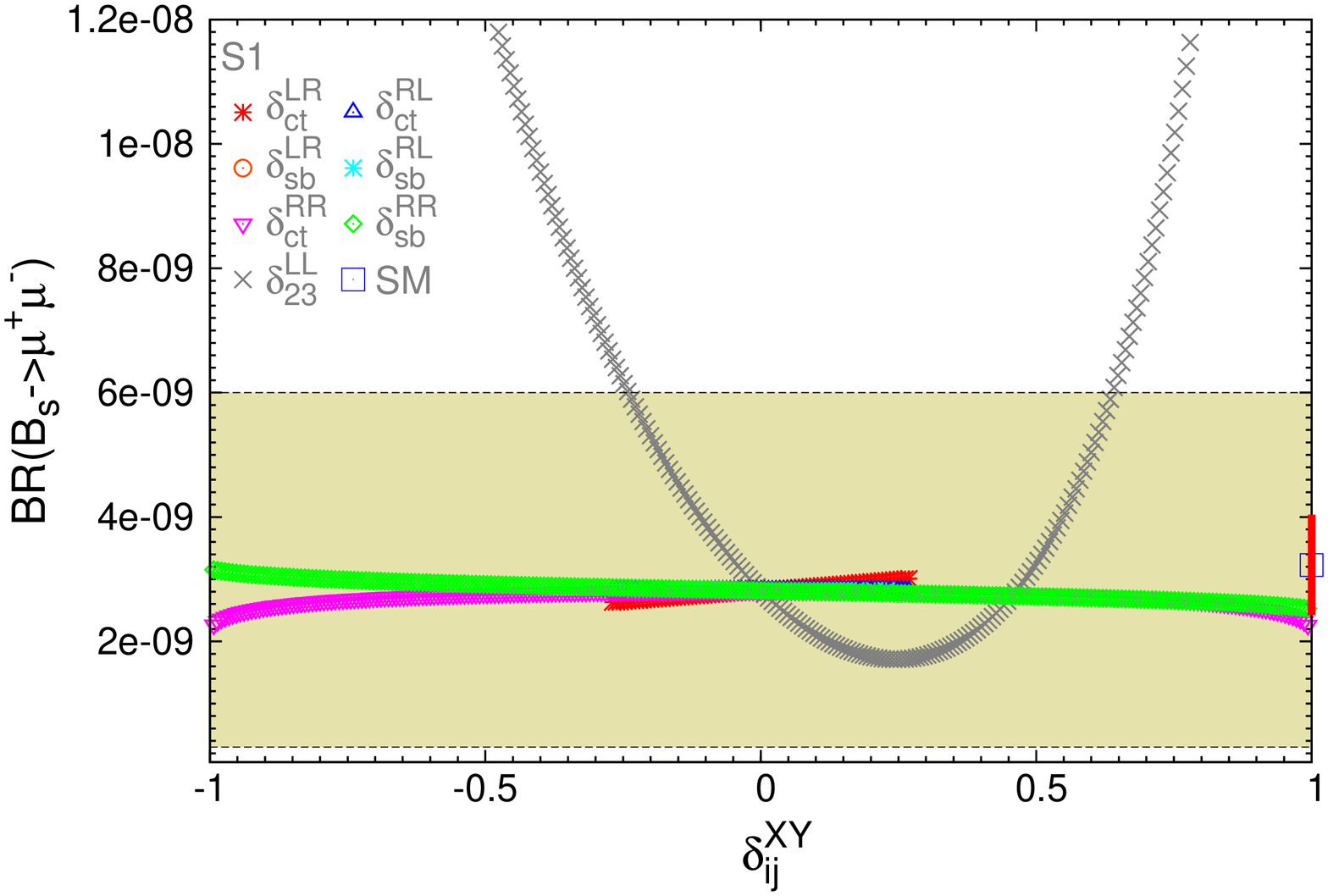}
\includegraphics[width=6.75cm,height=8cm,angle=270]{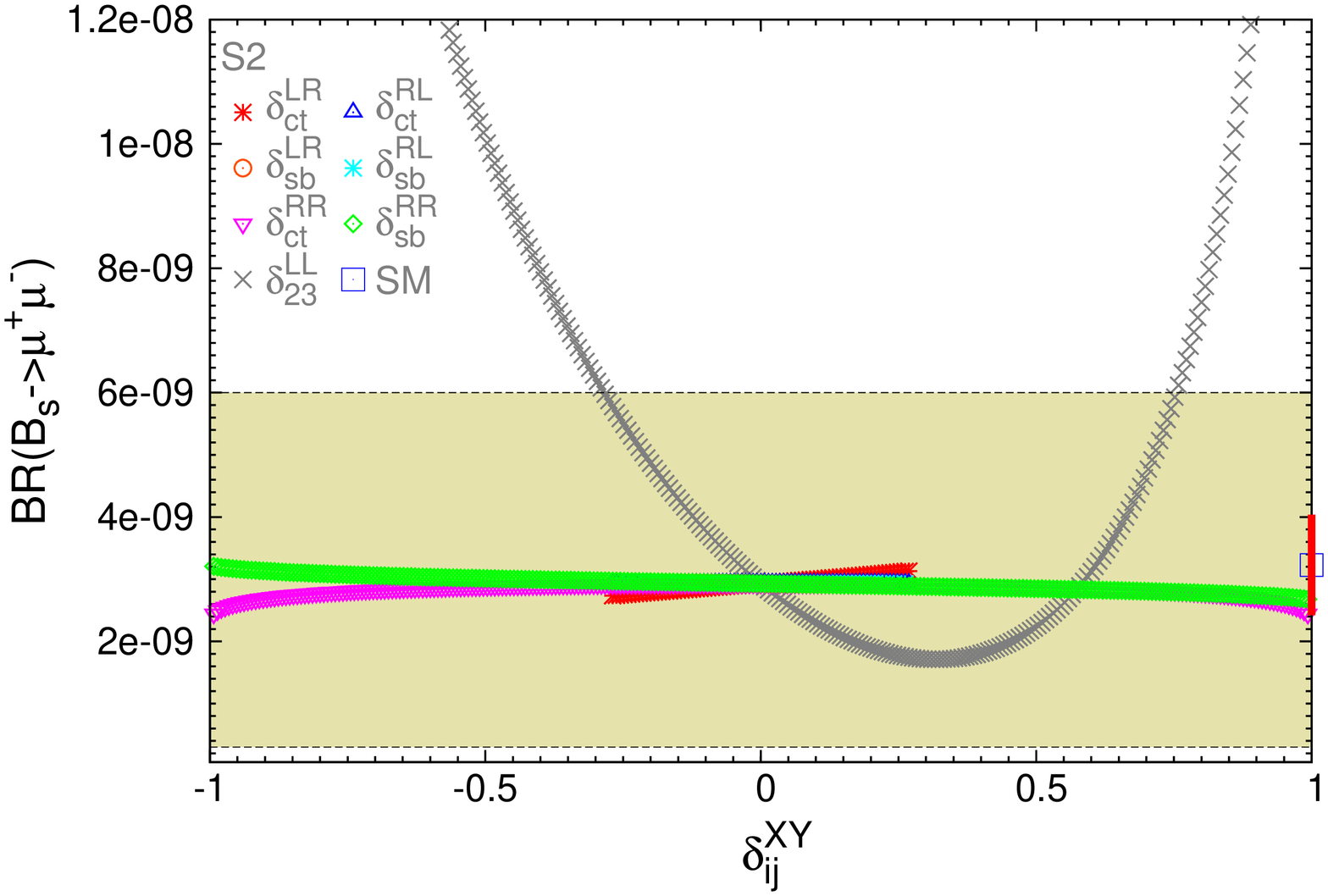}\\
\includegraphics[width=6.75cm,height=8cm,angle=270]{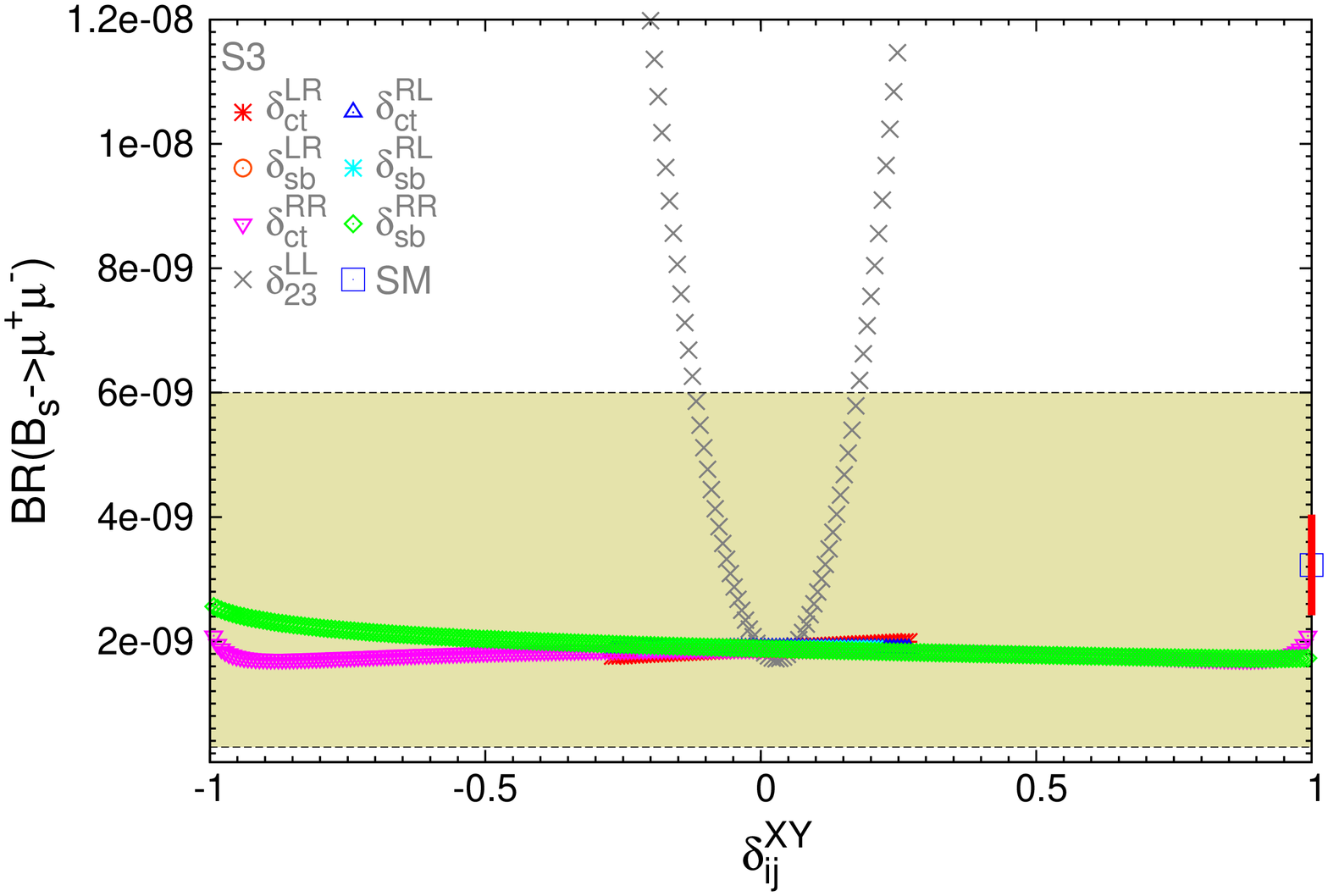}
\includegraphics[width=6.75cm,height=8cm,angle=270]{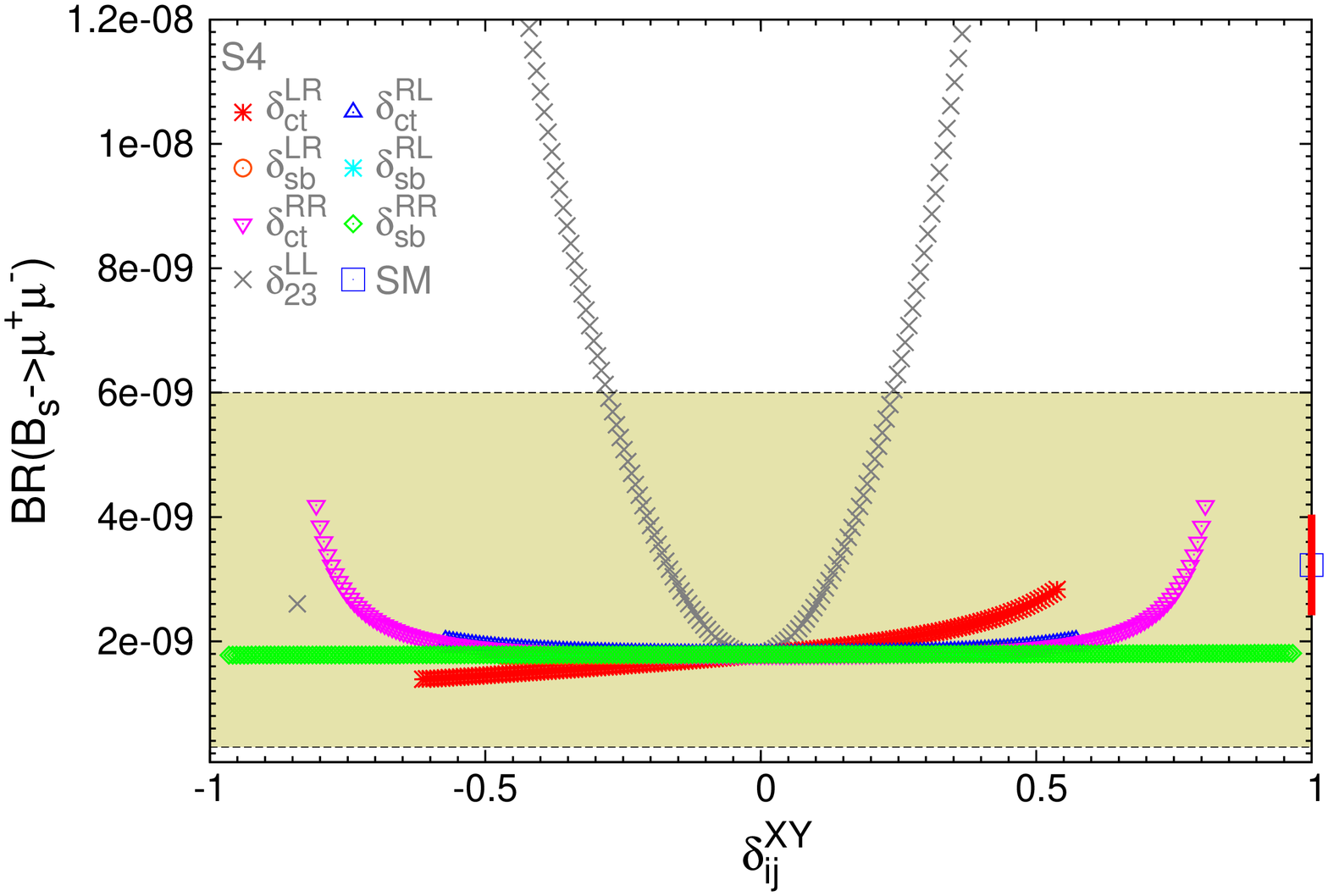}\\
\includegraphics[width=6.75cm,height=8cm,angle=270]{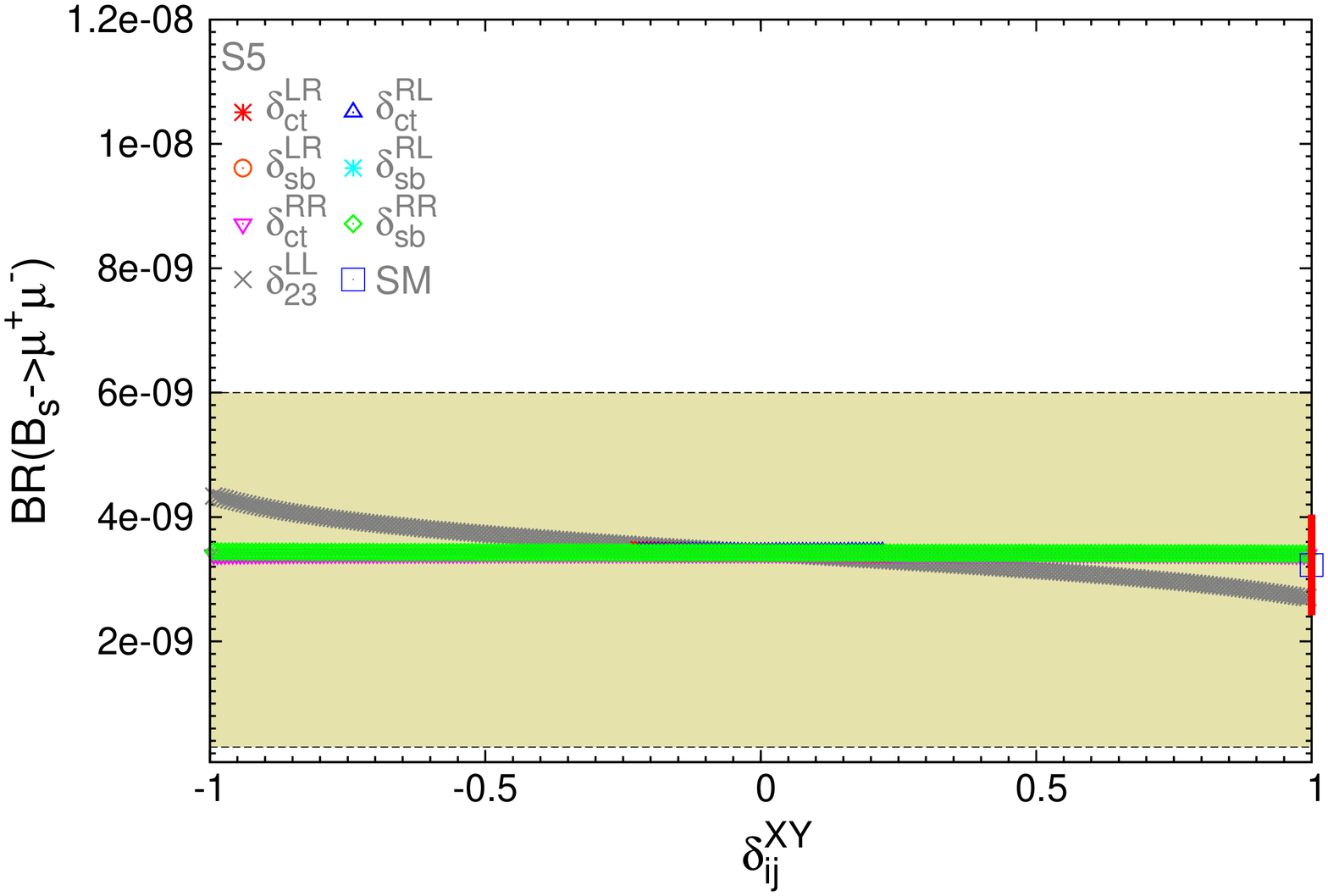}
\includegraphics[width=6.75cm,height=8cm,angle=270]{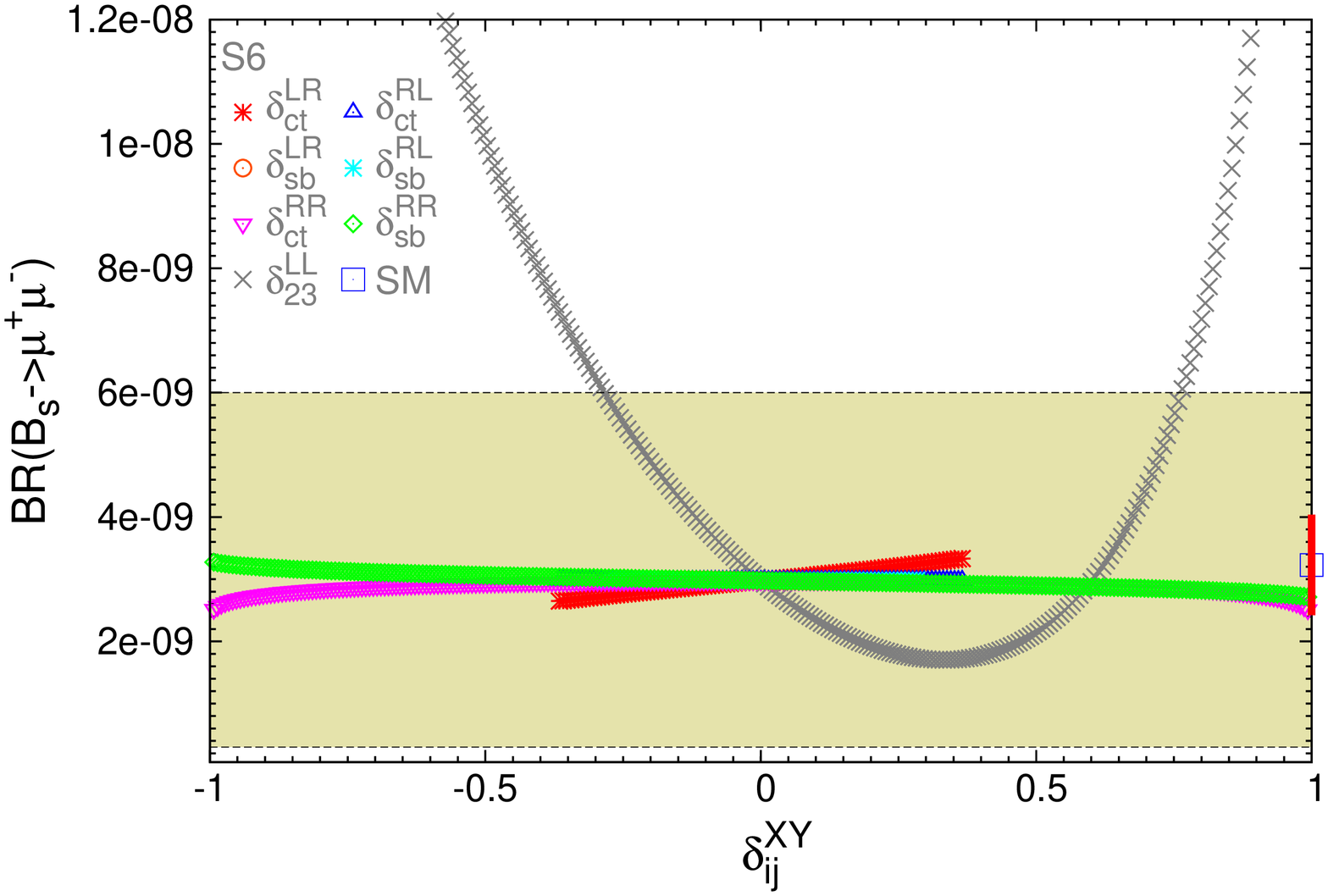}
\caption{Sensitivity to the NMFV deltas in \bmm\ for the points S1\ldots S6.
  The experimental allowed $3\sigma$ area is the horizontal
  colored band. The SM prediction and the theory uncertainty $\Dtheo(\bmm)$
  (red bar) is displayed on the right axis.}  
\label{fig:bmm}
\end{figure}

\begin{figure}[htb!] 
\centering
\includegraphics[width=6.75cm,height=8cm,angle=270]{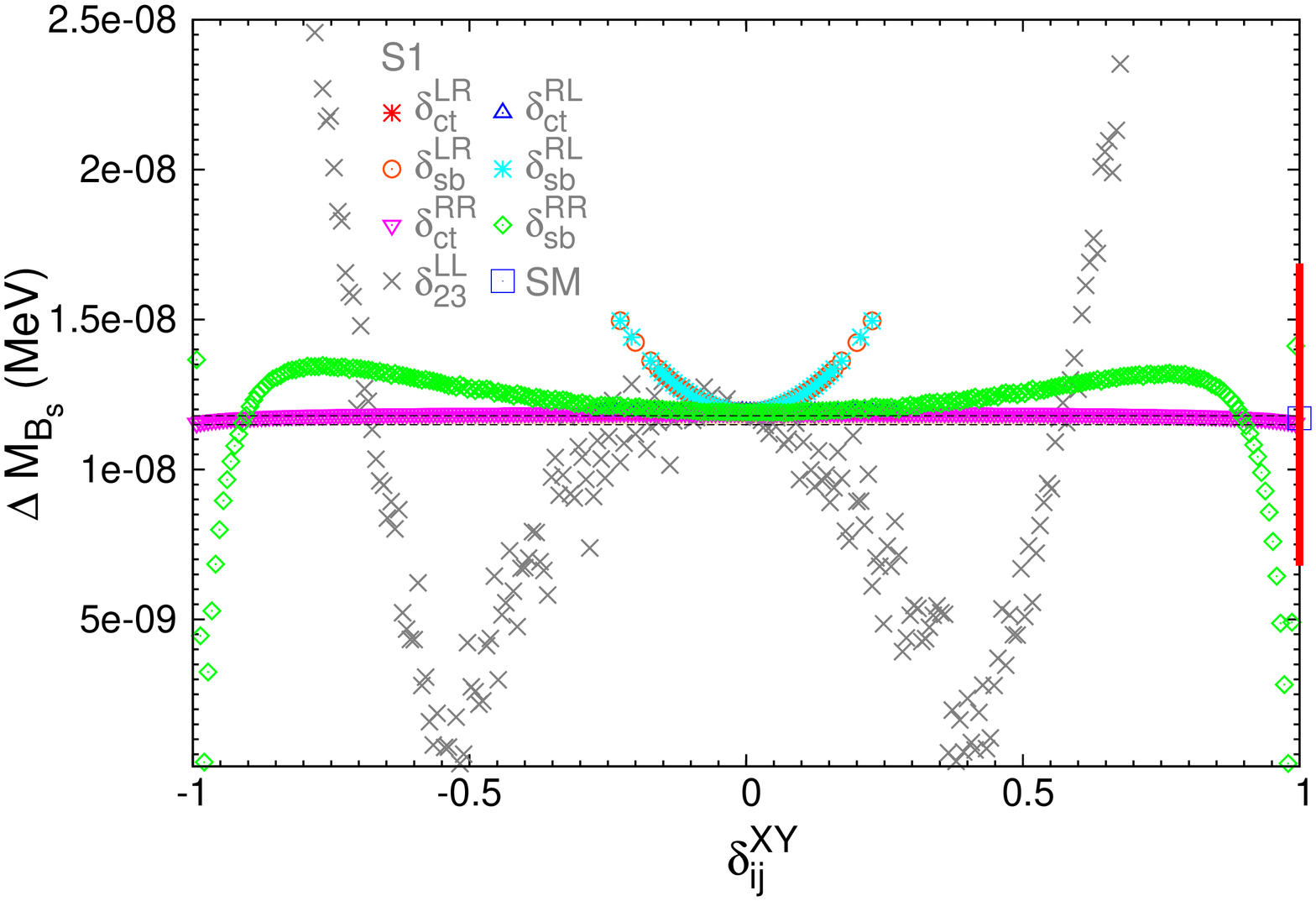}
\includegraphics[width=6.75cm,height=8cm,angle=270]{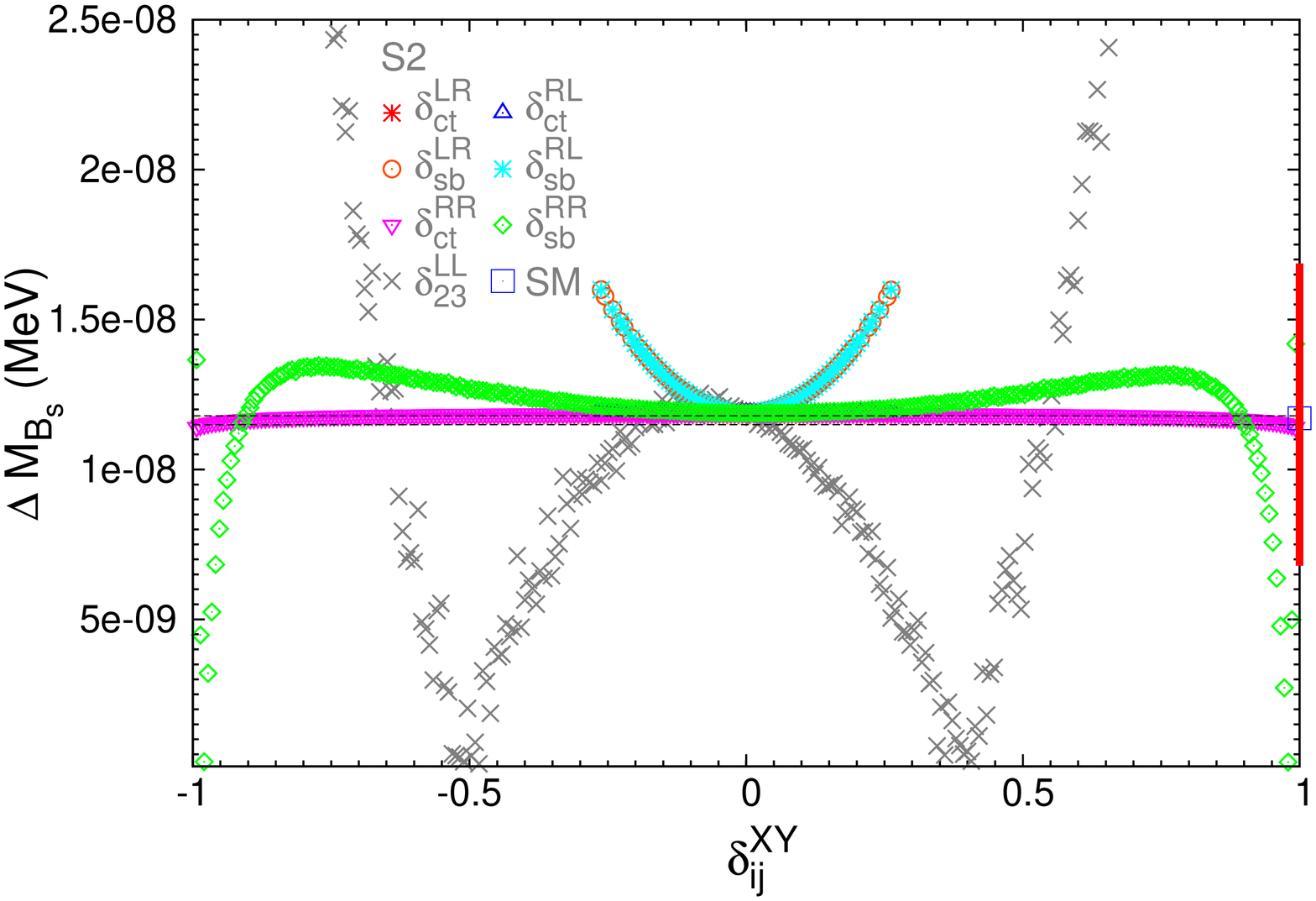}\\
\includegraphics[width=6.75cm,height=8cm,angle=270]{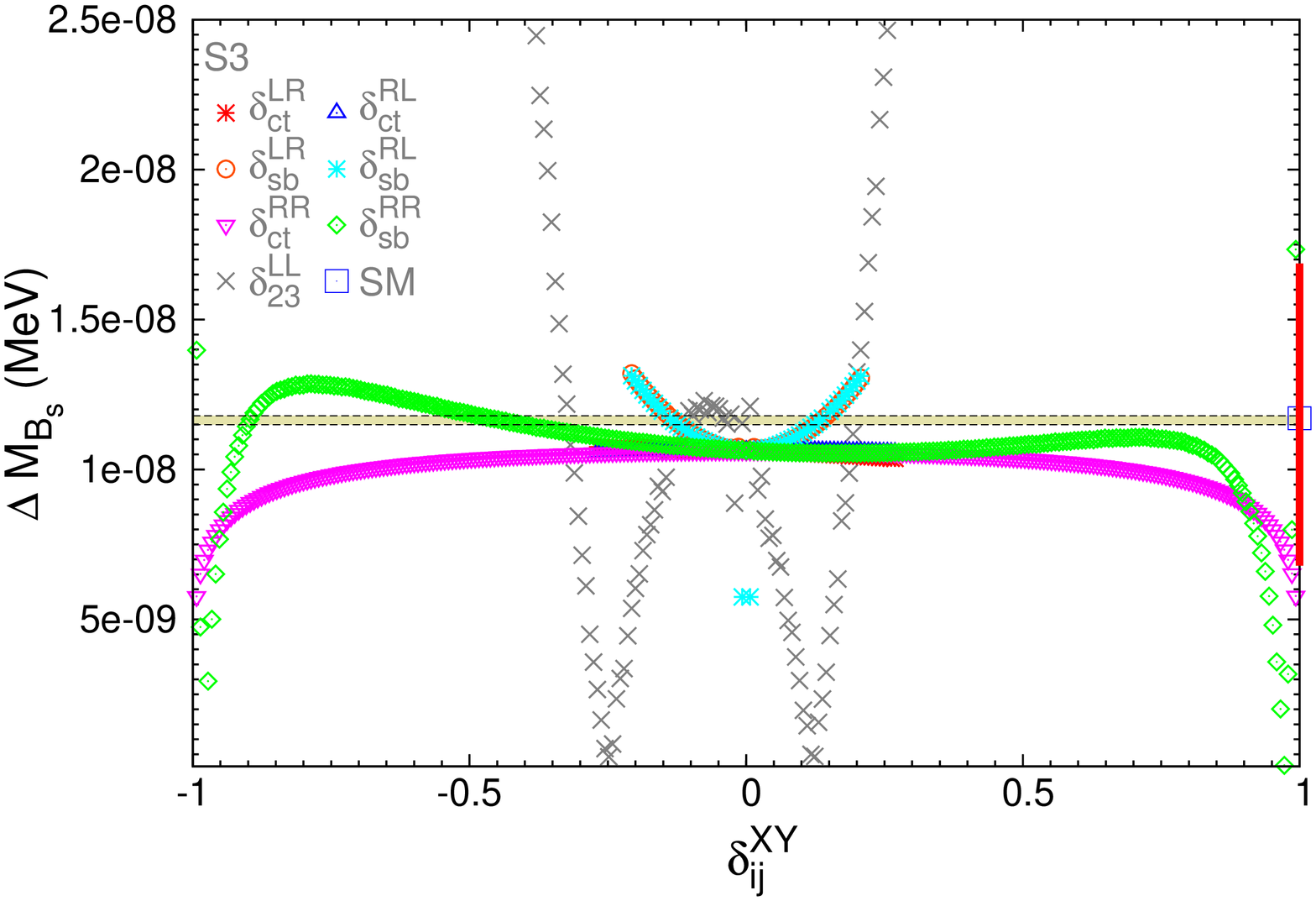}
\includegraphics[width=6.75cm,height=8cm,angle=270]{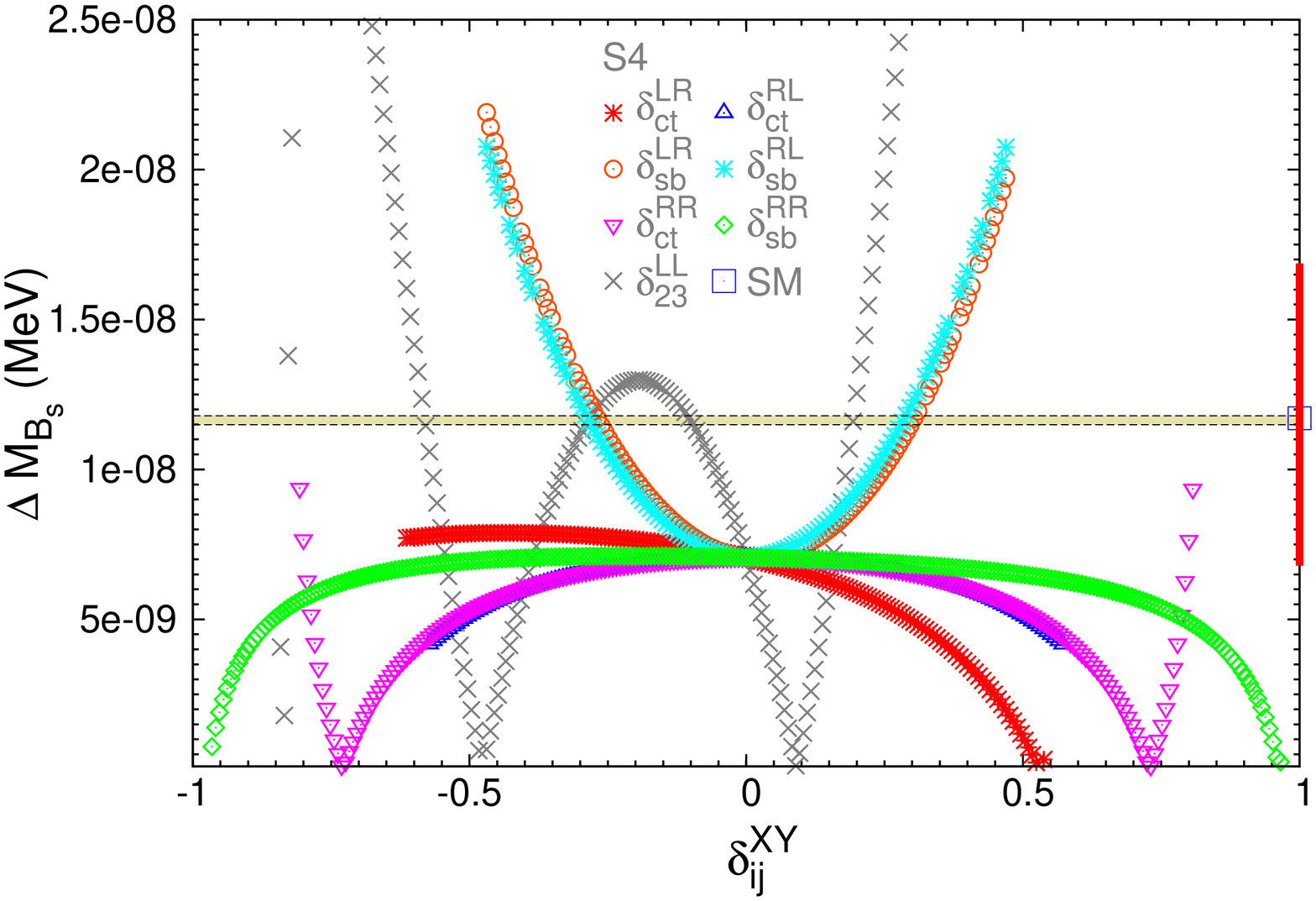}\\
\includegraphics[width=6.75cm,height=8cm,angle=270]{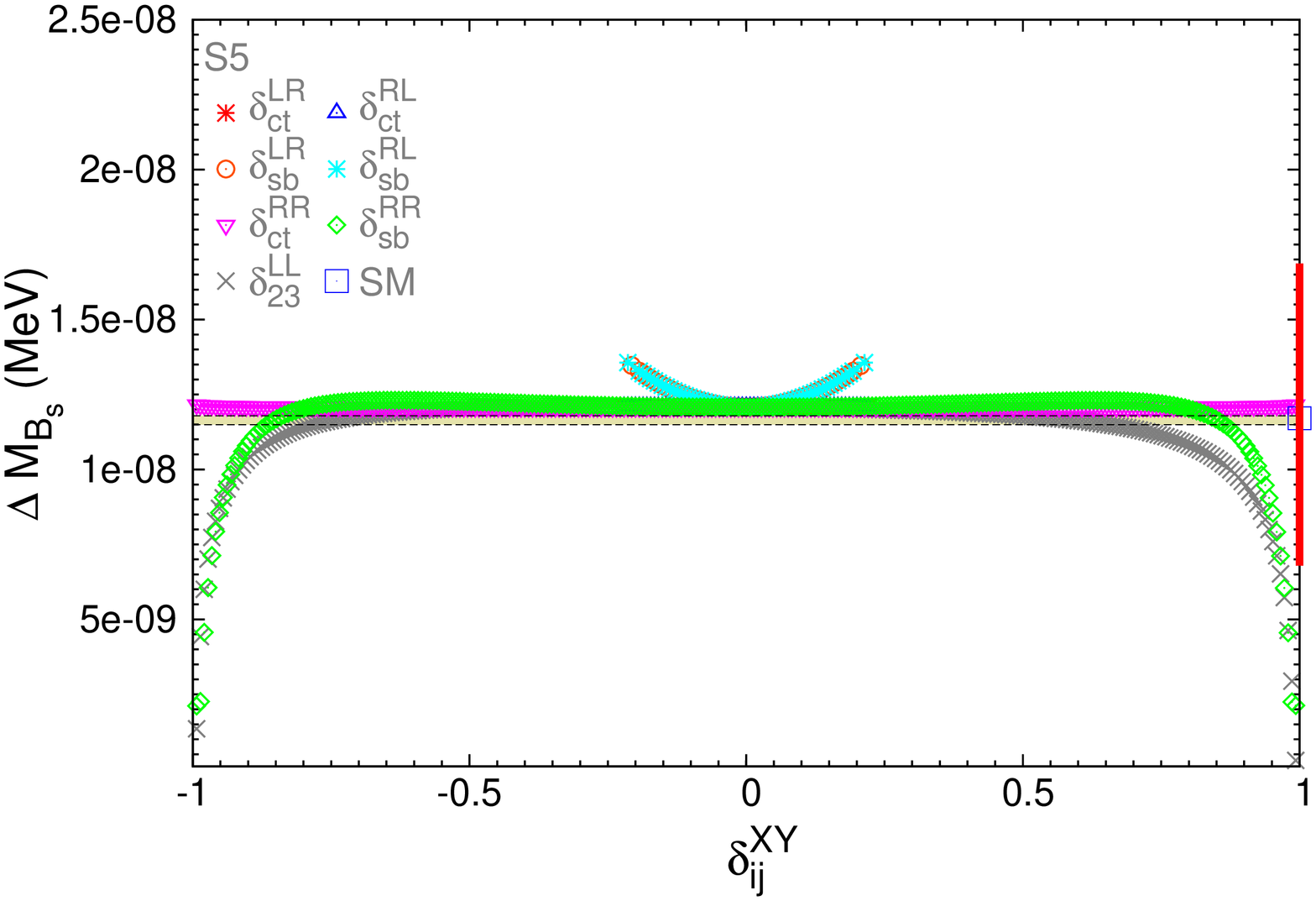}
\includegraphics[width=6.75cm,height=8cm,angle=270]{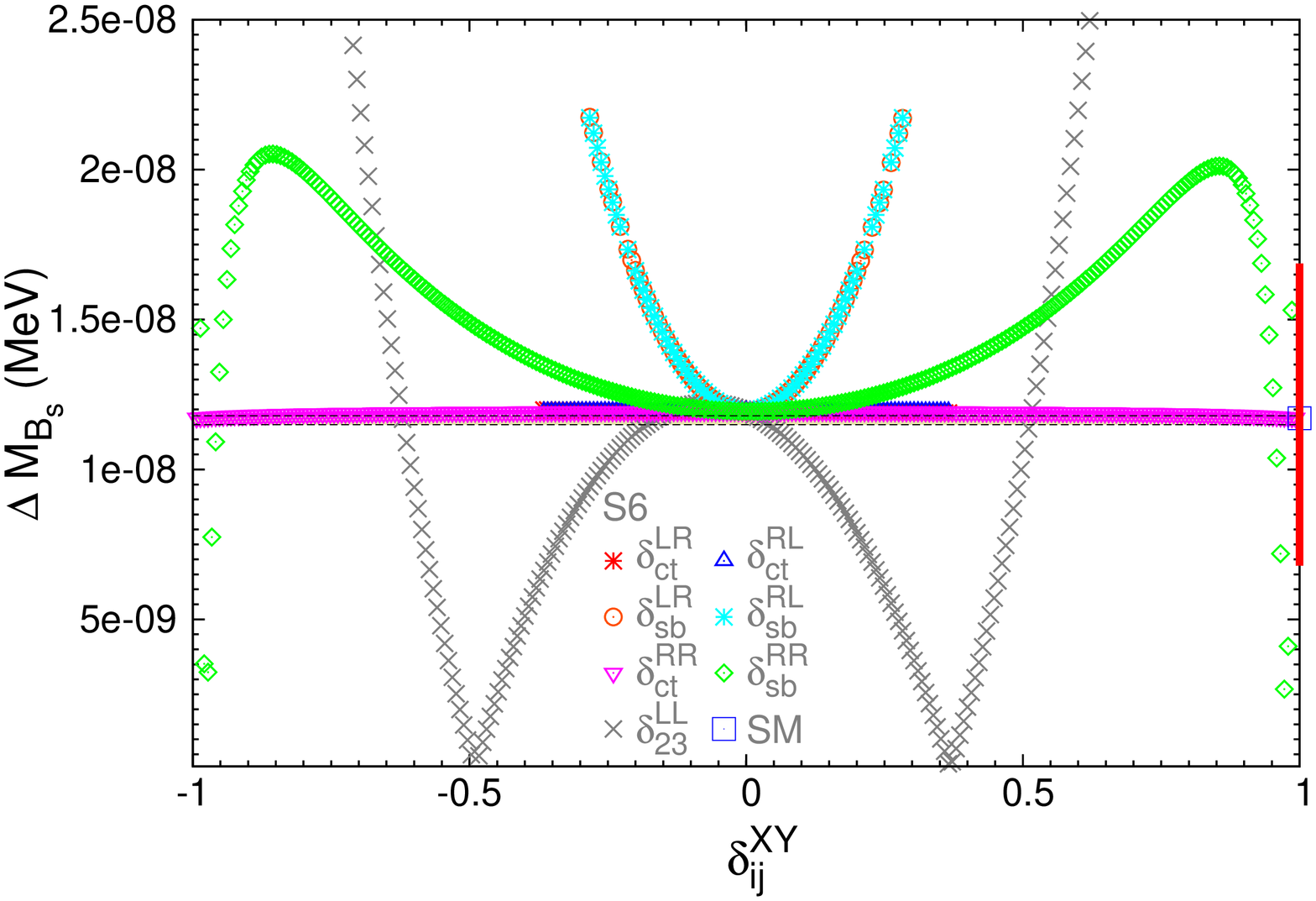}
\caption{Sensitivity to the NMFV deltas in \dmbs\ for the points S1\ldots S6.
  The experimental allowed $3\sigma$ area is the horizontal
  colored band. The SM prediction and the theory uncertainty $\Dtheo(\dmbs)$
  (red bar) is displayed on the right axis.}  
\label{fig:dmbs}
\end{figure}

In \reffis{fig:bsg} -- \ref{fig:dmbs} we show the results for the
three flavor observables discussed in \refses{sec:bsg} --
\ref{sec:dmbs}. The results are shown for the points S1\ldots S6, see
\refta{tab:spectra}, where the various \deABij\ are varied individually.
We have also included in the right vertical axis of these figures, for
comparison, the respective SM prediction in \refeqs{bsgamma-SM},
(\ref{bsmumu-SM}), and (\ref{deltams-SM}). The red error bars displayed are
the corresponding $3\,\si$ SM uncertainties 
(called $\De^{\rm theo}$).
The shadowed horizontal bands in all cases,  \bsg\ , \bmm\  and
\dmbs, are their corresponding experimental measurements 
in \refeqs{bsgamma-exp}, (\ref{bsmumu-exp}) and (\ref{deltams-exp}),
expanded with $3\,\si_{\rm exp}$ errors. 
In order to assess the total uncertainty the SM errors are also applied to
the MSSM predictions. If this error bar is outside the experimental band
the point can be regarded as excluded by the experimental
measurement. It should be noted that the theory uncertainties can be
larger in the MSSM than in the SM. However, estimates are much more
complicated than in the SM and strongly dependent on the chosen SUSY
parameters. Therefore we simply apply the SM uncertainty with $3\,\si$
errors. 

Regarding the explored intervals for the deltas in the following
\reffis{fig:bsg} -- \ref{fig:dmbs}, these will be $-1 \leq \deABij
\leq 1$, as discussed above. However, in some cases these
intervals are smaller: in computing the MSSM spectra with
non-vanishing $\deABij$ the code does not accept points that lead either
to too low MSSM masses, excluded by experiment, or even  non-physical
negative squared masses. This is, for instance, the case of
$\del{LR}{ij}$ and $\del{RL}{ij}$ with $ij=sb$ and $ij=ct$ that, as we
can see in \reffis{fig:bsg} -- \ref{fig:dmbs}, are explored in smaller
intervals since outside of them they lead to
negative squared scalar masses. In particular, the contributions from
the deltas leading to too low $\Mh$ will be studied further in the
following sections.

The analysis for \bsg\ is shown in \reffi{fig:bsg} in the scenarios 
S1\ldots S6. In the MFV case (i.e. for all $\deABij = 0$) we see
that all points, except S4, are in agreement
with experimental data. Only a very small variation with 
\del{LR}{ct}, \del{RL}{ct}, \del{RR}{ct}, \del{RR}{sb} (except for S4)
is observed. A clear dependence on \del{LL}{23} can be seen, placing bounds of
\order{0.1} on this NMFV parameter in all five scenarios, S1, S2, S3, S5
and S6. A very strong 
variation with \del{LR}{sb} and \del{RL}{sb} is found, which are
restricted to very small values $\leq$ \order{0.01} by the \bsg\ measurement.
In scenario S4 the strong variation with \del{LL}{23} or
\del{LR}{sb} can bring the prediction into agreement with the
experimental data. Turning the argument around, the scenario S4, which
appears to be excluded by the \bsg\ measurement is actually a valid
scenario for certain values of \del{LL}{23} and \del{LR}{sb}.

The results for \bmm\ are shown in \reffi{fig:bmm} for the six
scenarios. All \deABij\ show a relatively small impact, except for
\del{LL}{23}. From these plots we find the following allowed
intervals for \del{LL}{23}: $S1:(-0.3,0.7)$ , $S2:(-0.3,0.8)$,
$S3:(-0.1,0.2)$, $S4:(-0.3,0.3)$, $S6:(-0.3,0.8)$. Therefore, bounds
on this parameter ranging between $\sim - 0.1$ and  
$\sim + 0.8$ can be set in all scenarios except in S5 where we do not
get any constraint. This scenario 
is characterized by a very large value of $\MA = 1000 \gev$ and a
relatively small value of $\tb = 10$, leading to a strong suppression of
the contributions to \bmm. 

The predictions for \dmbs\ in the six scenarios are shown in
\reffi{fig:dmbs}. While the experimental precision is very high the
theoretical error is quite large, and the 
bounds on the \deABij\ are mainly given by the SM uncertainty in the
\dmbs\ prediction. All six scenarios for all $\deABij = 0$ are in
agreement with the experimental data, once the SM uncertainty is taken
into account. Except for S4, which is sensitive to all deltas, the
other points are nearly insensitive to \del{RR}{ct}, \del{LR}{ct} and
\del{RL}{ct}, therefore we do not get any additional bound for them in
the allowed range from this observable.
An important variation can be observed for 
\del{LR}{sb} and \del{RL}{sb}. However, due to the MSSM particle mass
restrictions commented above which shortened the allowed intervals,
hardly any new bounds are placed by \dmbs, except in S4 
and S6. Some sensitivity is found for \del{RR}{sb}, especially in S4 and
S6 where $|\del{RR}{sb}|$ is bounded by $\leq \order{0.5}$.  The
strongest variation is found for \del{LL}{23}, where due to the
particular 'W-shape' dependence, both 
intermediate and large values can be excluded.  
 
The overall allowed intervals for the seven \deABij\ in the six
scenarios and considering the three observables together, \bsg, \bmm
and \dmbs, can be found in \refta{tab:boundsS1S6}. From this table we
then conclude on the strongest bounds that can be 
obtained from the combination of all three $B$-physics observables. 

As a general comment, the main restrictions to the deltas come from
\bsg\ and in some cases from \dmbs\ and not yet from the young
measurement \bmm. 
The most restricted deltas are \del{LR}{sb} and \del{RL}{sb} that can
reach values at most of \order{0.01}, then \del{LL}{23}, \del{LR}{ct}
and \del{RL}{ct} that can be at most of \order{0.1}, with the first one
being slightly more restricted than the last two, and finally the less
restricted deltas are \del{RR}{ct} and \del{RR}{sb} that in general can
reach up to the largest explored values of \order{1}. Special
attention deserves scenario S4, where, as mentioned above, setting
$\deABij = 0$ leads to experimentally excluded predictions. Only non-zero
values of \del{LR}{sb} can reconcile this scenario with experimental
data. Consequently, assuming only {\em one} $\deABij$ different from zero
leads to an ``excluded'' scenario for all the other $\deABij$ as shown
in \refta{tab:boundsS1S6}.

It can also be seen that larger constraints in the ``$sb$ sector'' than in
the ``$ct$ sector'' are obtained, since the $B$-physics
observables are in general 
more sensitive to mixing among $b$-type squarks. We also see that the 
\del{LR}{} become more restricted than the others, since they involve the
trilinear couplings that provide in general large corrections.  

Regarding the comparison of our results with previous studies, we
conclude that the bounds on the squark mixing deltas that we find here
for the scenarios S1-S6 are more relaxed than in the set of benchmark
scenarios that were analyzed in \cite{mhNMFV} before the LHC started
operation. The scenarios investigated in the pre-LHC era contained
relatively light scalar quarks (now excluded), leading to relatively
large radiative corrections from NMFV effects. After the so far
unsuccessful search for beyond SM physics at the LHC, scalar quark
masses (in particular those of the first and second generation) have
substantially higher lower bounds. Benchmark scenarios that take this
into account (as our S1-S6) naturally permit larger values for the NMFV
deltas.

\renewcommand{\arraystretch}{1.1}
\begin{table}[H]
\begin{center}
\resizebox{9.0cm}{!} {
\begin{tabular}{|c|c|c|} \hline
 & & Total allowed intervals \\ \hline
$\delta^{LL}_{23}$ & \begin{tabular}{c}  S1 \\ S2 \\ S3 \\ S4 \\ S5 \\ S6 \end{tabular} &  
\begin{tabular}{c} 
(-0.27:0.28) \\ (-0.23:0.23) \\ (-0.12:0.06) (0.17:0.19) \\ excluded \\ (-0.83:-0.78) (-0.14:0.14) \\ (-0.076:0.14) \end{tabular} \\ \hline
$\delta^{LR}_{ct}$  & \begin{tabular}{c}  S1 \\ S2 \\ S3 \\ S4 \\ S5 \\ S6 \end{tabular}    
& \begin{tabular}{c} 
(-0.27:0.27) \\ (-0.27:0.27) \\ (-0.27:0.27) \\ excluded \\ (-0.22:0.22) \\ (-0.37:0.37) \end{tabular}   \\ \hline
$\delta^{LR}_{sb}$  & \begin{tabular}{c}  S1 \\ S2 \\ S3 \\ S4 \\ S5 \\ S6 \end{tabular}    & 
\begin{tabular}{c} 
(-0.0069:0.014) (0.12:0.13) \\ (-0.0069:0.014) (0.11:0.13) \\ (-0.0069:0.014) (0.11:0.13) \\ (0.076:0.12) (0.26:0.30) \\ (-0.014:0.021) (0.17:0.19) \\ (0:0.0069) (0.069:0.076) \end{tabular}  \\ \hline
$\delta^{RL}_{ct}$  & \begin{tabular}{c}  S1 \\ S2 \\ S3 \\ S4 \\ S5 \\ S6 \end{tabular}    & 
\begin{tabular}{c}
(-0.27:0.27) \\ (-0.27:0.27) \\ (-0.27:0.27) \\ excluded \\ (-0.22:0.22) \\ (-0.37:0.37) \end{tabular} 
  \\ \hline
$\delta^{RL}_{sb}$  & \begin{tabular}{c}  S1 \\ S2 \\ S3 \\ S4 \\ S5 \\ S6 \end{tabular}    &
\begin{tabular}{c} (-0.034:0.034) \\ (-0.034:0.034) \\ (-0.034:0.034) \\ excluded \\ (-0.062:0.062) \\ (-0.021:0.021) \end{tabular} 
  \\ \hline
$\delta^{RR}_{ct}$ & \begin{tabular}{c}  S1 \\ S2 \\ S3 \\ S4 \\ S5 \\ S6 \end{tabular}   & \begin{tabular}{c} 
(-0.99:0.99) \\ (-0.99:0.99) \\ (-0.98:0.97) \\ excluded \\ (-0.99:0.99) \\ (-0.96:0.94)  \end{tabular}    \\ \hline
$\delta^{RR}_{sb}$  & \begin{tabular}{c}  S1 \\ S2 \\ S3 \\ S4 \\ S5 \\ S6 \end{tabular}    &
\begin{tabular}{c}  (-0.96:0.96) \\ (-0.96:0.96) \\ (-0.96:0.94) \\ excluded \\ (-0.97:0.97) \\ (-0.97:-0.94) (-0.63:0.64) (0.93:0.97)
\end{tabular}    \\ \hline
\end{tabular}}  
\end{center}
\caption{Present allowed intervals on the squark mixing parameters
  $\delta^{AB}_{ij}$ for the selected S1-S6 MSSM points defined in
  \refta{tab:spectra}. 
}
\label{tab:boundsS1S6}
\end{table}
\renewcommand{\arraystretch}{1.55}


\subsection{Framework 1: effects on Higgs boson masses}

\begin{figure}[htb!] 
\centering
\includegraphics[width=7.10cm,height=8cm,angle=270]{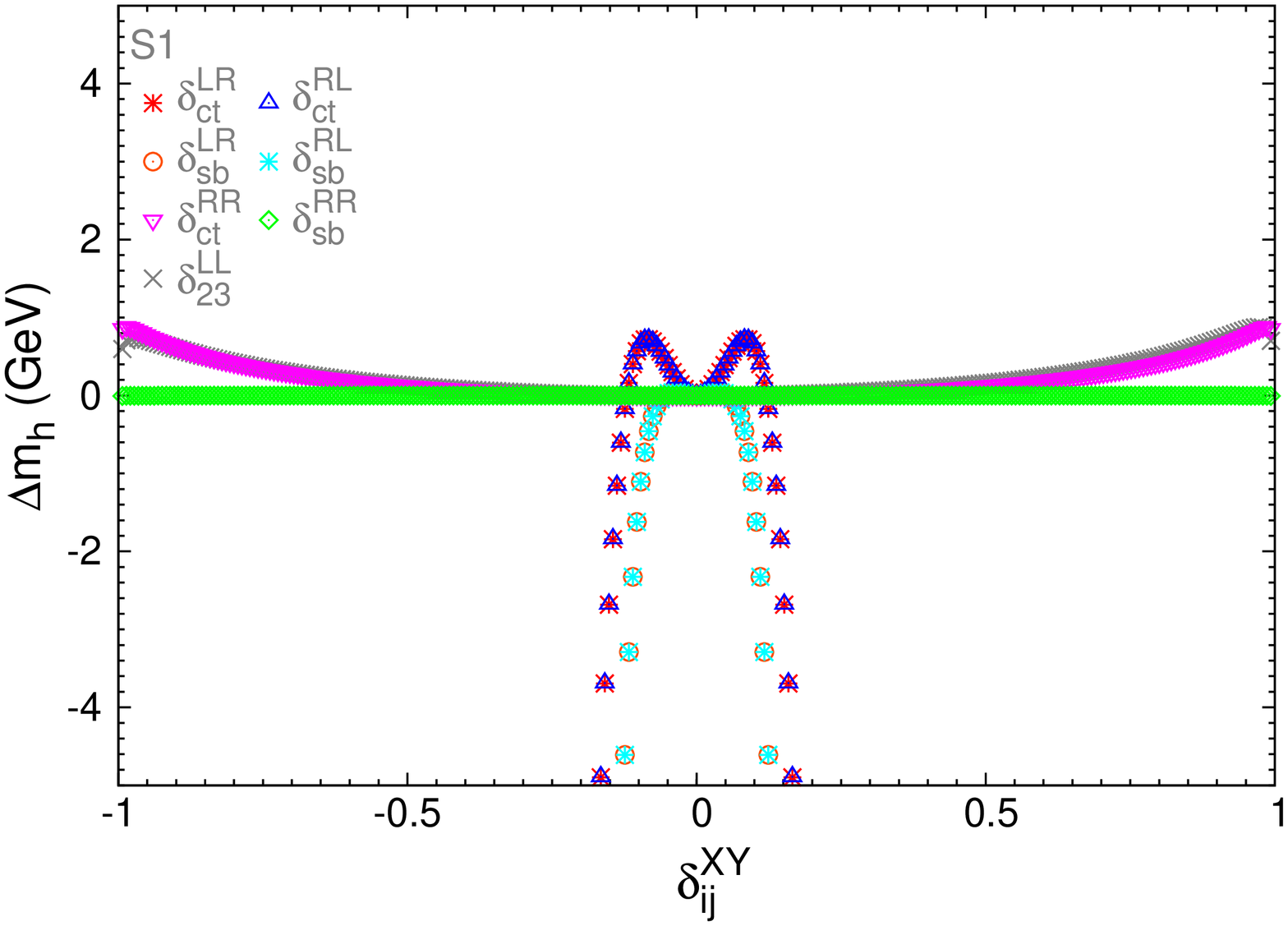}
\includegraphics[width=7.10cm,height=8cm,angle=270]{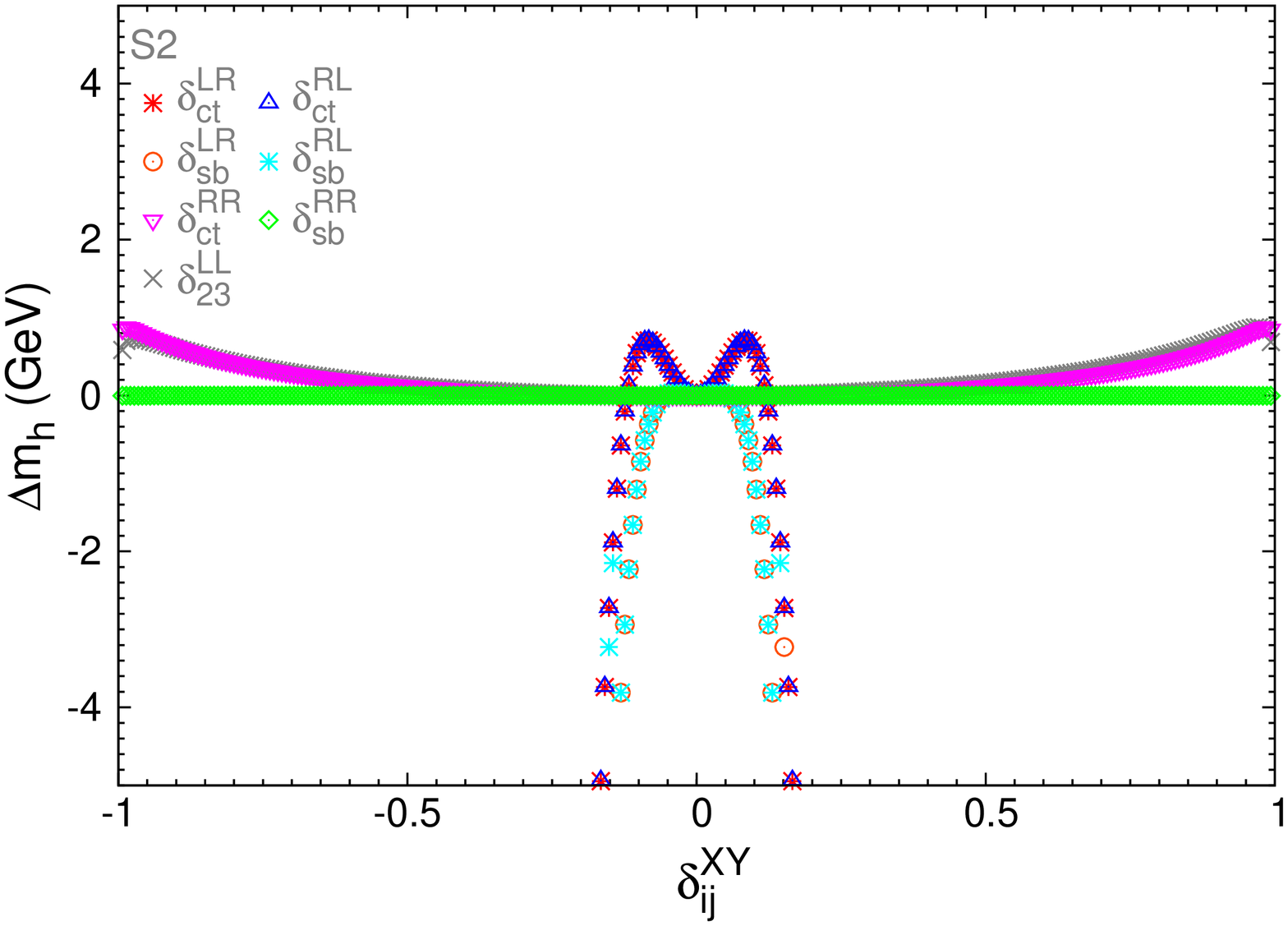}\\
\includegraphics[width=7.10cm,height=8cm,angle=270]{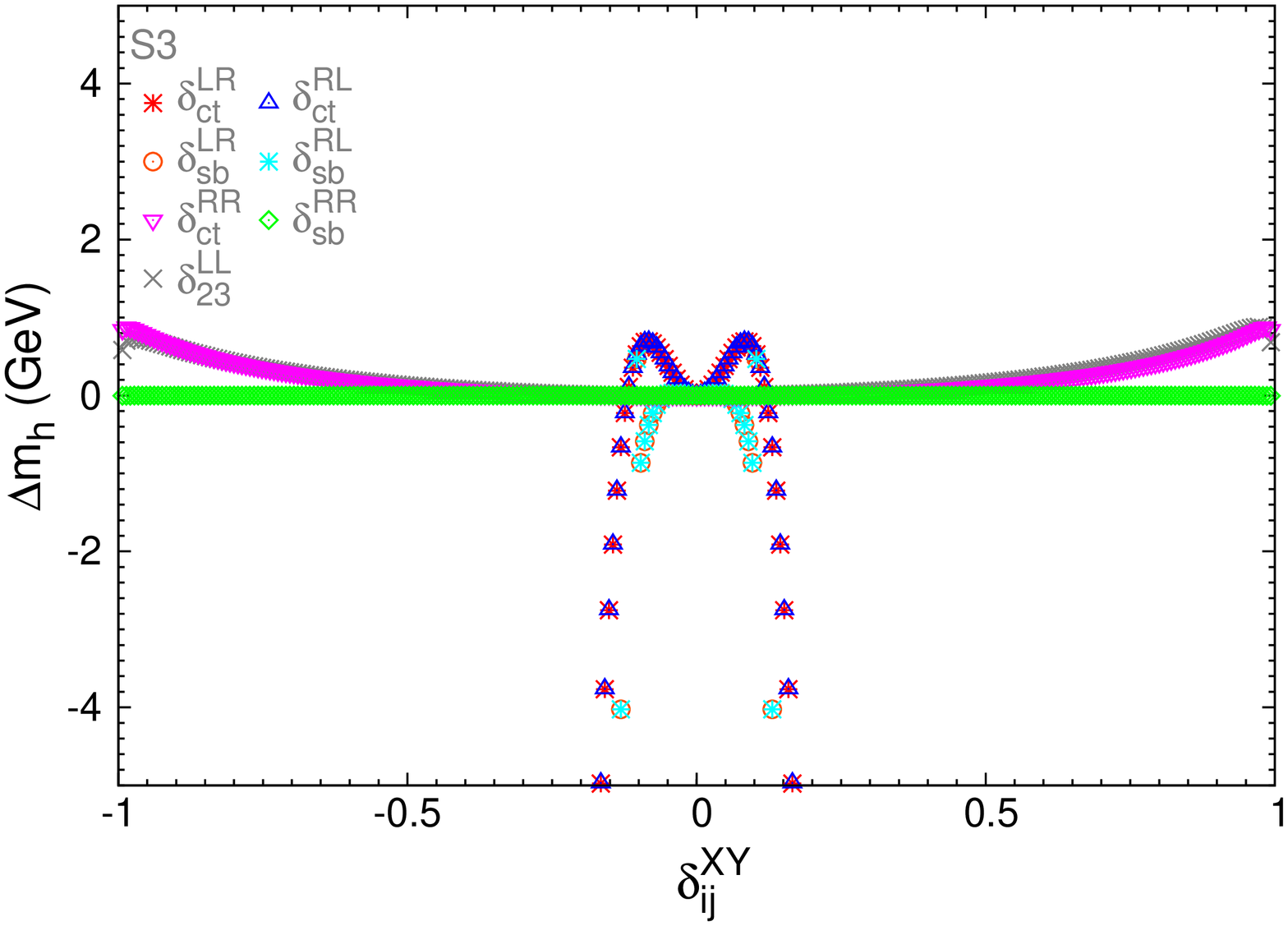}
\includegraphics[width=7.10cm,height=8cm,angle=270]{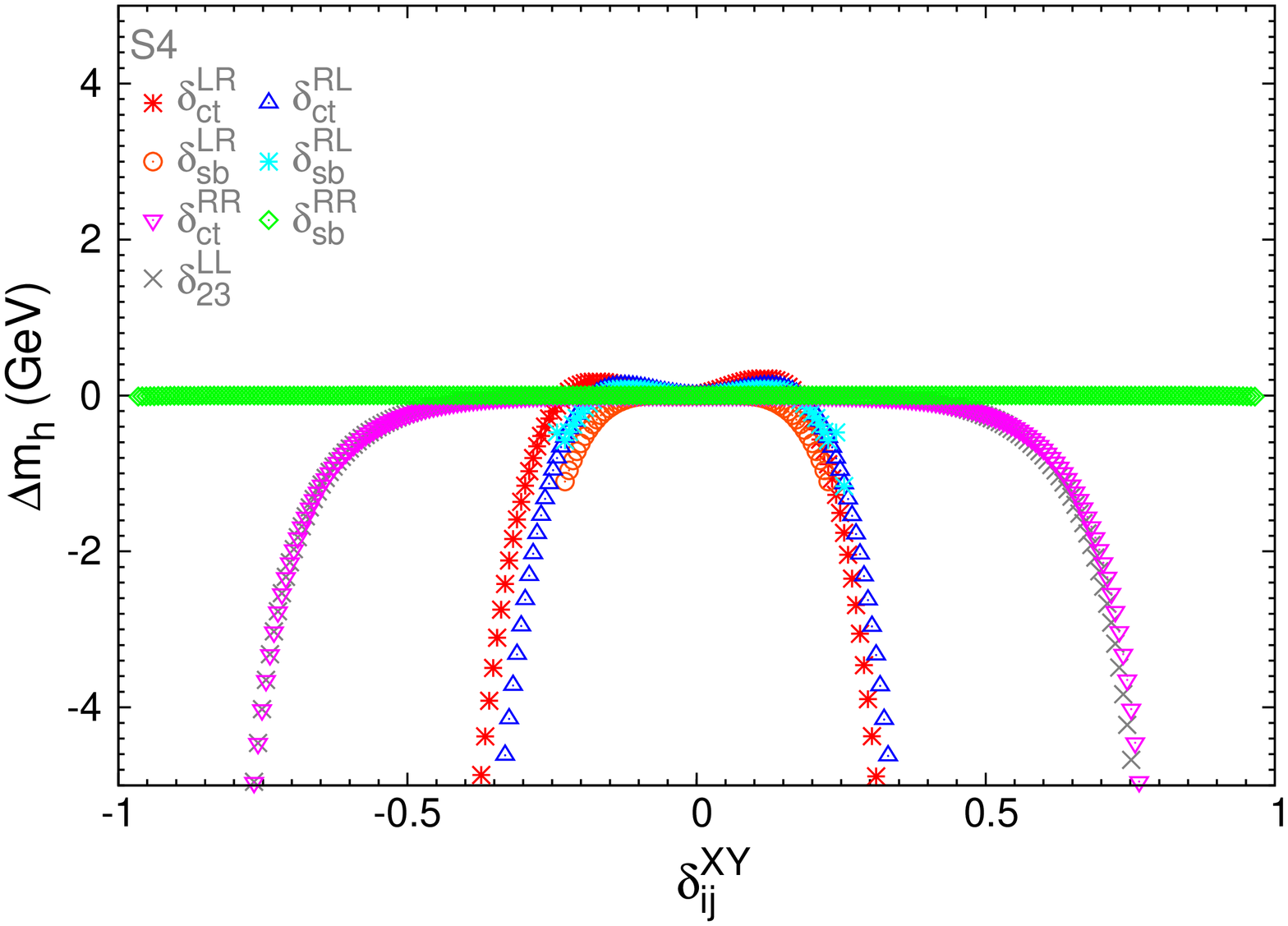}\\
\includegraphics[width=7.10cm,height=8cm,angle=270]{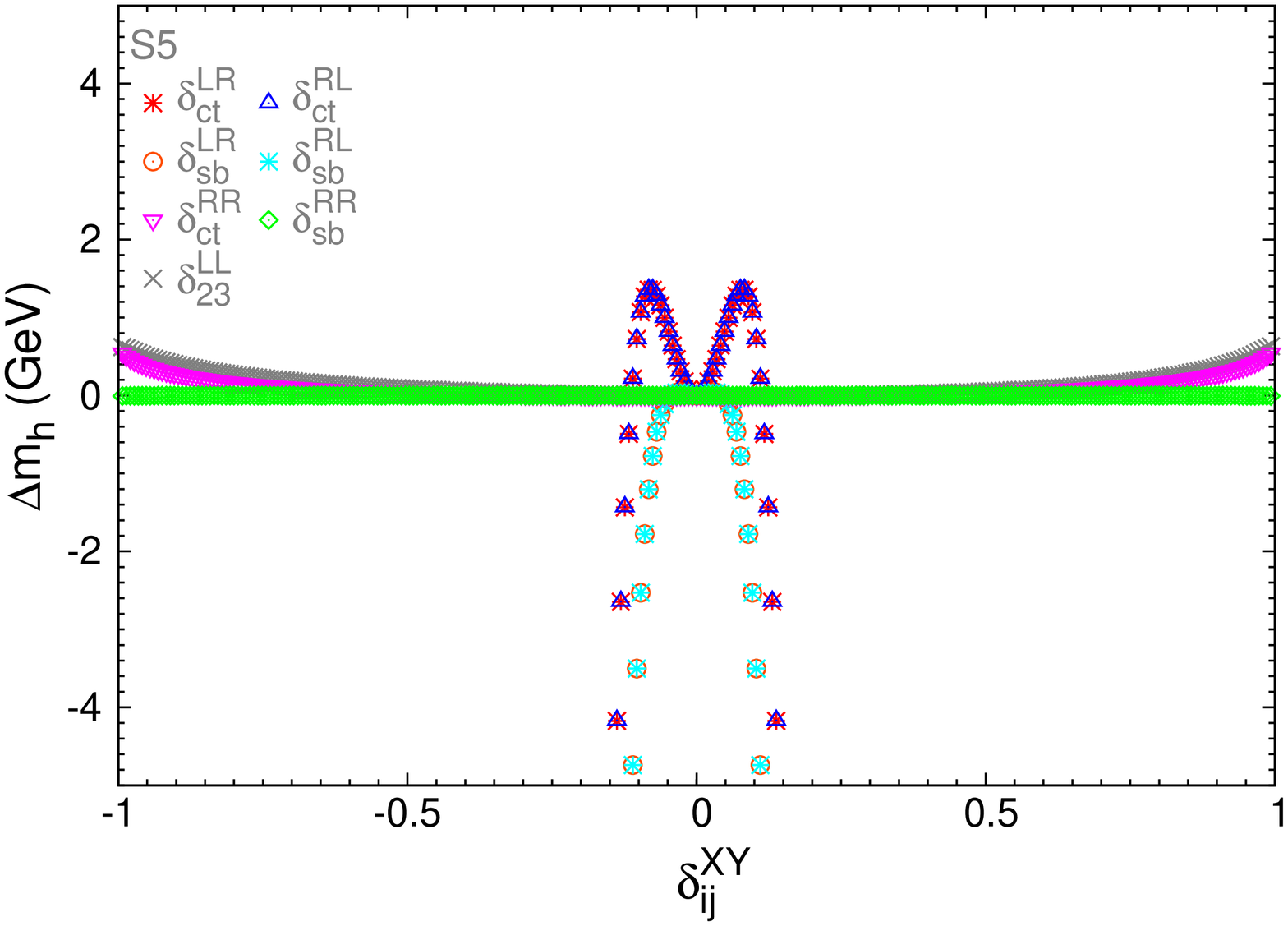}
\includegraphics[width=7.10cm,height=8cm,angle=270]{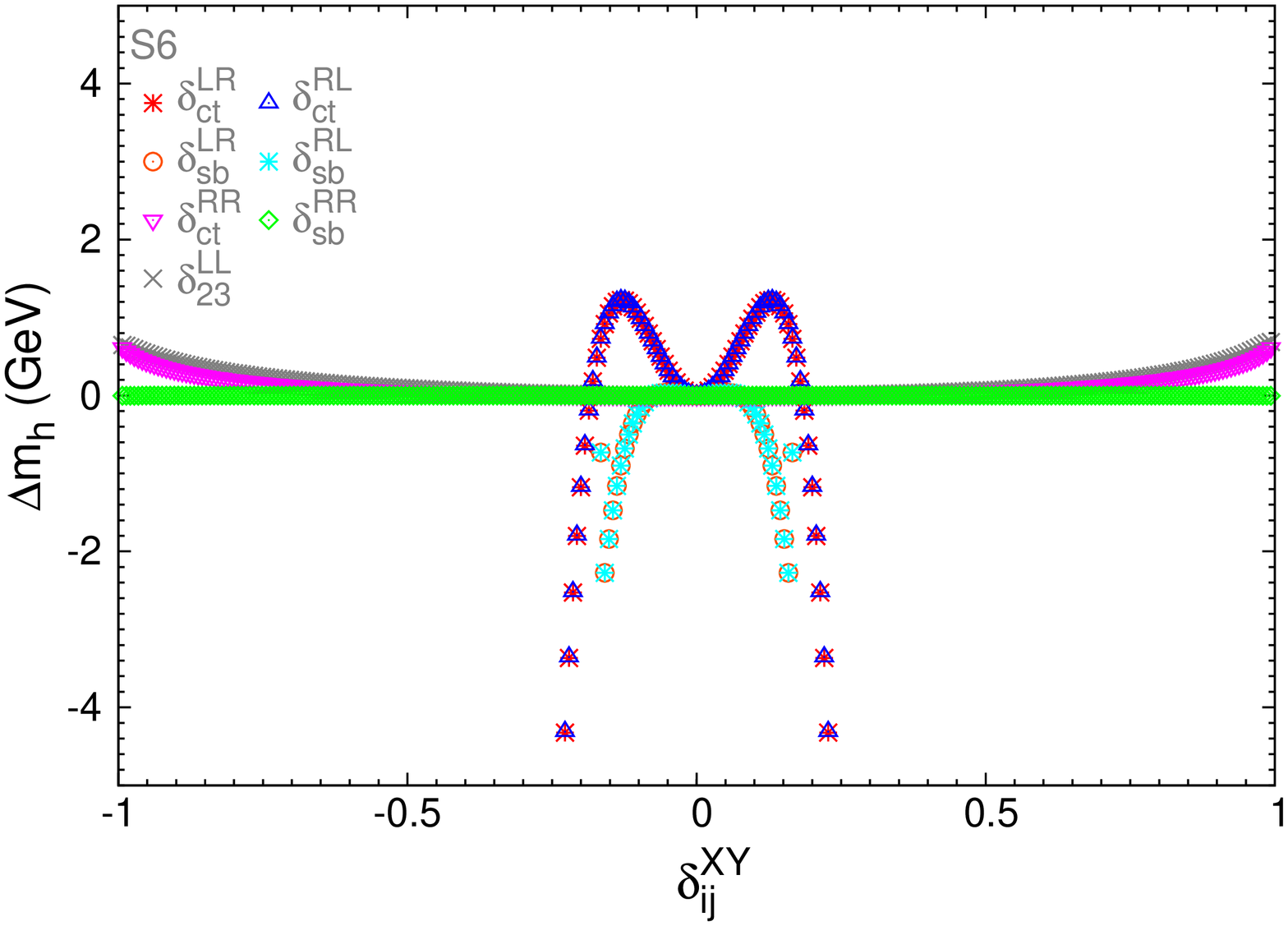}
\caption{One-loop corrections to $\Mh$ in the scenarios S1\ldots S6.
}
\label{fig:h0}
\end{figure}

\begin{figure}[htb!] 
\centering
\includegraphics[width=7.10cm,height=8cm,angle=270]{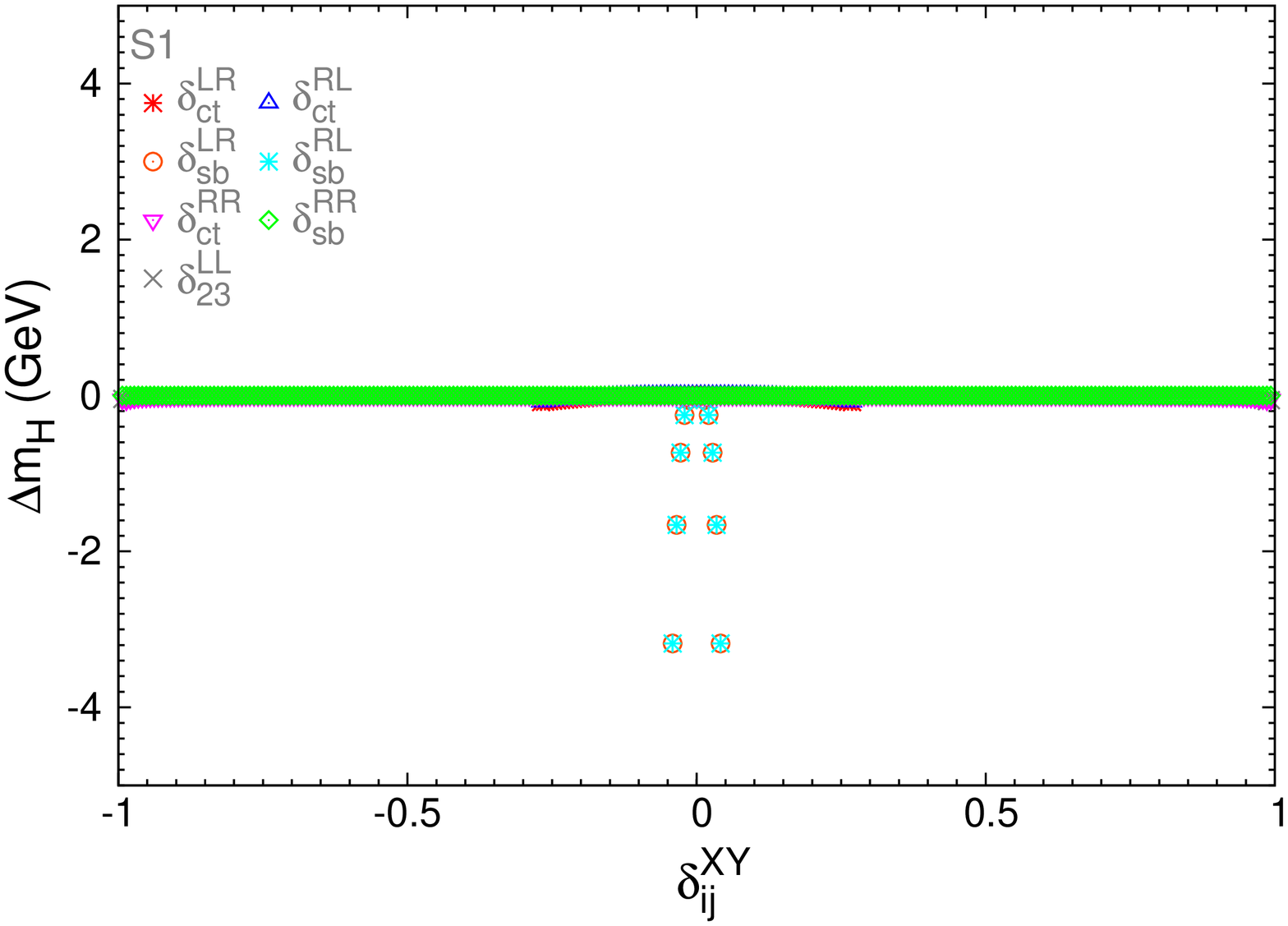}
\includegraphics[width=7.10cm,height=8cm,angle=270]{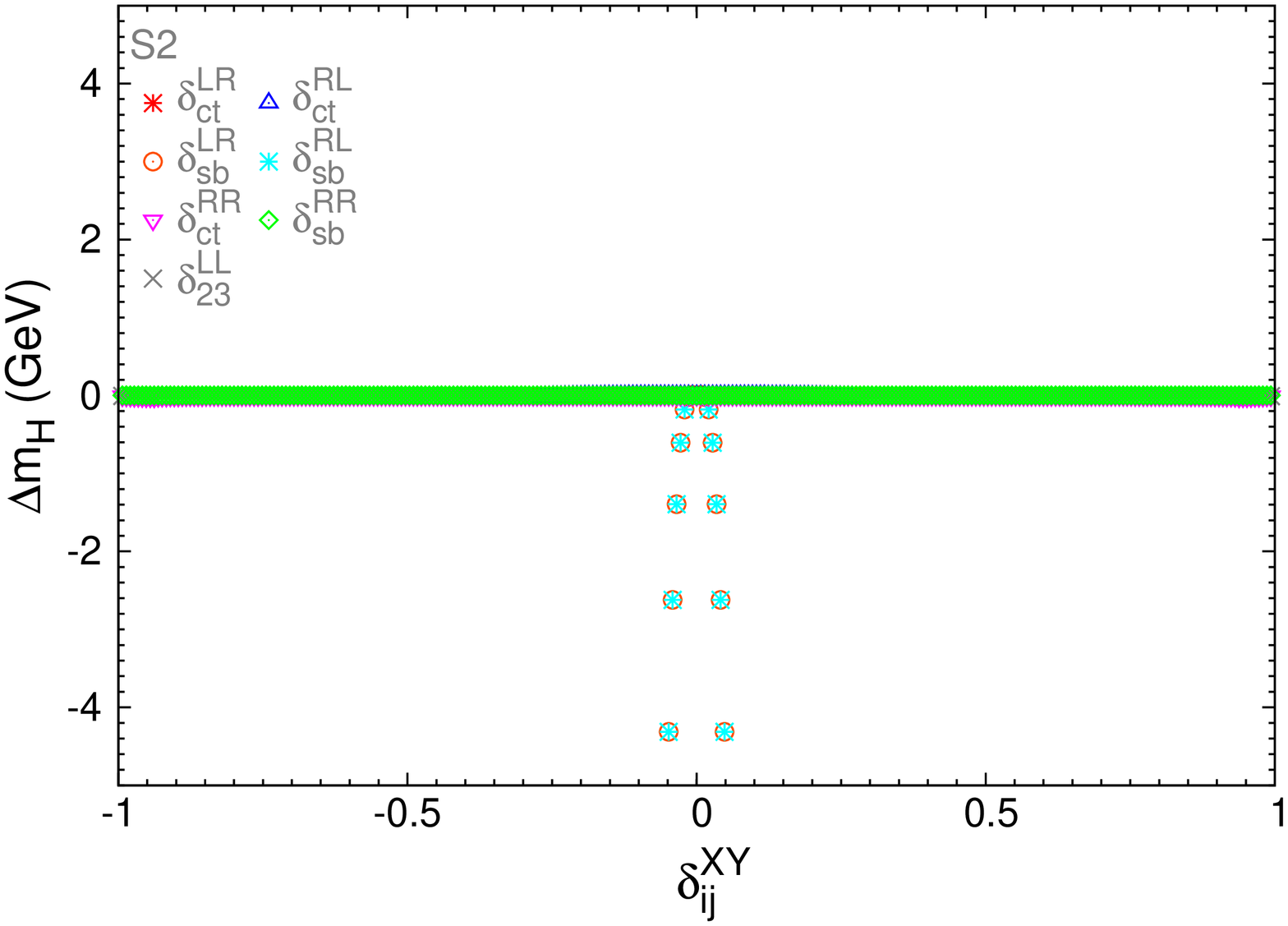}\\
\includegraphics[width=7.10cm,height=8cm,angle=270]{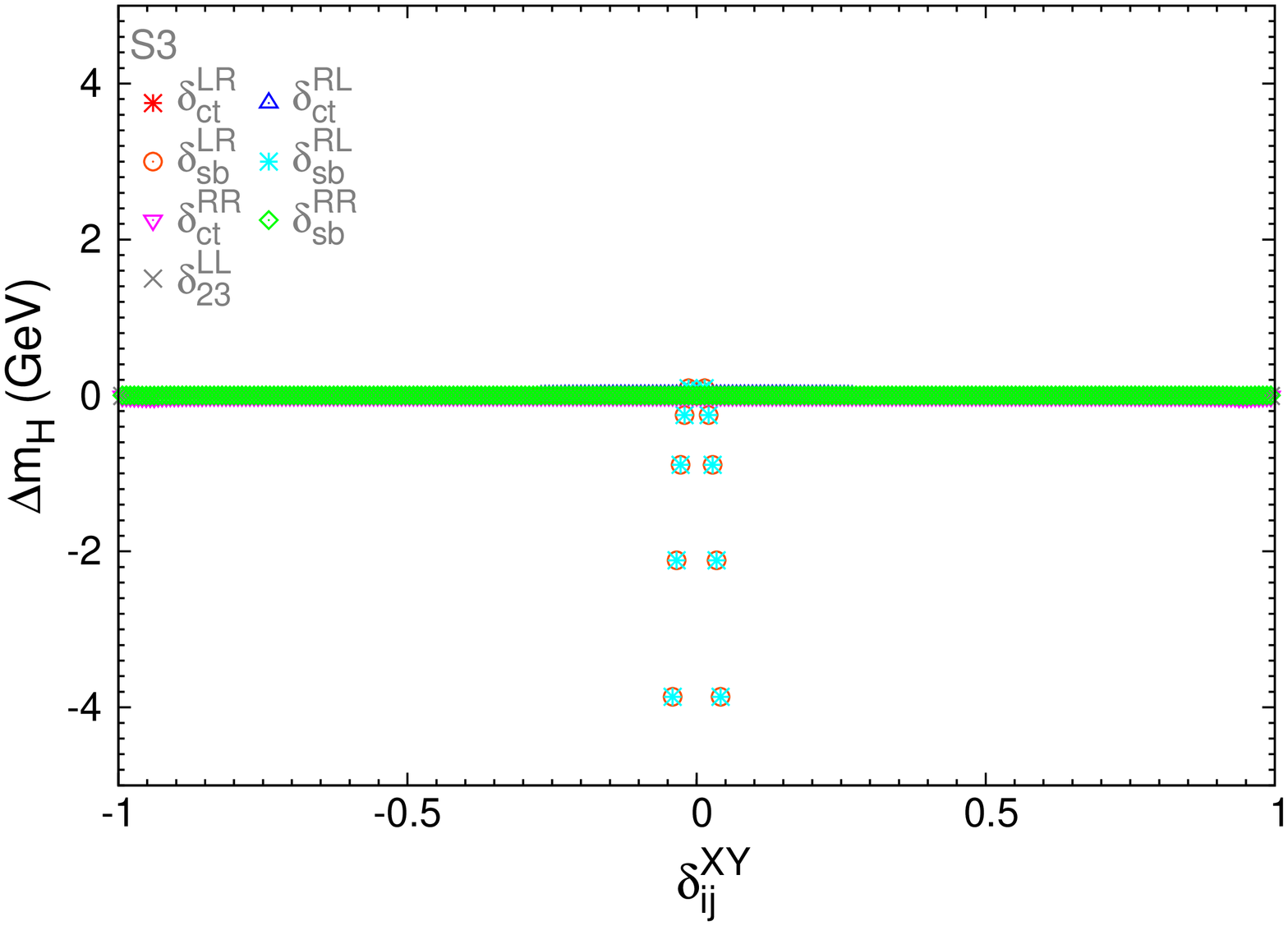}
\includegraphics[width=7.10cm,height=8cm,angle=270]{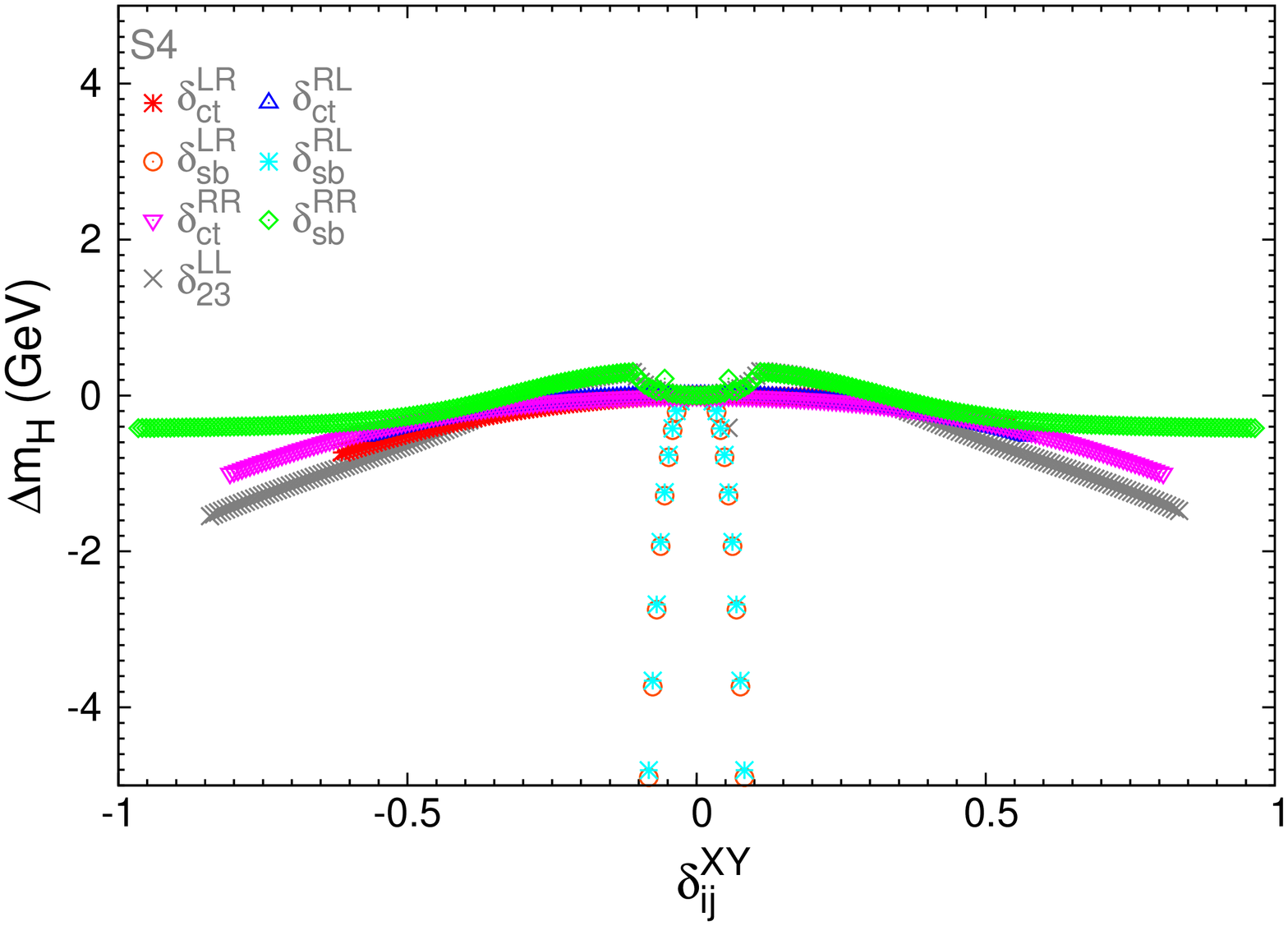}\\
\includegraphics[width=7.10cm,height=8cm,angle=270]{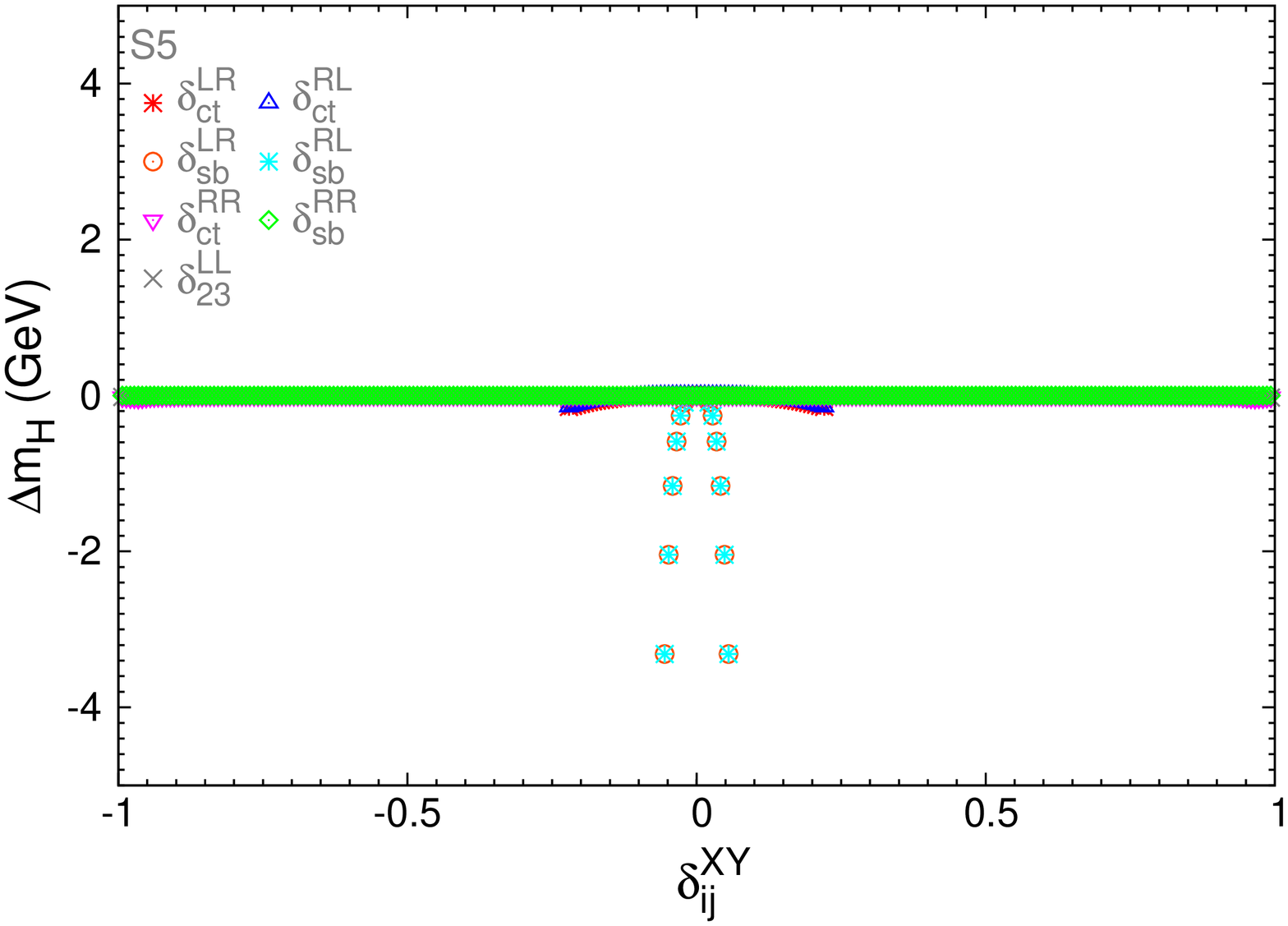}
\includegraphics[width=7.10cm,height=8cm,angle=270]{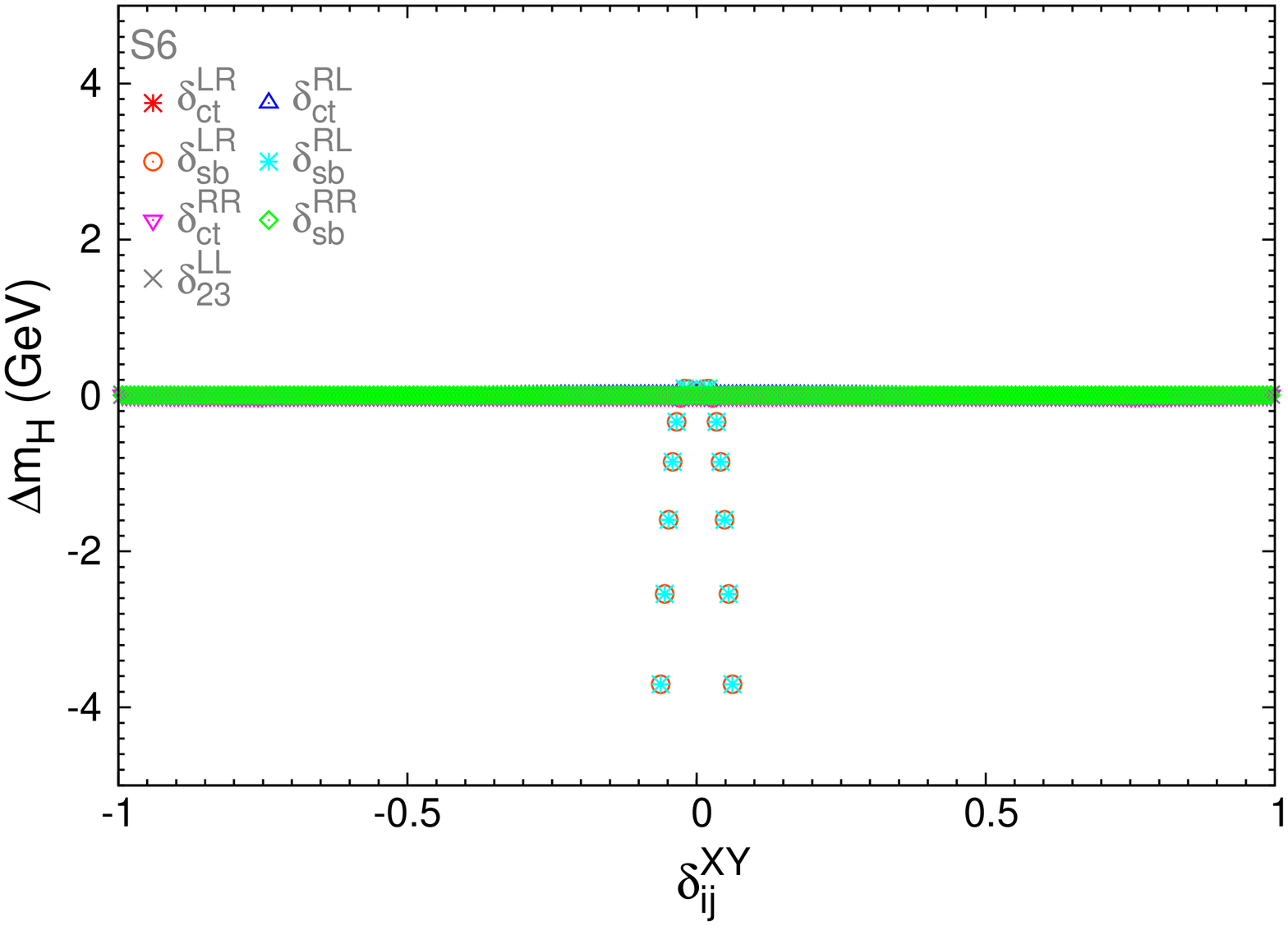}
\caption{One-loop corrections to $\MH$ in the scenarios S1\ldots S6.
}
\label{fig:H0}
\end{figure}

\begin{figure}[htb!] 
\centering
\includegraphics[width=7.10cm,height=8cm,angle=270]{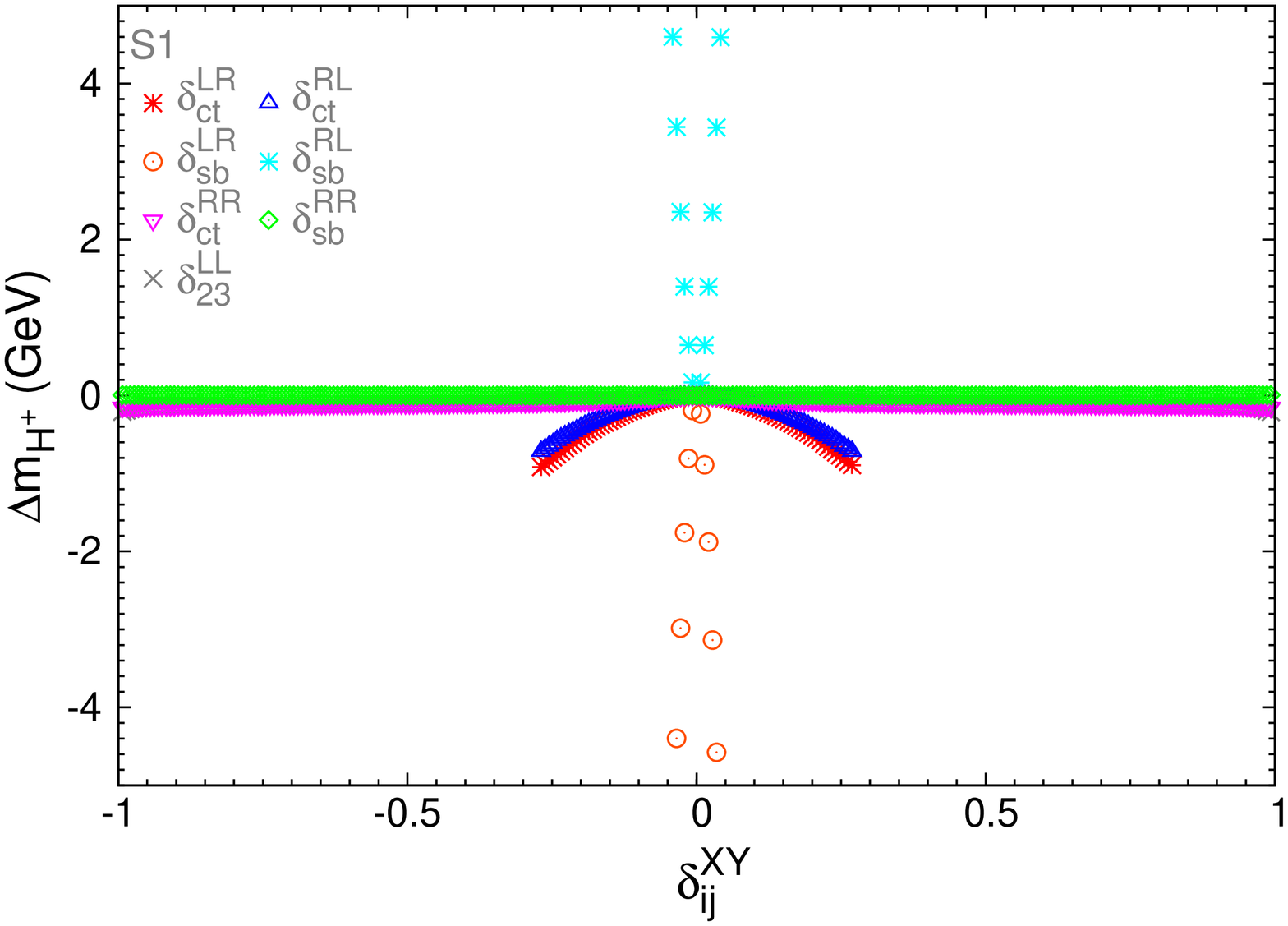}
\includegraphics[width=7.10cm,height=8cm,angle=270]{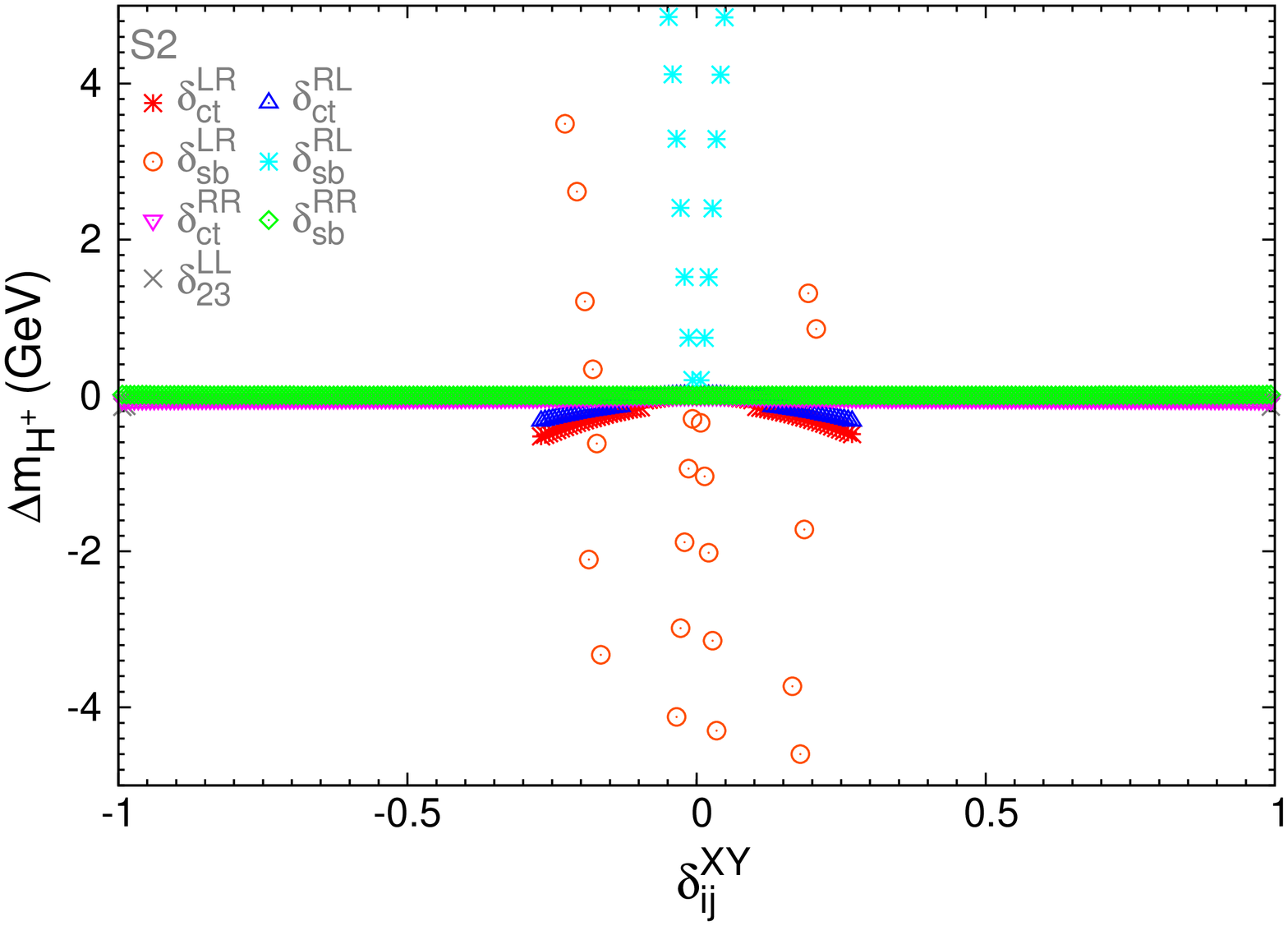}\\
\includegraphics[width=7.10cm,height=8cm,angle=270]{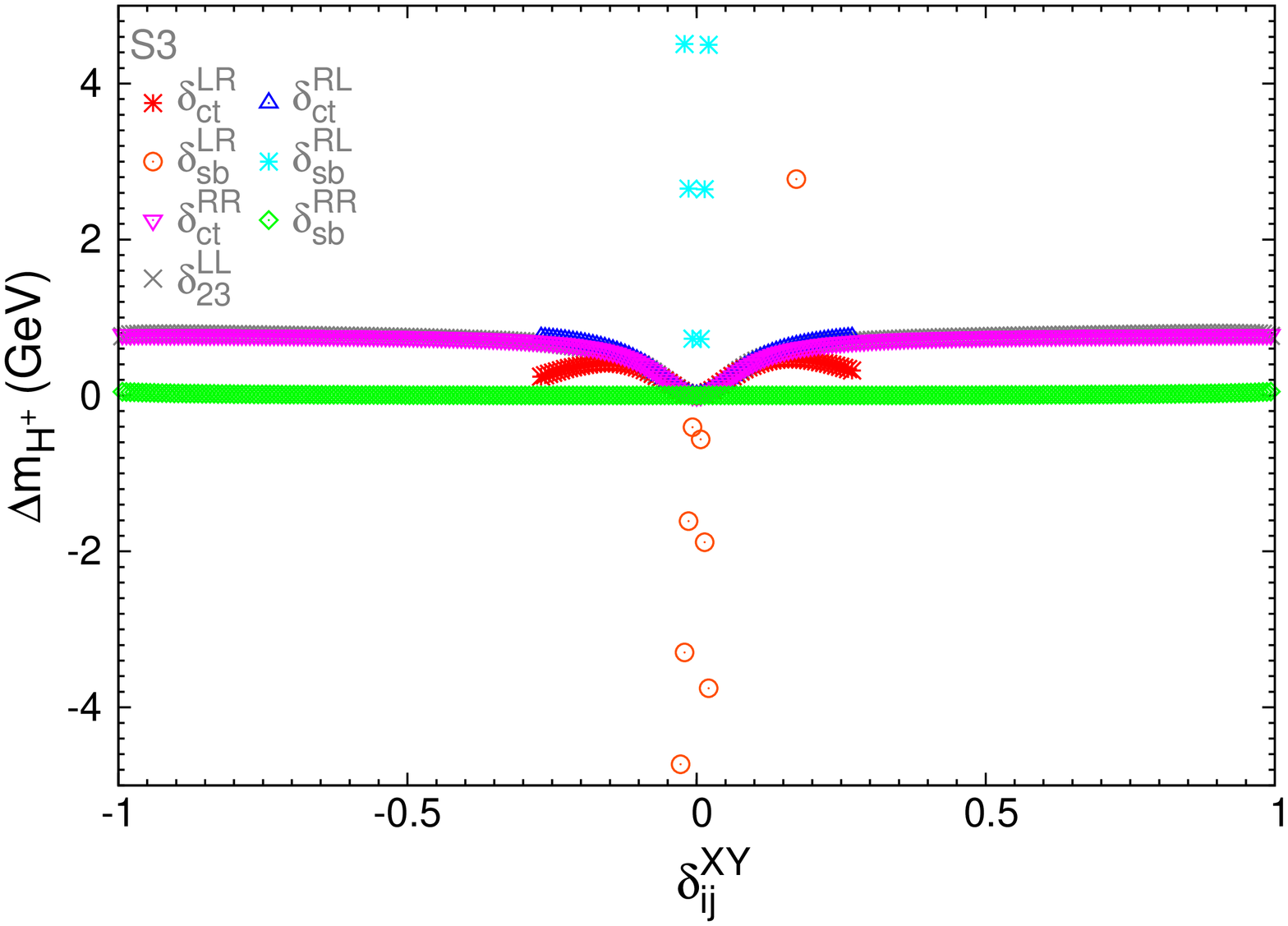}
\includegraphics[width=7.10cm,height=8cm,angle=270]{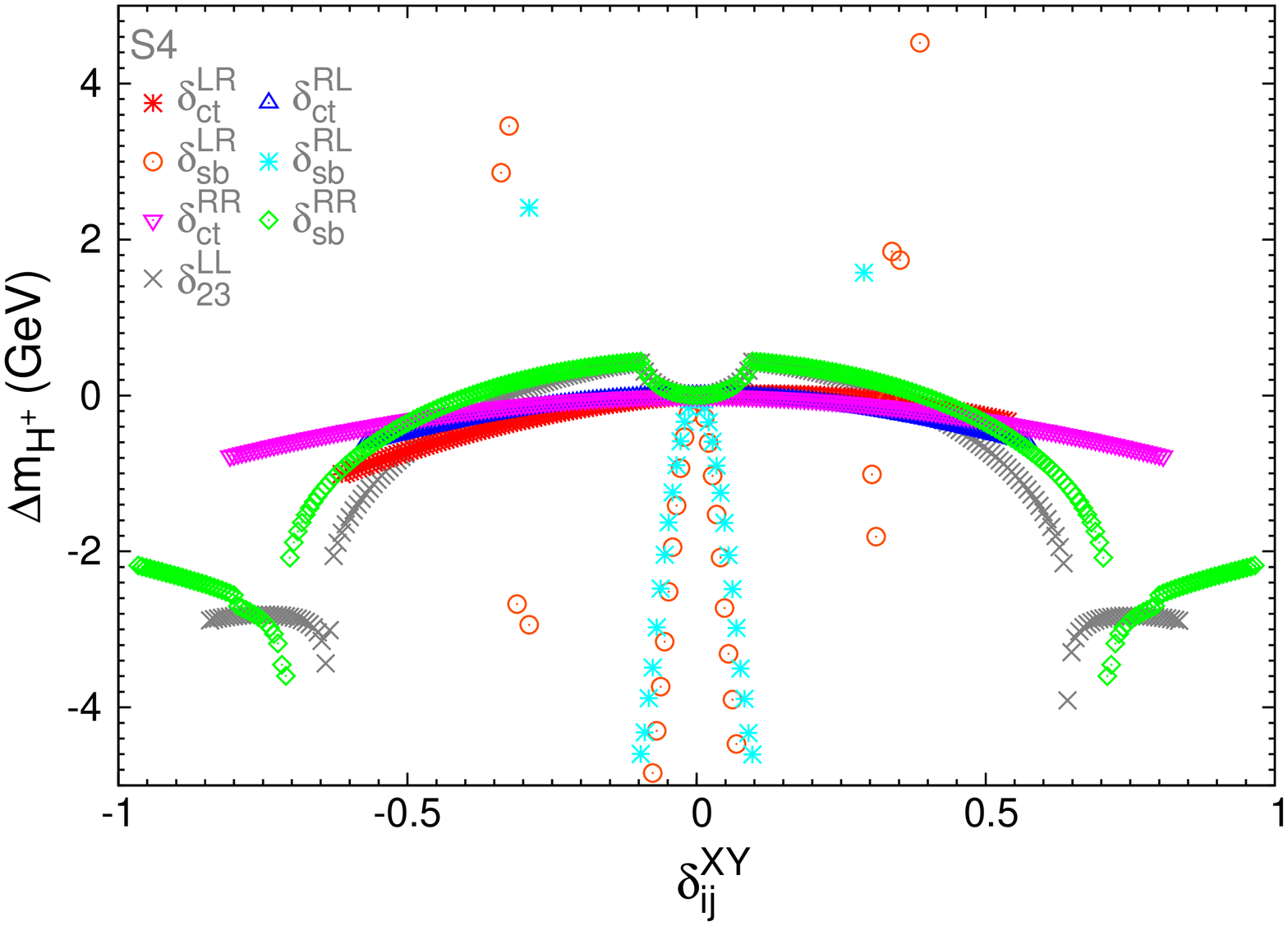}\\
\includegraphics[width=7.10cm,height=8cm,angle=270]{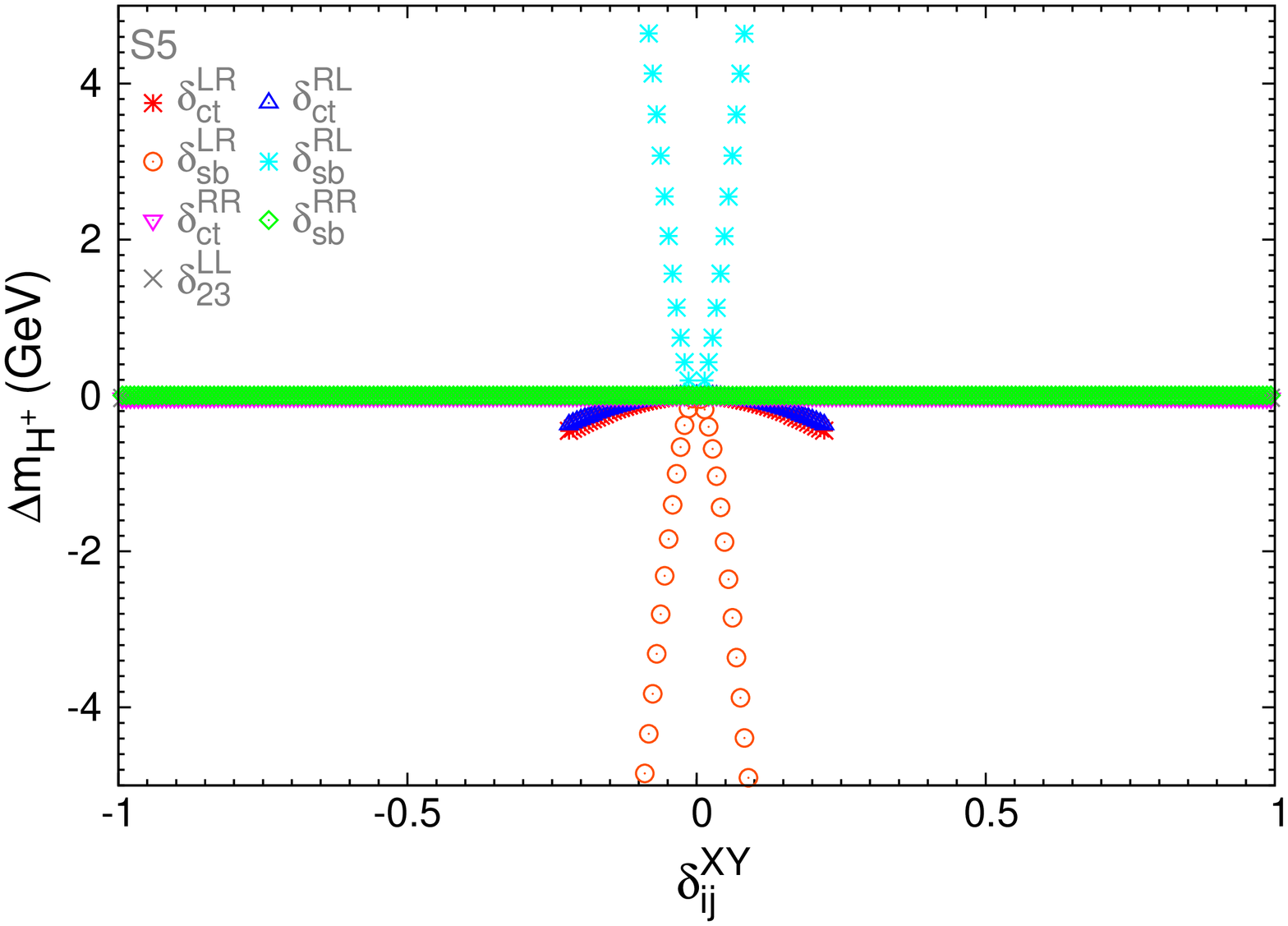}
\includegraphics[width=7.10cm,height=8cm,angle=270]{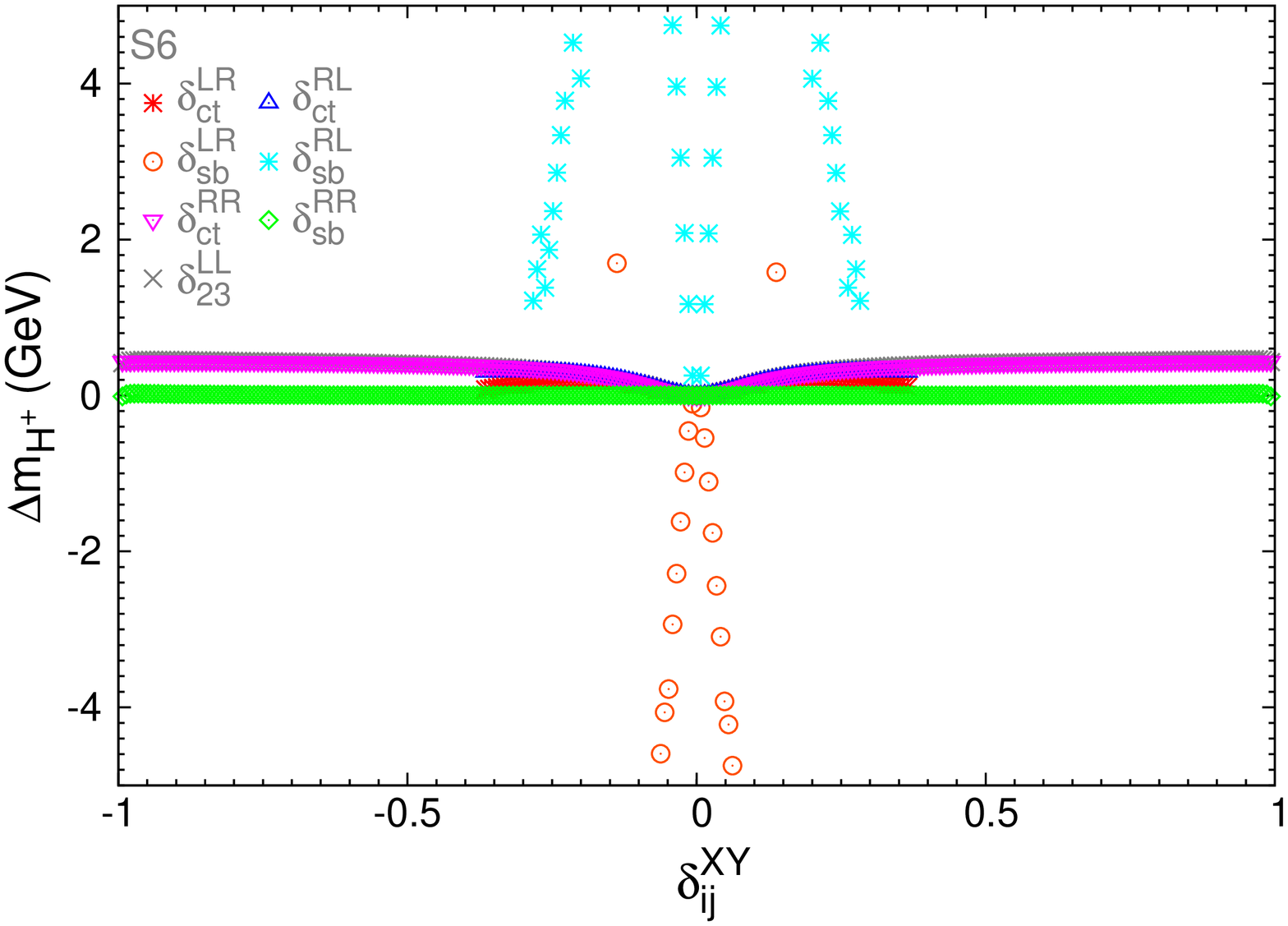}
\caption{One-loop corrections to $\MHp$ in the scenarios S1\ldots S6.
}
\label{fig:Hp}
\end{figure}

In this section we discuss the one-loop NMFV effects on the Higgs boson
masses. A more detailed description of the computation of these
one-loop NMFV effects in terms of one-loop diagrams and the
corresponding corrections to the involved self-energies can be found
in \citere{mhNMFV}. We are interested here mainly in the differences
between the predictions within NMFV and MFV.
We show, in \reffis{fig:h0}, \ref{fig:H0} and \ref{fig:Hp},
\begin{align}
\De m_\phi := M_\phi^{\rm NMFV} - M_\phi^{\rm MFV}, \quad \phi = h, H, H^\pm
\label{dmh}
\end{align}
as a function of \deABij\ in the scenarios S1\ldots S6.

We start our investigation with $\De\mh$ in \reffi{fig:h0}. 
Bounds on the \deABij\ can in principle only be placed by the $\Mh$
prediction, since this is the only mass parameter that has been
measured experimentally so far. 
It should be noted that the value of $\Mh^{\rm NMFV}$ depends
strongly on the MFV SUSY parameters, in particular on $\Xt$ (where 
$\mt\Xt$ is the off-diagonal entry in the scalar top mass matrix).
Consequently, delta values that produce an $\Mh^{\rm NMFV}$ value
slightly outside the allowed range, see \refeqs{MHexp}, 
(\ref{Mhintr}), could be brought in agreement with
experimental data by a small change in the scenario (e.g.\ by slightly
changing the $\Xt$ parameter).

As can be seen in \reffi{fig:h0}, 
a negligible variation is found for \del{RR}{sb} in all scenarios.
An enhancement of $\Mh$ by up to $1 \gev$ is found for \del{LL}{23} and
\del{RR}{ct} once the largest considered values of \order{1} are
reached. However, whereas these are possible for \del{RR}{ct}, such
large values are excluded in the \del{LL}{23} case, as we have seen in
\refta{tab:boundsS1S6}. The only 
exception here is scenario S4, where \del{LL}{23} and \del{RR}{ct} lead
to a sizable reduction of $\Mh$ once values larger than $\pm 0.5$ are
reached. The remaining \del{LR,RL}{ij} have a larger impact on the $\Mh$
prediction. Again the corresponding trilinear couplings involved
play a relevant role here. 
Small \del{LR,RL}{ct} values lead to
an enhancement of up to $1 \gev$, and 
larger values of \order{0.1} yield a large negative contribution to
$\Mh$ (i.e.\ an effect similar to the dependence on $\Xt$ can be
observed). Consequently, bounds of \order{0.2} can be placed on
\del{LR,RL}{ct}, predicting $\Mh$ values that are 
outside the allowed range, see \refeqs{MHexp}, (\ref{Mhintr}).
Similar bounds can be derived for
\del{LR,RL}{bs}, however, these are in general weaker than the
previous bounds found from the $B$-physics observables, as can be
seen in \refta{tab:boundsS1S6}. 
The strong sensitivity to $LR$ and $RL$ parameters can be
understood due to the relevance of the $\cA_{ij}$-terms in these
Higgs mass corrections. 
It can be seen in the Feynman rules (i.e.\ see the coupling of two squarks
and one/two Higgs bosons in Appendix~A of \citere{mhNMFV}) that the
$\cA_{ij}$-terms enter directly into the couplings, and in some cases,
as in the couplings of down-type squarks to the $\cp$-odd Higgs boson,
enhanced by $\tb$. Therefore, considering the relationship between the
$\cA_{ij}$-terms and these $LR$ and $RL$ parameters, as is shown in
\refeqs{eq:SCKM-entries}, (\ref{v2Au}) and (\ref{v1Ad}), 
the strong sensitivity to these parameters can be
understood. A similar strong sensitivity to $\de^{LR}_{ct}$ in $\De\mh$
has been found in \cite{Cao1}.

The predictions for $\De\mH$ and $\De\mHp$ are shown in \reffis{fig:H0}
and \ref{fig:Hp}. In general, only \del{LR}{sb} and \del{RL}{sb} lead to
sizable effects in $\MH$ and $\MHp$, where large (negative for $\MH$,
and both negative and positive for $\MHp$) contributions 
are found for delta values exceeding $\sim 0.05$. However, since
these masses are mainly determined by the overall MSSM Higgs boson mass
scale, $\MA$, no strong conclusions (or bounds stronger than from the
$\Mh$ prediction) can be drawn. 
On the other hand, these corrections will become relevant {\em after} 
a possible discovery of these heavy Higgs bosons.

\subsection{Framework 2:}

The main goal of this part is to investigate how the upper
bounds on the deltas can be placed from the corrections induced for the
light MSSM Higgs boson mass.
In order to explore the variation of these bounds for different choices
in the MSSM parameter space, we investigate the four qualitatively
different scenarios {\bf (a)}, {\bf (b)}, {\bf (c)} and {\bf (d)}
defined in \refeqs{Sa}, 
(\ref{Sb}), (\ref{Sc}) and (\ref{Sd}), respectively. 
As explained above, the idea is to explore generic scenarios that are
compatible with present data, in particular with the measurement of
a Higgs boson mass, which we interpret as the mass of the light
$\cp$-even Higgs boson in the MSSM (for all $\deABij = 0$), and the present
experimental measurement of $(g-2)_\mu$. Taking these experimental
results into account, we have re-analyzed the full set of bounds for
the single deltas that are 
extracted from the requirement that the corrections to $\Mh$ do not
exceed $125.6 \pm 5 \gev$%
\footnote{This is, allowing for a slightly larger
interval according to our discussion after \refeq{dmh}.}
~as a function of the
two most relevant parameters in our framework 2: the generic
SUSY mass scale $\msusy$ $(\equiv \mQCD)$ and $\tb$. 
In order to find $\Mh$ around $125.6 \gev$ for $\deABij = 0$ the
scale $\mQCD$ as well as the trilinear couplings have been
chosen to sufficiently high values, see \refse{sec:f2}.
Alternatively one could choose
scenarios with a light Higgs boson mass {\em not} in agreement with the
experimental data and explore the regions of $\deABij$ that reconcile
the $\Mh$ prediction with the experimental data. However, we will not
pursue this alternative here.

\begin{figure}[ht!]
\begin{center}
\includegraphics[width=7.10cm,height=8cm]{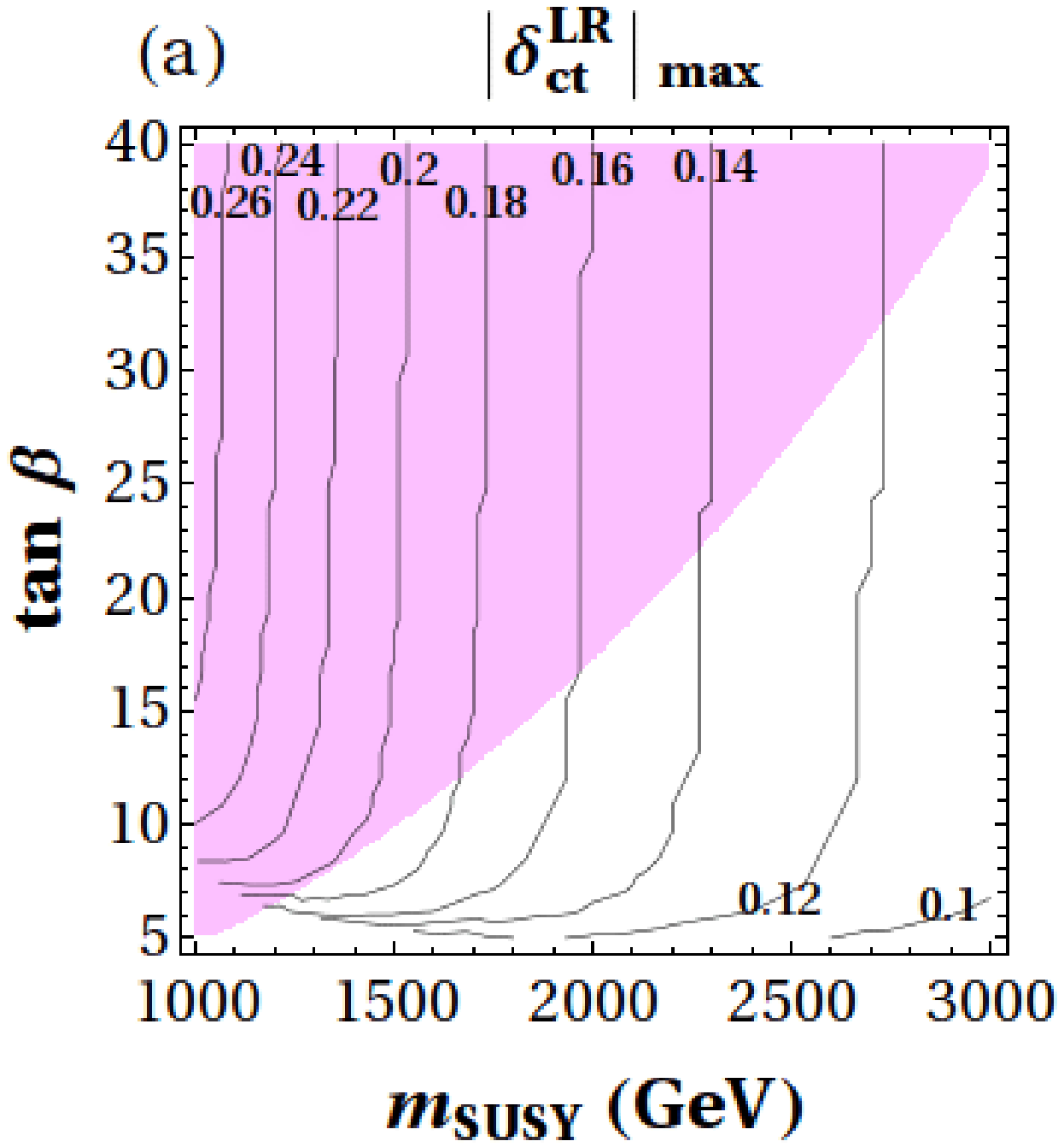}
\includegraphics[width=7.10cm,height=8cm]{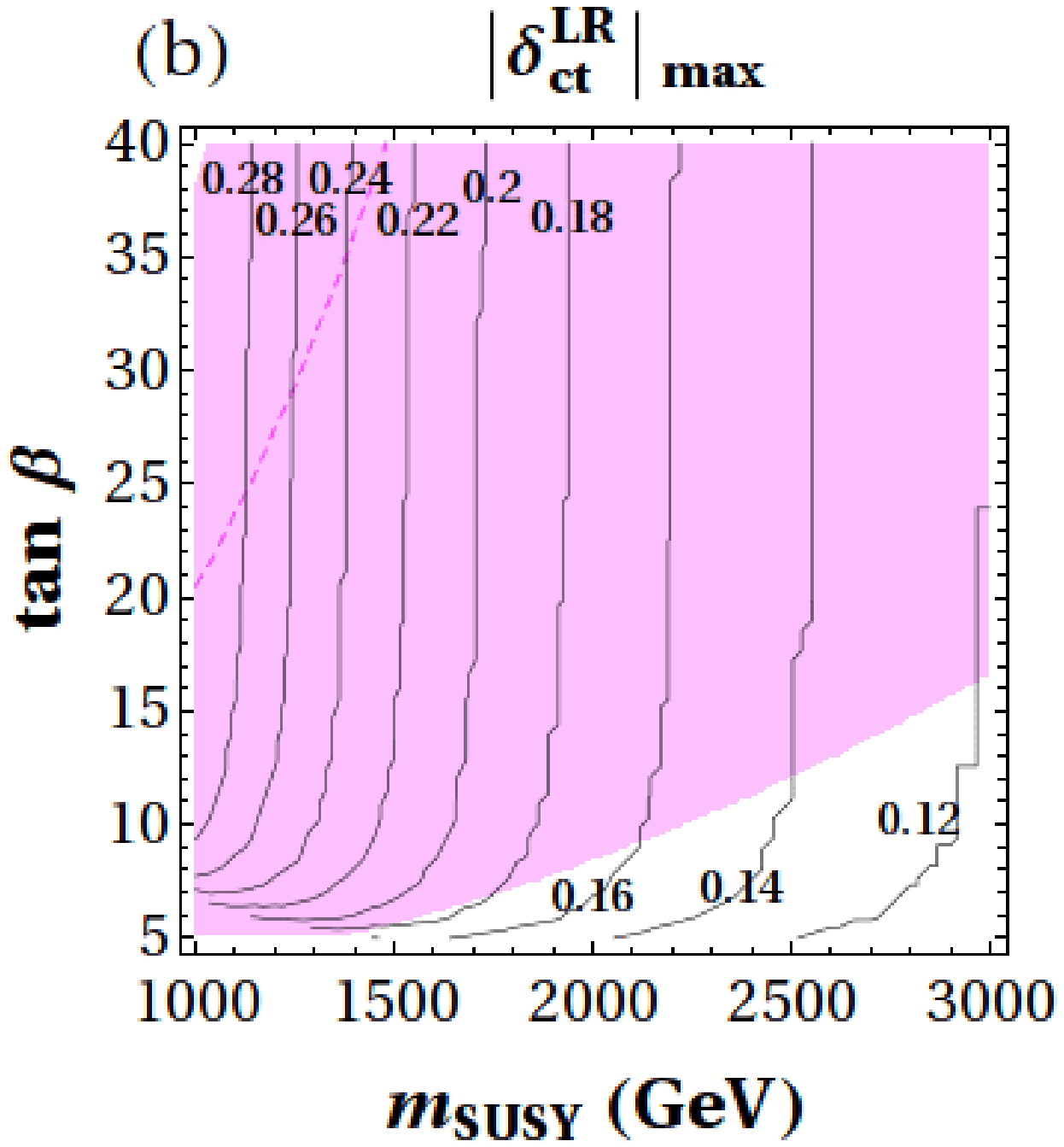}\\[3em]
\includegraphics[width=7.10cm,height=8cm]{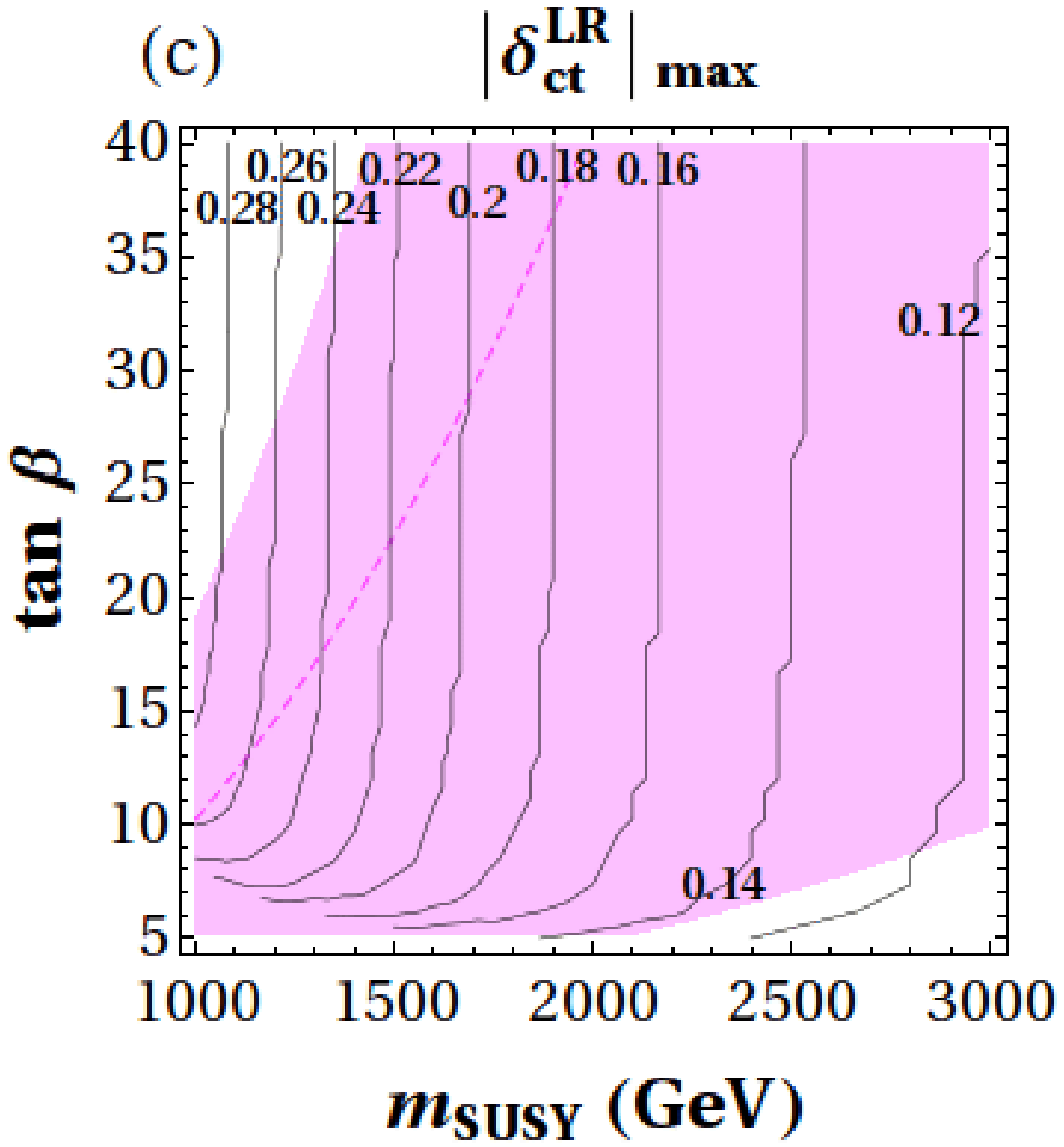}
\includegraphics[width=7.10cm,height=8cm]{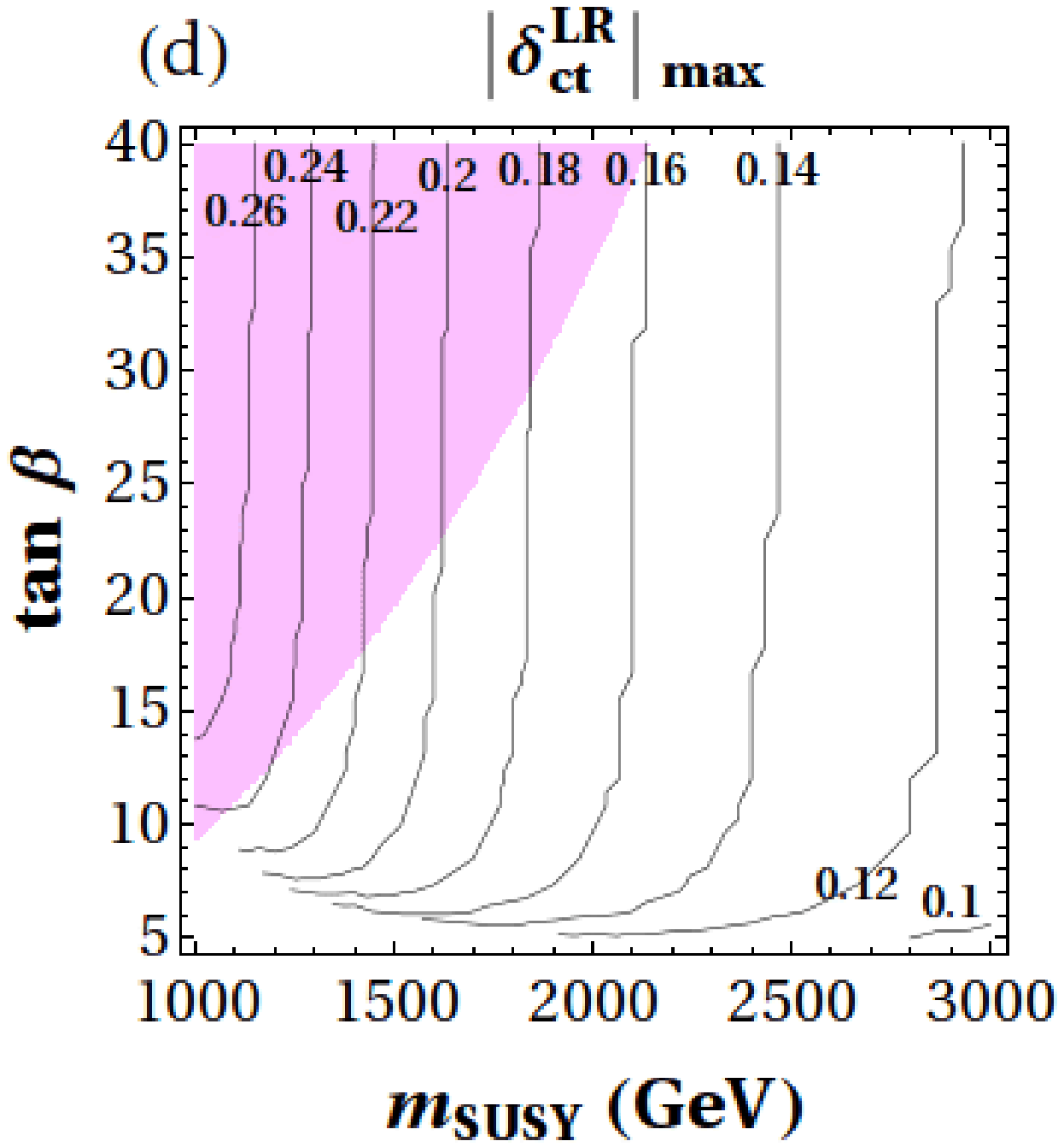}
\end{center}
\caption{ Contourlines in the 
($\msusy$, $\tb$) plane of maximum squark mixing
  $|\del{LR}{ct}|_{\rm max}$ that are allowed by the requirement that
  the correction to $\Mh$ does not exceed $\pm 5 \gev$ 
  for the scenarios {\bf (a)}, {\bf (b)}, {\bf (c)} and
  {\bf (d)} of our framework 2 The shaded (pink)
  areas are the regions leading to a $(g-2)_\mu^{\rm SUSY}$
  prediction in the $(3.2,57.2) \times 10^{-10}$ interval. The interior
  pink dashed contourline corresponds to 
  $(g-2)_\mu^{\rm SUSY}$ exactly 
  at the central value of the discrepancy 
  $(g-2)_\mu^{\rm exp}-(g-2)_\mu^{\rm SM}=30.2 \times 10^{-10}$ .
}
\label{msusytb-LRct}
\vspace{1em}
\end{figure}

We present the numerical results of our analysis in framework~2  
in \reffi{msusytb-LRct}, where we restrict ourselves to the
analysis of \del{LR}{ct} and \del{RL}{ct}, which are the only parameters showing a strong
impact on $\Mh$, apart from \del{LR}{sb} and \del{RL}{sb} that are
strongly restricted by $B$-physics observables, see the
previous subsection. Furthermore, almost identical 
results are obtained for \del{LR}{ct} and \del{RL}{ct}, and
consequently, we restrict ourselves to one of those parameters.
In each plot we show the resulting contourlines in the 
($\msusy$, $\tb$) plane of maximum allowed value of $|\del{LR}{ct}|$,
i.e.\ the ones that do not lead to contributions to $\Mh$ 
outside $125.6 \pm 5 \gev$.
The shaded areas in pink are the regions leading to a
$(g-2)_\mu^{\rm SUSY}$ prediction, from the SUSY one-loop contributions,
in the allowed interval of $(3.2,57.2) \times 10^{-10}$. The
interior pink dashed contourline corresponds to  $(g-2)_\mu^{\rm SUSY}$
exactly at the central value of the discrepancy 
$(g-2)_\mu^{\rm exp}-(g-2)_\mu^{\rm SM}=30.2 \times 10^{-10}$. 
As in the previous framework~1, we use here again
\fh~\cite{feynhiggs,mhiggsAEC}  to evaluate 
$\Mh$ and {\tt SPHENO}~\cite{Porod:2003um} to evaluate  
$(g-2)_\mu$ (where \fh\ gives very similar results). 
Due to the different relations between the SUSY-QCD and the SUSY-EW
scales in our 
four scenarios the pink shaded areas differ substantially in the four
plots. In particular in scenario {\bf (d)}, where we have set 
$\mEW := \mQCD$ only relatively small values of $\msusy$ yield a good
prediction of $(g-2)_\mu^{\rm SUSY}$.

One can observe in \reffi{msusytb-LRct} that the bounds on 
$|\del{LR}{ct}|$
depend only weakly on the chosen scenario, such that they can be
regarded as relatively general. For $\msusy \sim 1 \tev$ bounds around
$|\del{LR}{ct}| \lsim 0.28$ are found, whereas for $\msusy \sim 3 \tev$
only $|\del{LR}{ct}| \lsim 0.12$ is allowed. For most of the parameter
space the results are nearly independent of $\tb$. 
Only for $\tb \lsim 7$ 
smaller bounds for smaller $\msusy$ values are reached.
The results are consistent with previous findings, i.e.\ large SUSY mass
scales, leading to larger intergenerational mixing terms (and in
particular $A$-terms) lead to larger effects and thus to smaller allowed 
\deABij. Comparing the obtained contours, which depend on $\mQCD$, with
the $(g-2)_\mu$ preferred regions, which depend on $\mEW$,
slightly
smaller $|\del{LR}{ct}|_{\rm max}$ values as in {\bf (c)} or 
slightly larger ones as in {\bf (d)} are 
favored. However, this just reflects the choice of the hierarchy between
these two fundamental mass scales used in the respective scenario.


\section{Conclusions}
\label{sec:conclusions}
We presented an up-to-date comparison of the predictions for flavor and
Higgs observables based on NMFV parameters in the MSSM with the current
experimental data.
The flavor observables include \bsg, \bmm\ and \dmbs. In the Higgs
sector we evaluated the corrections to the light and heavy
$\cp$-even Higgs masses as well as to the charged Higgs boson mass.
Within the MSSM the calculations were performed at the full one-loop
level with the full (s)quark flavor structure, i.e.\ not relying on the
mass insertion or other approximations. 

In the first part we analyzed six representative scenarios which are in
agreement with current bounds on the SUSY and Higgs searches at the
LHC. We derived the most up-to-date bounds on $\deABij$ within
these six scenarios from flavor observables, 
thus giving an idea of the overall size of these
parameters taking the latest experimental bounds into account. 
The corresponding contributions indicate which level of higher-order
corrections are possible and allowed by the inclusion of NMFV. In
particular in the case of the light Higgs boson we find that
the prediction of $\Mh$ can lead to additional new constraints on
the deltas, specifically on \del{LR}{ct} and \del{RL}{ct}.
This is due to the fact that 
$\cA_{ij}$-terms enter directly into the couplings, 
creating a strong sensitivity to these parameters.

In the second part we analyzed four different two-dimensional scenarios,
which are characterized by universal scales for the SUSY electroweak
scale, $\mEW$, that determines the masses of the scalar leptons and
electroweak particles, and for
the SUSY QCD scale, $\mQCD$, that determines the masses of the scalar
quarks. As additional free parameter we kept $\tb$. 
Within this simplified model it is possible to analyze the
behavior of the corrections to $\Mh$, where at the same time agreement with
the anomalous magnetic moment of the muon, \gmt\ is required. 
We demanded that the correction to $\Mh$ does not yield values
outside $125.6 \pm 5 \gev$,
leading to new improved bounds on \del{LR}{ct} and \del{RL}{ct}, 
whereas no limits on the other \deABij can be obtained.
The limits on $|\del{LR}{ct}|$ turn out to be relatively
independent on the choice of the scenario. For $\mQCD \sim 1 (3) \tev$ bounds of
$|\del{LR}{ct}| \lsim 0.28 (0.12)$ were found. These bounds
on \del{LR}{ct} and \del{RL}{ct} are genuine from Higgs physics
and do not have competitive bounds from $B$-physics observables.


\subsection*{Acknowledgments}

The work of S.H.\ was supported 
by the Spanish MICINN's Consolider-Ingenio 2010
Program under grant MultiDark CSD2009-00064. 
The work of M.H.\ and M.A.-C.\ was partially supported by the
European Union FP7 ITN INVISIBLES (Marie Curie Actions, PITN- GA-2011-
289442), by the CICYT through the project FPA2012-31880,  
by the Spanish Consolider-Ingenio 2010 Programme CPAN (CSD2007-00042) 
and by the Spanish MINECO's ``Centro de Excelencia Severo Ocho''
Programme under grant SEV-2012-0249.



\end{document}